\documentclass[runningheads]{llncs}
\usepackage{graphicx}

\usepackage{tikz}
\usepackage{comment}
\usepackage{amsmath,amssymb} 
\usepackage{color}
\usepackage{graphicx}
\usepackage{amsmath}
\usepackage{amssymb}
\usepackage{booktabs}
\usepackage{scalerel}
\usepackage{comment}
\usepackage{bm}
\usepackage{tabu}
\usepackage{hyperref}
\usepackage{mathrsfs}
\usepackage{yfonts}
\usepackage[ruled,noend]{algorithm2e}

\SetAlCapNameFnt{\small}
\SetAlCapFnt{\small}

\usepackage{multirow}
\usepackage{setspace}
\usepackage{subcaption}
\usepackage[font=small, labelfont=bf]{caption}
\usepackage{wrapfig}
\usepackage[accsupp]{axessibility}  

\usepackage[width=122mm,left=12mm,paperwidth=146mm,height=193mm,top=12mm,paperheight=217mm]{geometry}

\usepackage[capitalize]{cleveref}
\crefname{section}{Sec.}{Secs.}
\Crefname{section}{Section}{Sections}
\Crefname{table}{Table}{Tables}
\crefname{table}{Tab.}{Tabs.}

\makeatletter
\DeclareRobustCommand\onedot{\futurelet\@let@token\@onedot}
\def\@onedot{\ifx\@let@token.\else.\null\fi\xspace}

\def\eg{\emph{e.g}\onedot} 
\def\ie{\emph{i.e}\onedot} 
 
\def\etc{\emph{etc}\onedot} 
\def\wrt{w.r.t\onedot} 
\def\etal{\emph{et al}\onedot}
\makeatother

\newcommand{\norm}[1]{\lVert#1\rVert}

\SetKwInOut{Input}{Input}
\SetKwInOut{Output}{Output}
\newcommand{\ccirc}{\mathbin{\mathchoice
  {\xcirc\scriptstyle}
  {\xcirc\scriptstyle}
  {\xcirc\scriptscriptstyle}
  {\xcirc\scriptscriptstyle}
}}
\newcommand{\xcirc}[1]{\vcenter{\hbox{$#1\circ$}}}

\DeclareMathOperator{\argmin}{argmin}
\DeclareMathOperator{\mymin}{min}

\begin{document}

\pagestyle{headings}
\mainmatter
\def\ECCVSubNumber{3484}  

\title{Unfolded Deep Kernel Estimation for \\ Blind Image Super-resolution} 


\titlerunning{Unfolded Deep Kernel Estimation for Blind Image Super-resolution}
%
\author{Hongyi Zheng, Hongwei Yong, Lei Zhang\thanks{Corresponding author.}}
\authorrunning{H. Zheng \etal}
%
\institute{Dept. of Computing, The Hong Kong Polytechnic University
  \email{\{cshzheng,cshyong,cslzhang\}@comp.polyu.edu.hk}}
\maketitle

\begin{abstract}
  Blind image super-resolution (BISR) aims to reconstruct a high-resolution image from its low-resolution counterpart degraded by unknown blur kernel and noise. Many deep neural network based methods have been proposed to tackle this challenging problem without considering the image degradation model. However, they largely rely on the training sets and often fail to handle images with unseen blur kernels during inference. Deep unfolding methods have also been proposed to perform BISR by utilizing the degradation model. Nonetheless, the existing deep unfolding methods cannot explicitly solve the data term of the unfolding objective function, limiting their capability in blur kernel estimation. In this work, we propose a novel unfolded deep kernel estimation (UDKE) method, which, for the first time to our best knowledge, explicitly solves the data term with high efficiency. The UDKE based BISR method can jointly learn image and kernel priors in an end-to-end manner, and it can effectively exploit the information in both training data and image degradation model. Experiments on benchmark datasets and real-world data demonstrate that the proposed UDKE method could well predict complex unseen non-Gaussian blur kernels in inference, achieving significantly better BISR performance than state-of-the-art. The source code of UDKE is available at: \url{https://github.com/natezhenghy/UDKE}.
  \keywords{blind image super-resolution, blur kernel estimation, unfolding method}
\end{abstract}

\section{Introduction}
\label{sec:intro}
Blind image super-resolution (BISR), which aims to reconstruct a high-resolution (HR) image from its low-resolution (LR) counterpart without knowing the degradation kernel and noise, is a very challenging computer vision problem~\cite{yang2019deep}. The degradation process from an HR image to an LR image can be expressed as:
\begin{equation}
  \mathbf{Y} = (\mathbf{K}\circledast\mathbf{X})\downarrow_{s} + \mathbf{n}
  \label{eq:degradation_model}
\end{equation}
where $\mathbf{X}$ is the HR image, $\mathbf{Y}$ is its observed LR counterpart, $\mathbf{K}$ is the blur kernel, $\circledast$ is the 2D convolution operator, $\downarrow_{s}$ is the downsampling operator with scaling factor $s$, and $\mathbf{n}$ is the additive white Gaussian noise.

A variety of classical methods have been proposed to tackle the BISR problem~\cite{michaeli2013nonparametric,he2009soft,shao2015simple}. The interpolation-based methods, such as bilinear and bicubic interpolation, are efficient to implement, whereas they have poor results for BISR. Model based methods employ a degradation model (\eg, Eq.~\ref{eq:degradation_model}) to constrain the fidelity between the predicted SR image and the LR input, and exploit image priors to regularize the solution. Some representative methods include Maximum a Posterior~\cite{capel2000super}, recurrence prior~\cite{glasner2009super}, \etc. Learning based methods aim to learn image priors and mappings between the LR input and HR image from the training data, \eg, dictionary learning~\cite{wang2012semi} and patch based learning~\cite{begin2004blind}.

With the rapid development of deep learning, the deep neural network (DNN) based methods have become prevalent in the research of super-resolution (SR) and shown highly competitive performance~\cite{dong2014learning}. However, most of the existing DNN based methods focus on the non-blind SR tasks, where the degradation process is simply assumed to be bicubic downsampling~\cite{zhang2018image}, or direct downsampling after blurred by fixed isotropic Gaussian kernels~\cite{zhang2018learning}. In real-world applications, however, the image degradation process is much more complex due to the unknown varying blur kernels and the corrupted noise, and these non-blind SR methods often fail. Therefore, DNN based BISR methods have been later proposed. Zhou~\etal~\cite{zhou2019kernel} analyzed blur kernels in real LR images by using the dark channel priors~\cite{pan2016blind}, and built a BISR blur kernel dataset. The KGAN (Kernel-GAN)~\cite{bell2019blind} employs a generative adversarial network (GAN), which is trained online during the inference stage, to estimate the blur kernel with some presumed priors, \eg, Gaussian prior. However, these methods rely heavily on training data and they do not consider the LR image degradation process. They often fail to handle images degraded with unseen kernels during inference.

To address the limitations of the above purely data-driven methods, some deep unfolding methods have been proposed to encode the degradation model into the learning process. With the input LR image $\mathbf{Y}$, the objective function of deep unfolding methods can be generally depicted as:
\begin{equation}
  \mymin_{\{\mathbf{K},\mathbf{X}\}}\tfrac{1}{2\sigma^2}\norm{(\mathbf{K}\circledast\mathbf{X})\downarrow_{s}-\mathbf{Y}}_{\scaleto{2}{4pt}}^{\scaleto{2}{4pt}}+\lambda_{\mathbf{X}}\bm{\psi}(\mathbf{X})+\lambda_{\mathbf{K}}\bm{\phi}(\mathbf{K})
  \label{eq:objective}
\end{equation}
where $\bm{\psi}$ and $\bm{\phi}$ represent the priors on $\mathbf{X}$ and $\mathbf{K}$, $\lambda_{\mathbf{X}}$ and $\lambda_{\mathbf{K}}$ are the balance parameters, and $\sigma$ is the noise level. Eq.~\ref{eq:objective} can be divided into two components, namely data term ($\norm{(\mathbf{K}\circledast\mathbf{X})\downarrow_{s}-\mathbf{Y}}_{\scaleto{2}{4pt}}^{\scaleto{2}{4pt}}$) and prior term ($\lambda_{\mathbf{X}}\bm{\psi}(\mathbf{X})+\lambda_{\mathbf{K}}\bm{\phi}(\mathbf{K})$). Typical deep unfolding methods include IKC~\cite{gu2019blind} and DAN~\cite{luo2020unfolding}, which employ an iterative framework to unfold Eq.~\ref{eq:objective} and perform BISR. Nevertheless, the data term is difficult to solve under the deep learning framework, and these methods do not solve the data term explicitly in kernel estimation, which limits their BISR performance (please refer to Sec.~\ref{problems} for more discussions).

In this work, we propose a novel unfolded deep kernel estimation method, namely UDKE, by explicitly solving the data term under the deep learning framework. Based on UDKE, we implement a BISR framework, which, to our best knowledge, is the first deep unfolding framework that fully unfolds the objective function in Eq.~\ref{eq:objective}. UDKE effectively and efficiently encodes the knowledge of the image degradation model into the DNN architecture and learns priors of images and blur kernels jointly in an end-to-end manner. By explicitly solving the objective function of BISR with learned priors during inference, it can efficiently estimate the unseen complex non-Gaussian blur kernels, surpassing existing kernel estimation methods by a large margin.

We extensively evaluate the proposed UDKE based BISR framework on multiple BISR benchmarks as well as real-world data. It records new state-of-the-art BISR performance, while costs only 1\% the inference time of leading online-learning based methods (\eg, DIP-FKP~\cite{liang2021flow}).

\section{Related Work}
\textbf{Traditional BISR methods.} Traditional BISR methods can be categorized into model-based methods~\cite{he2009soft,michaeli2013nonparametric,shao2015simple,liu2020blind,wang2005patch} and learning based methods~\cite{begin2004blind,corduneanu2005learning}. The former adopts an image degradation model and image priors to estimate the desired HR image. He~\etal~\cite{he2009soft} proposed a soft Maximum a Posteriori based method to alternatively perform blur kernel estimation and HR image reconstruction. Michaeli~\etal~\cite{michaeli2013nonparametric} proposed a non-parametric BISR model that exploits the inherent recurrence property of image patches. Shao~\etal~\cite{shao2015simple} employed the convolution consistency prior to estimate the blur kernel. These methods follow the constraints of the degradation model and have good interpretability; however, their BISR performance is usually limited because of the relatively weak handcrafted priors.

Equipped with a training dataset, learning based methods aim to learn from it more effective image priors and/or LR-to-HR image mappings. Begin~\etal~\cite{begin2004blind} designed a framework to estimate camera parameters from the LR image, and estimate the HR image. Corduneanu~\etal~\cite{corduneanu2005learning} proposed a spatial-variant BISR method, which learns a set of linear blur filters from the neighboring pixels. Liu~\etal~\cite{liu2020blind} developed a sparse representation based method for BISR, which utilizes the image self-similarity prior to learn an over-complete dictionary to represent the HR image. These methods, by learning from external training data, exhibit better BISR performance than model-based methods; however, their performance will drop a lot when the degradation parameters (\eg, blur kernel) of test data are much different from that of the training data.

\textbf{Direct DNN based BISR methods.}
DNN based methods have become the mainstream of BISR research, outperforming the traditional learning-based methods by a large margin. Many DNN based BISR methods directly perform BISR without performing blur kernel estimation. CinCGAN~\cite{yuan2018unsupervised} converts the LR image with unknown degradation into the bicubic degradation domain and then performs non-blind SR. Degradation GAN~\cite{bulat2018learn} learns the degradation process implicitly via a GAN to assist SR task. DASR~\cite{wang2021unsupervised} learns abstract representations of various degradations, and then adopts a DNN to perform the SR task. Kligler~\etal~\cite{bell2019blind} proposed the KGAN method, which trains a GAN on the LR image to estimate the blur kernel based on Gaussian prior and patch recurrence property. Liang~\etal~\cite{liang2021flow} enhanced KGAN with flow-based prior, which works well when the blur kernel follows Gaussian assumption. These methods do not consider the image degradation process and largely rely on the training dataset in model learning. Their performance would deteriorate when encounter unseen degradation parameters (\eg, blur kernel) in inference.

\textbf{Deep unfolding BISR methods.}
To address the limitations of direct DNN based BISR methods, a few deep unfolding BISR methods have been proposed. These methods share an iterative framework to unfold the objective function in Eq.~\ref{eq:objective}. By alternatively estimating the blur kernel and the super-resolved image, they aim to utilize the image degradation model to assist the BISR task. The early deep unfolding methods~\cite{barbu2009training,samuel2009learning} utilize the maximum a posterior framework to perform the image denoising task. Zhang~\etal~\cite{zhang2020deep} proposed a deep unfolding framework for non-blind SR by using the Half Quadratic Splitting algorithm to unfold the objective function. For BISR, Gu~\etal~\cite{gu2019blind} proposed the IKC method, which adopts a DNN to iteratively correct the blur kernel estimation in an implicit dimension-reduced space. Luo~\etal\cite{luo2020unfolding} proposed a DAN approach, which iteratively estimates the blur kernel and super-resolved image with the help of conditional residual block.

However, all the previous unfolding deep methods do not explicitly solve the data term in kernel estimation, thus they do not fully unfold the objective. This limits their capability to estimate complex unseen kernels, and they fail to address the limits of direct DNN based methods properly. Actually, their performance might be even worse than those direct DNN based methods when encounter unseen kernels during inference (\eg, IKC and DAN only work on Gaussian kernels). We propose an effective and efficient kernel estimation method by explicitly solving the data term and hence truly unfolding the whole objective function under the deep learning framework. Our proposed method can estimate more complex unseen non-Gaussian blur kernels in inference.


\section{Methodology}
\subsection{Problems of previous deep unfolding BISR methods}
\label[sec]{problems}
As described in Eq.~\ref{eq:objective}, the objective function of deep unfolding methods can be divided into the data term ($\norm{(\mathbf{K}\circledast\mathbf{X})\downarrow_{s}-\mathbf{Y}}_{\scaleto{2}{4pt}}^{\scaleto{2}{4pt}}$) and the prior term ($\lambda_{\mathbf{X}}\bm{\psi}(\mathbf{X})+\lambda_{\mathbf{K}}\bm{\phi}(\mathbf{K})$). According to~\cite{zhang2020deep}, the data term enforces physical constraints on the image degradation process, and it should be solved explicitly to enable an unfolding method to estimate unseen (different from those in training) blur kernels during inference. However, all the previous deep unfolding methods~\cite{gu2019blind,luo2020unfolding} employ DNNs to estimate the blur kernel implicitly without solving the data term explicitly. Therefore, they do not fully unfold the objective function to utilize the information embedded in the image degradation model. As a result, most of these methods simply assume Gaussian blur kernels in BISR and they have limited generalization capability to more complex non-Gaussian kernels.

The major reason that the previous methods do not explicitly unfold the data term lies in that the available solutions, which are developed in traditional unfolding methods, are hard to be incorporated into the deep learning framework. In traditional methods, the data term can be solved by either numerical methods or analytical methods. The numerical methods such as the Alternating Direction Method of Multipliers~\cite{boyd2011distributed} solve the data term iteratively. Such iterative methods work well in traditional unsupervised BISR methods; however, they are too time-consuming to use in deep learning framework, which requires training on a large amount of data. On the other hand, the analytical methods such as the Least Squares Method (LSM)\cite{abdi2007method} can provide an analytical solution of the data term. However, the image-to-column (im2col) operation required in LSM would increase the memory-overhead by thousands of times, which is not acceptable in deep learning framework. One way to waive the im2col operation is to transform the original problem into the frequency domain by the Fast Fourier Transform. Such methods have been used in the non-blind SR task~\cite{zhang2020deep}, where the blur kernel is known. In BISR, however, the support set of the unknown blur kernels is much smaller than that of images, making the explicit solution in frequency domain hard to achieve.

In this work, we investigate deeply this challenging problem, and propose an effective yet efficient method to explicitly solve the data term with minimal memory overhead under the deep learning framework.

\subsection{Unfolded deep kernel estimation based BISR framework}
\begin{figure}[!t]
  \begin{center}
    \includegraphics[width=\linewidth]{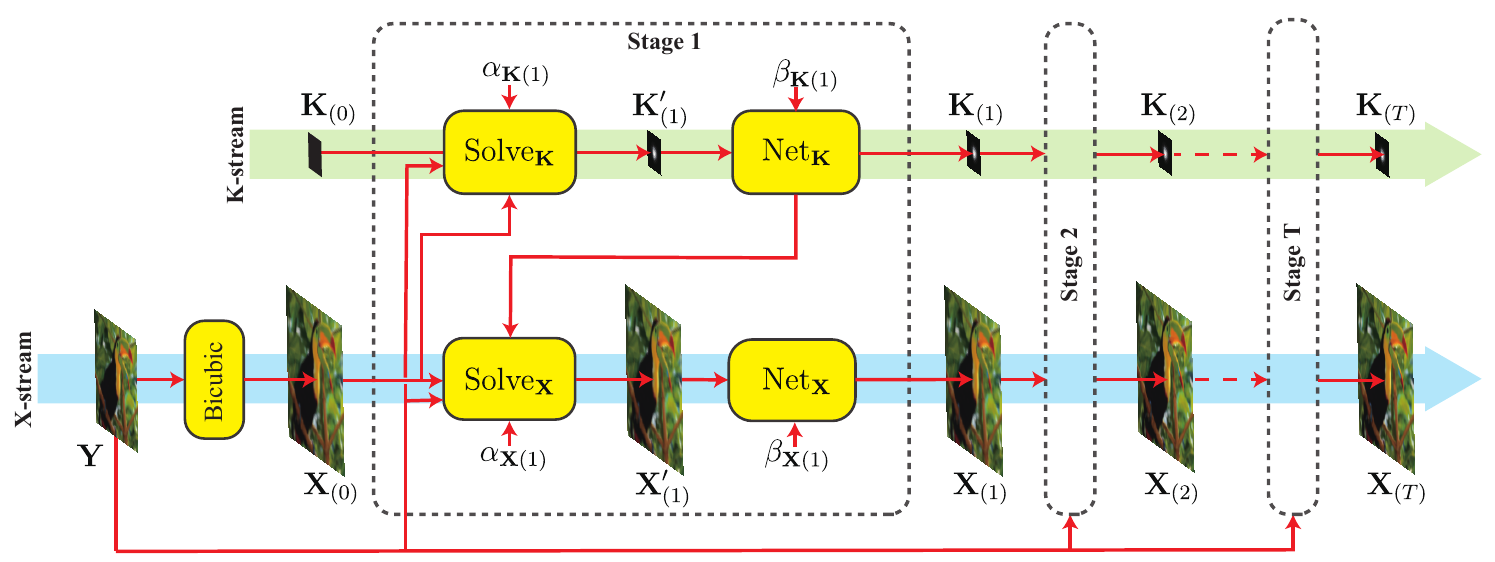}
  \end{center}
  \caption{The overall architecture of our UDKE based BISR framework.}
  \label{fig:framework}
  \vspace{-5mm}
\end{figure}

The framework of our proposed unfolded deep kernel estimation (UDKE) based BISR method is shown in Fig.~\ref{fig:framework}. Suppose we are given $N$ training triplets $\{\mathbf{Y}_i, \mathbf{Y}_i^{gt}, \mathbf{K}_i^{gt}\}$, where $\mathbf{Y}_i$ is the $i^{th}$ observed LR image, and $\mathbf{Y}_i^{gt}$ and $\mathbf{K}_i^{gt}$ are the ground-truth HR image and ground-truth blur kernel, respectively. In order to accommodate the unfolding objective in Eq.~\ref{eq:objective} into an end-to-end training framework, we rewrite it into a bi-level optimization problem as follows:
\begin{subequations}
  \begin{gather}
    \mymin_{\{\theta_{\bm{\psi}},\theta_{\bm{\phi}}\}}~\tfrac{1}{N}\textstyle \sum\nolimits_{i=1}^{N}L_{\mathbf{X}}(\mathbf{X}_i,\mathbf{Y}_i^{gt})+\gamma L_{\mathbf{K}}(\mathbf{K}_i,\mathbf{K}_i^{gt})
    \label{eq:bi-level_1}\\
    \vspace{+2mm}
    \text{s.t.}\ \{\mathbf{K}_i,\mathbf{X}_i\}=\argmin_{\mathbf{K},\mathbf{X}}\tfrac{1}{2\sigma_i^2}\norm{(\mathbf{K}\circledast \mathbf{X})\downarrow_{s}-\mathbf{Y}_i}_{\scaleto{2}{4pt}}^{\scaleto{2}{4pt}}+\lambda_{\mathbf{K}}\bm{\phi}(\mathbf{K})+\lambda_{\mathbf{X}}\bm{\psi}(\mathbf{X})\label{eq:bi-level_2}
  \end{gather}
\end{subequations}
where $\mathbf{X}_i$ and $\mathbf{K}_i$ are the predicted HR image and blur kernel; $L_{\mathbf{X}}(\cdot,\cdot)$ and $L_{\mathbf{K}}(\cdot,\cdot)$ are the loss functions, $\gamma$ is a trade-off parameter; $\theta_{\bm{\psi}}$ and $\theta_{\bm{\phi}}$ denote the parameters of deep priors (\ie, DNNs) $\bm{\psi}$ and $\bm{\phi}$.

In the above bi-level optimization problem, Eq.~\ref{eq:bi-level_1} describes its backward pass, where the parameters of deep priors $\theta_{\bm{\psi}}$ and $\theta_{\bm{\phi}}$ are updated based on the losses $L_{\mathbf{X}}$ and $L_{\mathbf{K}}$. Eq.~\ref{eq:bi-level_2} describes the forward pass of the framework, which takes the LR observation $\mathbf{Y}_i$ as input to estimate the blur kernel $\mathbf{K}_i$ and the HR image $\mathbf{X}_i$ with the learned deep priors. For the convenience of expression, we omit the subscript ``$i$'' in the following development. Eq.~\ref{eq:bi-level_2} can be split into the following two sub-problems by the Half Quadratic Splitting (HQS) algorithm:
\begin{subequations}
  \begin{align}
    \vspace{+2mm}
    \begin{split}
      \mymin_{\{\mathbf{X},\mathbf{X}^\prime\}}\tfrac{1}{2\sigma^2}\norm{(\mathbf{K}\circledast\mathbf{X}^\prime)\downarrow_{s}-\mathbf{Y}}_{\scaleto{2}{4pt}}^{\scaleto{2}{4pt}}
      +\lambda_{\mathbf{X}}\bm{\psi}(\mathbf{X})+\tfrac{\mu_{\mathbf{X}}}{2}\norm{\mathbf{X}-\mathbf{X}^\prime}_{\scaleto{2}{4pt}}^{\scaleto{2}{4pt}}
      \label{HQS_1}
    \end{split} \\
    \begin{split}
      \mymin_{\{\mathbf{K},\mathbf{K}^\prime\}}\tfrac{1}{2\sigma^2}\norm{(\mathbf{K}^\prime\circledast\mathbf{X})\downarrow_{s}-\mathbf{Y}}_{\scaleto{2}{4pt}}^{\scaleto{2}{4pt}}
      +\lambda_{\mathbf{K}}\bm{\phi}(\mathbf{K})+\tfrac{\mu_{\mathbf{K}}}{2}\norm{\mathbf{K}-\mathbf{K}^\prime}_{\scaleto{2}{4pt}}^{\scaleto{2}{4pt}}\label{HQS_2}
    \end{split}
  \end{align}
\end{subequations}
where $\mathbf{X}^\prime$ and $\mathbf{K}^\prime$ are auxiliary variables; $\mu_{\mathbf{X}}$ and $\mu_{\mathbf{K}}$ are penalty parameters.

Eqs.~(\ref{HQS_1}) and~(\ref{HQS_2}) can be solved iteratively. Particularly, in the $t$-th iteration we can solve the auxiliary variables $\mathbf{K}_{(t)}^\prime$ and  $\mathbf{X}_{(t)}^\prime$ with analytical solutions, while the two prior DNNs, denoted by $\text{Net}_\mathbf{K}$ and $\text{Net}_\mathbf{X}$, apply the learned priors to solve the variables $\mathbf{K}_{(t)}$ and $\mathbf{X}_{(t)}$:
\begin{subequations}
  \begin{alignat}{4}
    \mathbf{K}_{(t)}^\prime & =\text{Solve}_{\mathbf{K}}(\mathbf{Y},\mathbf{K}_{( t\text{-}1)},\mathbf{X}_{(t\text{-}1)},\alpha_{\mathbf{K}})\makebox[0pt][l]{}\label{eq:unfold-1}                                                                                                                                       \\
                            & =\argmin_{\mathbf{K}^\star}\tfrac{1}{2}\norm{(\mathbf{K}^\star\circledast \mathbf{X}_{(t\text{-}1)})\downarrow_{s}-\mathbf{Y}}_{\scaleto{2}{4pt}}^{\scaleto{2}{4pt}}+\tfrac{\alpha_{\mathbf{K}}}{2}\norm{\mathbf{K}^\star-\mathbf{K}_{(t\text{-}1)}}_{\scaleto{2}{4pt}}^{\scaleto{2}{4pt}}
    \notag                                                                                                                                                                                                                                                                                                               \\
    \vspace{+2mm}
    \mathbf{K}_{(t)}        & =\text{Net}_{\mathbf{K}}(\mathbf{K}_{(t)}^\prime,\beta_{\mathbf{K}})\makebox[0pt][l]{} =\argmin_{\mathbf{K}^\star}~\bm{\phi}(\mathbf{K}^\star)+\tfrac{\beta_{\mathbf{K}}}{2}\norm{\mathbf{K}_{(t)}^\prime-\mathbf{K}^\star}_{\scaleto{2}{4pt}}^{\scaleto{2}{4pt}}
    \label{eq:unfold-2}                                                                                                                                                                                                                                                                                                  \\
    \vspace{+2mm}
    \mathbf{X}_{(t)}^\prime & =\text{Solve}_{\mathbf{X}}(\mathbf{Y},\mathbf{K}_{( t)},\mathbf{X}_{(t\text{-}1)},\alpha_{\mathbf{X}})\makebox[0pt][l]{}\label{eq:unfold-3}                                                                                                                                                \\
                            & =\argmin_{\mathbf{X}^\star}\tfrac{1}{2}\norm{(\mathbf{K}_{(t)}\circledast \mathbf{X}^\star)\downarrow_{s}-\mathbf{Y}}_{\scaleto{2}{4pt}}^{\scaleto{2}{4pt}}+\tfrac{\alpha_{\mathbf{X}}}{2}\norm{\mathbf{X}^\star-\mathbf{X}_{(t\text{-}1)}}_{\scaleto{2}{4pt}}^{\scaleto{2}{4pt}}
    \notag                                                                                                                                                                                                                                                                                                               \\
    \vspace{+2mm}
    \mathbf{X}_{(t)}        & =\text{Net}_{\mathbf{X}}(\mathbf{X}_{(t)}^\prime,\beta_{\mathbf{X}})=\argmin_{\mathbf{X}^\star}~\bm{\psi}(\mathbf{X}^\star)+\tfrac{\beta_{\mathbf{X}}}{2}\norm{\mathbf{X}_{(t)}^\prime-\mathbf{X}^\star}_{\scaleto{2}{4pt}}^{\scaleto{2}{4pt}}
    \label{eq:unfold-4}
  \end{alignat}
  \label{HQS3}%
\end{subequations}
where $\{\alpha_{\mathbf{K}},\alpha_{\mathbf{X}},,\beta_{\mathbf{K}}\beta_{\mathbf{X}}\}=\{\mu_{\mathbf{K}}\sigma^2,\mu_{\mathbf{X}}\sigma^2,\tfrac{\mu_{\mathbf{K}}}{\lambda_{\mathbf{K}}},\tfrac{\mu_{\mathbf{X}}}{\lambda_{\mathbf{X}}}\}$.

\vspace{2pt}

The architecture of our UKDE based BISR framework is built from the unfolded equations Eqs.~(\ref{eq:unfold-1})$\sim$(\ref{eq:unfold-4}). It has two branches. The kernel estimation branch corresponds to Eqs.~\ref{eq:unfold-1}$\sim$\ref{eq:unfold-2} and is represented as the K-stream in Fig.~\ref{fig:framework}, which will be discussed in Section.~\ref{sec:solve-k}. Eqs.~\ref{eq:unfold-3}$\sim$\ref{eq:unfold-4} super-resolve the HR image with the estimated kernel, and they are represented as the X-stream in Fig.~\ref{fig:framework}, which will be discussed in Section~\ref{sec:solve-x}.

\subsection{K-stream: unfolded explicit kernel estimation}
\label{sec:solve-k}
In this section, we elaborate in detail the proposed novel kernel estimation method, which corresponds to the K-stream in Fig.~\ref{fig:framework}. The first step is to explicitly solve Eq.~\ref{eq:unfold-1} (data term), which is represented as $\text{Solve}_{\mathbf{K}}$ in Fig.~\ref{fig:framework}. It takes $\mathbf{Y}$ and the estimations of $\mathbf{K}$ and $\mathbf{X}$ in previous stage as inputs, then explicitly solves Eq.~\ref{eq:unfold-1} to update $\mathbf{K}^\prime$. As the dimension of $\mathbf{K}$ is much lower than that of $\mathbf{X}$, this system is over-determined and can be solved by the LSM method~\cite{abdi2007method}. Denote by $\bm{U}_a$ the im2col operator with block size $a$, and by $\bm{P}_{\frac{a-1}{2}}$ the circular padding operator with padding size of ${\frac{a-1}{2}}$. Let $\mathfrak{X} = \bm{U}_k(\bm{P}_{\frac{a-1}{2}}(\mathbf{X}))$ and $\mathfrak{Y} = \bm{U}_k(\mathbf{Y})$,
then Eq.~\ref{eq:unfold-1} can be written as:
\begin{equation}
  \argmin_{\mathbf{k}^\star}~\tfrac{1}{2}\norm{\mathfrak{M}_{s}\mathfrak{X}\mathbf{k}-\mathfrak{Y}}_{\scaleto{2}{4pt}}^{\scaleto{2}{4pt}}+\tfrac{\alpha_{\mathbf{K}}}{2}\norm{\mathbf{k}^\star-\mathbf{k}}_{\scaleto{2}{4pt}}^{\scaleto{2}{4pt}}
  \label{solveD_unfold}
\end{equation}
where $\mathbf{k}^\star=\text{vec}(\mathbf{K}^\star)$, $\mathbf{k}=\text{vec}(\mathbf{K})$, $\text{vec}(\cdot)$ is the vectorization operator, and $\mathfrak{M}_s$ is the matrix representation of the downsampling operator $\downarrow_{s}$ with scale factor $s$.

By taking the derivative of Eq.~\ref{solveD_unfold} \wrt. $\mathbf{k}^\star$ and letting the derivative be zero, we can obtain the closed-form solution of $\mathbf{K^\prime}$ as follows:
\begin{equation}
  \mathbf{K}^\prime=\text{vec}^{-1}\{(\mathfrak{X}^T\mathfrak{M}_s^T\mathfrak{M}_s\mathfrak{X}+\alpha_{\mathbf{K}}\mathbf{I})^{-1}(\mathfrak{X}^T\mathfrak{M}_s^T\mathfrak{Y}+\alpha_{\mathbf{K}}\mathbf{k})\}
  \label{eq:unfold_solve}
\end{equation}
where $\text{vec}^{-1}(\cdot)$ reverses the vectorization operator $\text{vec}(\cdot)$. However, $\mathfrak{X}$ has a size of $C\times h \times w \times k \times k$, which is much larger than $\mathbf{X}$ of size $C \times h \times w$, where $C$, $h$, $w$, and $k$ are channel number, height, width and blur kernel size of the super-resolved image. It is too memory-consuming to directly compute Eq.~\ref{eq:unfold_solve} in practice. In the rest of this section, we will elaborate the proposed memory-efficient solution to tackle this problem.

It can be seen that the size of $\mathfrak{X}^T\mathfrak{M}_s^T\mathfrak{M}_s\mathfrak{X}$ is $C\times k \times k \times k \times k$  and $k\ll h$, $k\ll w$. We propose an efficient solution to calculate $\mathfrak{X}^T\mathfrak{M}_s^T\mathfrak{M}_s\mathfrak{X}$ from $\mathbf{X}$ without storing $\mathfrak{X}$ or $\mathfrak{Y}$, reducing significantly the memory consumption. (Note that $\mathfrak{X}^T\mathfrak{M}_s^T\mathfrak{Y}$ can be regarded as a special case of $\mathfrak{X}^T\mathfrak{M}_s^T\mathfrak{M}_s\mathfrak{X}$ , where $\mathfrak{Y}=\mathfrak{M}_s\mathfrak{X}$). Denote by $\bm{U}_k$ the im2col operator with block size $k$, and by $\bm{P}_{\frac{k-1}{2}}$ the circular padding operator with padding size ${\frac{k-1}{2}}$. The element at $(x,y)$ in $\mathfrak{X}^T\mathfrak{M}_s^T\mathfrak{M}_s\mathfrak{X}$ can be calculated through dilated convolution between the $x^{th}$ and the $y^{th}$ im2col blocks in $\bm{U}_{k}\circ\bm{P}_{\frac{k-1}{2}}(\mathbf{X})$, where $\circ$ is the notation of function composition.

Unfortunately, calculating $h\times w$ elements in $\mathfrak{X}^T\mathfrak{M}_s^T\mathfrak{M}_s\mathfrak{X}$ requires $h\times w$ convolution operations, and each of them is based on a unique pair of kernel and feature maps, which is too time-consuming. Thus, we have to convert them into parallel operations to utilize the modern parallel computing library such as CUDA. This can be done by convolving $\bm{P}_{k-1}(\mathbf{X})$ with a series of dilated feature maps, each has a unique dilation pattern. Generally speaking, with the scale factor $s$, there are $s^2$ dilation patterns. We use 2-dimensional indices to arrange these dilated feature maps, denoted by $\hat{\mathbf{X}}^{(i,j)}$, by using the following rule:
\begin{equation}
  \hat{\mathbf{X}}^{(i,j)}:\begin{cases}
    \vspace{+2mm}
    \hat{\mathbf{X}}^{(i,j)}_{(x,y)}=\mathbf{X}_{(x,y)} & x\%i=0 \ \& \ y\%j=0 \\
    \hat{\mathbf{X}}^{(i,j)}_{(x,y)}=0                  & otherwise
  \end{cases}
  \label{eq:cases}
\end{equation}
where $\%$ is the modulo operator, $i=\{0,1...s-1\}$ and $j=\{0,1...s-1\}$. Convolution operations between $\bm{P}_{k-1}(\mathbf{X})$ and $\hat{\mathbf{X}}^{(i,j)}$ result in $s^2$ feature maps. Then we merge them into a single feature map, denoted by $\mathbf{F}$, with the help of a mapping function $\bm{f}$. $\bm{f}$ and $\mathbf{F}$ are defined as follows:
\begin{subequations}
  \begin{gather}
    \begin{split}
      \bm{f}:(x,y)\xrightarrow[]{}((\lfloor\frac{k-1}{2}\rfloor-x) \% s,(\lfloor\frac{k-1}{2}\rfloor-y) \%s)
    \end{split}\\
    \begin{split}
      \mathbf{F}: \mathbf{F}_{(x, y)}=(\hat{\mathbf{X}}^{\bm{f}(x,y)}\circledast\bm{P}_{k-1}(\mathbf{X}))_{(x,y)}
    \end{split}
  \end{gather}
  \label{eq:merge}
  \vspace{-5mm}
\end{subequations}

Eq.~\ref{eq:merge} costs $s^2$ operations to solve, which is still time-consuming when $s$ is large. With the help of pixel-shuffle operation, Eq.~\ref{eq:merge} can be further reduced into a constant number of operations as follows:
\begin{subequations}
  \begin{gather}
    \bm{g}:(x,y)\xrightarrow[]{}x\times s + y\\
    \vspace{+2mm}
    \mathbf{F}=\bm{S}_s^{\text{-}1}\circ\bm{M}_{\bm{g}\circ \bm{f}\circ \bm{g}^{-1}}\{\bm{S}_s(\mathbf{X})\circledast\bm{S}_s\circ\bm{P}_{k\text{-}1}(\mathbf{X})\}\label{eq:memory-efficient}
  \end{gather}
\end{subequations}
where $\bm{g}$ is a mapping function, $\bm{S}_s$ and $\bm{S}_s^{-1}$ are the pixel shuffle/un-shuffle operations with scale factor $s$, and $\bm{M}$ reorders channels of a matrix according to mapping ${\bm{g}\circ \bm{f}\circ \bm{g}^{-1}}$. The elements in the $x^{th}$ row of $\mathfrak{X}^T\mathfrak{M}_s^T\mathfrak{M}_s\mathfrak{X}$ will reside in the $x^{th}$ im2col block of $\bm{U}_{k}(\mathbf{F})$. Finally, $\mathfrak{X}^T\mathfrak{M}_s^T\mathfrak{M}_s\mathfrak{X}$ can be computed as:
\begin{equation}
  \mathfrak{X}^T\mathfrak{M}_s^T\mathfrak{M}_s\mathfrak{X}=\bm{R}\circ\bm{U}_k(\mathbf{F})
  \label{complete}
\end{equation}
where $\bm{R}(\mathbf{A})$ flips each row of matrix $\mathbf{A}$.

The memory-efficient solution described in Eqs.~\ref{eq:cases}$\sim$\ref{complete} can reduce the memory consumption of solving blur kernels by a factor of $\frac{h\times w}{k\times k}$. For example, to super-resolve an image to 2K resolution ($2048\times1024$) with $k=11$, it can save over $17000\times$ memory overhead.

The second step in the K-stream is to solve Eq.~\ref{eq:unfold-2} to apply deep priors on $\mathbf{K^\prime}$, which is done by a DNN ($\text{Net}_\mathbf{K}$ in Fig.~\ref{fig:framework}). The $\text{Net}_\mathbf{K}$ consists of 3 blocks, each of which is composed of two Convolutional (Conv) layers and one LeakyReLU layer. All Conv layers have 16 channels and all LeakyReLU layers have a negative slope of 0.01. A trailing ReLU layer is added to restrict the output estimation to be positive. The architecture of $\text{Net}_\mathbf{K}$ is illustrated in Fig.~\ref{fig:network}.

\subsection{X-stream: super-resolved image estimation}
\label{sec:solve-x}
The X-stream solves Eqs.~\ref{eq:unfold-3}$\sim$\ref{eq:unfold-4} to estimate the super-resolved image. Given the blur kernel estimated by UDKE, it reduces into a non-blind SR problem, which can be easily done in two steps. The first step takes $\mathbf{Y}$ and the estimations of $\mathbf{K}$ and $\mathbf{X}$ as inputs, then solves Eq.~\ref{eq:unfold-3} to update $\mathbf{X}^\prime$. According to~\cite{zhang2020deep}, the closed-form solution of $\mathbf{X}^\prime$ in Eq.~\ref{eq:unfold-3} can be derived by the Fast Fourier Transform~(FFT):
\begin{equation}
  \begin{split}
    \mathbf{X}^\prime=\tfrac{1}{\alpha_{\mathbf{X}}}\mathscr{F}^{-1}\{\mathcal{Z}-\mathcal{K}\odot(\frac{(\bar{\mathcal{K}\ }\odot \mathcal{Z})}{\alpha_{\mathbf{X}}+(\bar{\mathcal{K}\ }\odot\mathcal{K})})\}
  \end{split}
\end{equation}
where $\mathscr{F}(\cdot)$ and $\mathscr{F}^{-1}(\cdot)$ denote the 2D FFT and its inverse, $\mathcal{K} =\ \mathscr{F}(\mathbf{K})$, $\mathcal{X}^\star =\ \mathscr{F}(\mathbf{X}^\star)$, $\mathcal{Y} =\ \mathscr{F}(\mathbf{Y})$, $\mathcal{X} =\ \mathscr{F}(\mathbf{X})$, $\mathcal{Z}=\mathcal{K}\ccirc\mathcal{Y}+\alpha_{\mathbf{X}}\mathcal{X}$, $\bar{\mathcal{K}\ }$ is the complex conjugate of $\mathcal{K}$, $\odot$ and $\tfrac{\ \cdot \ }{\ \cdot \ }$ are the 2D Hadamard product and division, respectively.

\begin{figure}[!h]
  \captionsetup{font=small}
  \begin{center}
    \includegraphics[width=0.6\linewidth]{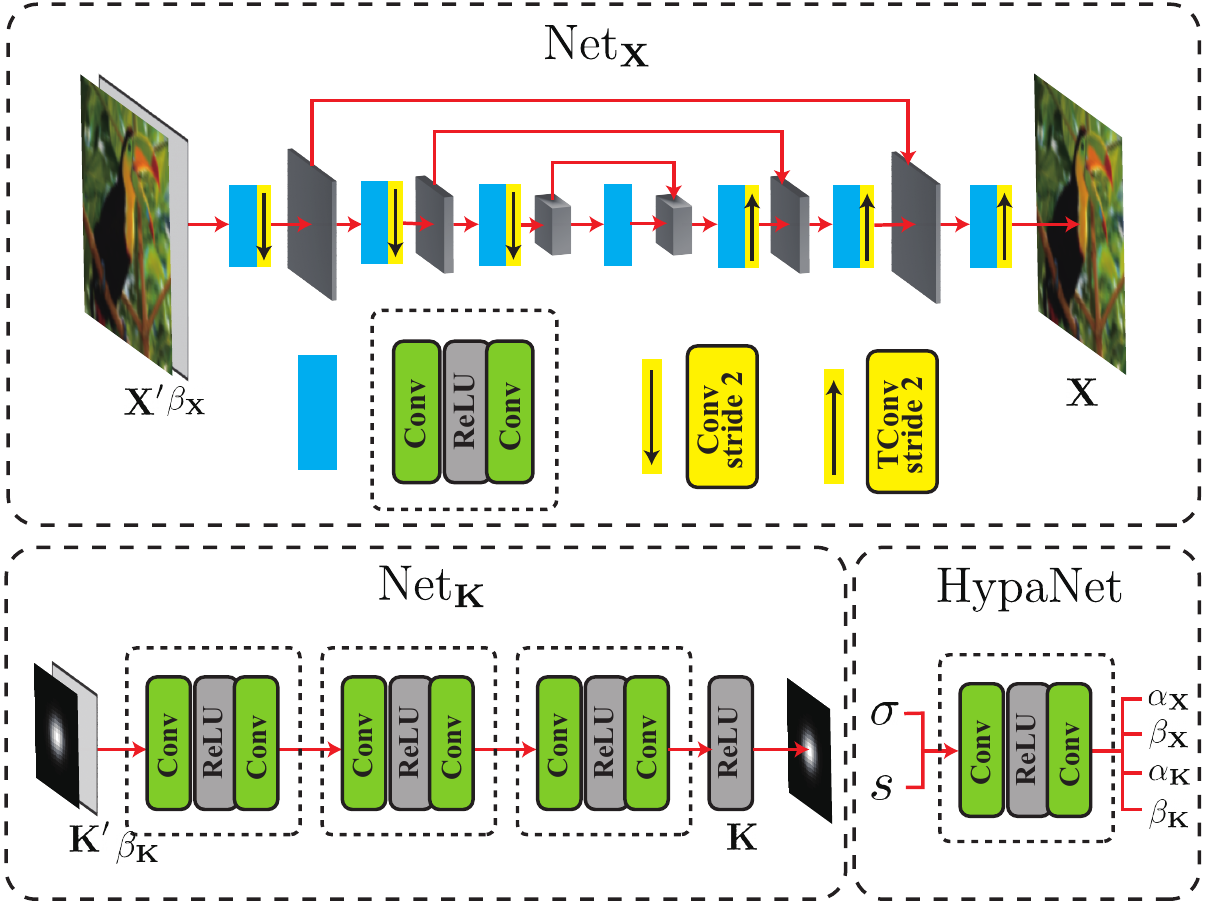}
  \end{center}
  \caption{Network architectures of $\text{Net}_{\mathbf{X}}$, $\text{Net}_{\mathbf{K}}$ and HypaNet.}
  \label{fig:network}
\end{figure}

\begin{figure}[!t]
  \begin{algorithm}[H]\small
    \caption{Overall unfolding process of our UDKE based BISR framework}\label{algorithm}
    \Input{LR image $\mathbf{Y}$, stages no. $T$, kernel size $k$, scale factor $s$, noise level $\sigma$}
    \Output{Predicted HR image $\mathbf{X}^{pred}$, predicted blur kernel $\mathbf{K}^{pred}$}
    $\mathbf{X}_0=\text{bic}_s\left(\mathbf{Y}\right)$, $\mathbf{K}_0={\frac{1}{k^2}}$\;
    \For{$t = 1,...,T$}
    {
    $\left\{\alpha_{\mathbf{X}\left(t\right)},\alpha_{\mathbf{K}\left(t\right)}, \beta_{\mathbf{X}\left(t\right)}, \beta_{\mathbf{K}\left(t\right)}\right\}=\text{HyperNet}_{\left(t\right)}\left(s,\sigma\right)$\;
    $\mathbf{K}_{\left(t\right)}^\prime=\text{Solve}_{\mathbf{K}}\left(\mathbf{Y},\mathbf{K}_{\left( t\text{-}1\right)},\mathbf{X}_{\left(t\text{-}1\right)},\alpha_{\mathbf{K}}\left(t\right)\right)$\;
    $\mathbf{K}_{\left(t\right)}=\text{Net}_{\mathbf{K}}\left(\mathbf{K}_{\left(t\right)}^\prime,\beta_{\mathbf{K}\left(t\right)}\right)$\;
    $\mathbf{X}_{\left(t\right)}^\prime=\text{Solve}_{\mathbf{X}}\left(\mathbf{Y},\mathbf{K}_{\left( t\right)},\mathbf{X}_{\left( t\text{-}1\right)},\alpha_{\mathbf{X}\left(t\right)}\right)$\;
    $\mathbf{X}_{\left(t\right)}=\text{Net}_{\mathbf{X}}\left(\mathbf{X}_{\left(t\right)}^\prime,\beta_{\mathbf{X}\left(t\right)}\right)$\;
    }
    $\mathbf{X}^{pred}=\mathbf{X}_T$\;
    $\mathbf{K}^{pred}=\mathbf{K}_T$\;
    \label{alg1}
  \end{algorithm}
  \vspace{-5mm}
\end{figure}

The second step solves Eq.~\ref{eq:unfold-4} and applies deep priors on $\mathbf{X^\prime}$, which is done by a DNN $\text{Net}_\mathbf{X}$. We adopt the U-Net~\cite{ronneberger2015u} as the DNN by following the SOTA non-blind SR method~\cite{zhang2020deep}. $\text{Net}_\mathbf{X}$ consists of 7 blocks. The first 3 blocks downsample the feature maps through strided convolution, and the last 3 blocks upsample the feature maps by transposed convolution. Each block consists of 4 residual units, while each of them consists of 2 Conv layers with ReLU and a skip connection. The channel numbers of Conv layers in the first 4 blocks are 64, 128, 256, 512, respectively. The architecture of $\text{Net}_\mathbf{X}$ is illustrated in Fig.~\ref{fig:network}.

\subsection{Summary of the unfolding process}
\label{sec:overall}
The K-stream and X-stream work alternatively to estimate the blur kernel $\mathbf{K}$ and HR image $\mathbf{X}$ for $T$ stages. The determination of $T$ will be discussed in Section~\ref{sec:implementation}. In the first stage, the input $\mathbf{X}_0$ is initialized by $\text{bic}_s(\mathbf{Y})$, where $\text{bic}_s$ is the bicubic upsampling operation with scale factor $s$, while all the elements of $\mathbf{K}_0$ are initialized to $\frac{1}{k^2}$. For each stage, a tiny 2-layer fully connected network, called HypaNet, which takes $\sigma$ and $s$ as inputs, is introduced to predict the hyperparameters. The architecture of HypaNet is illustrated in Fig.~\ref{fig:network}. Algorithm~\ref{alg1} depicts the overall unfolding process.

\begin{figure*}[!t]
  \begin{center}
    \includegraphics[width=0.6\linewidth]{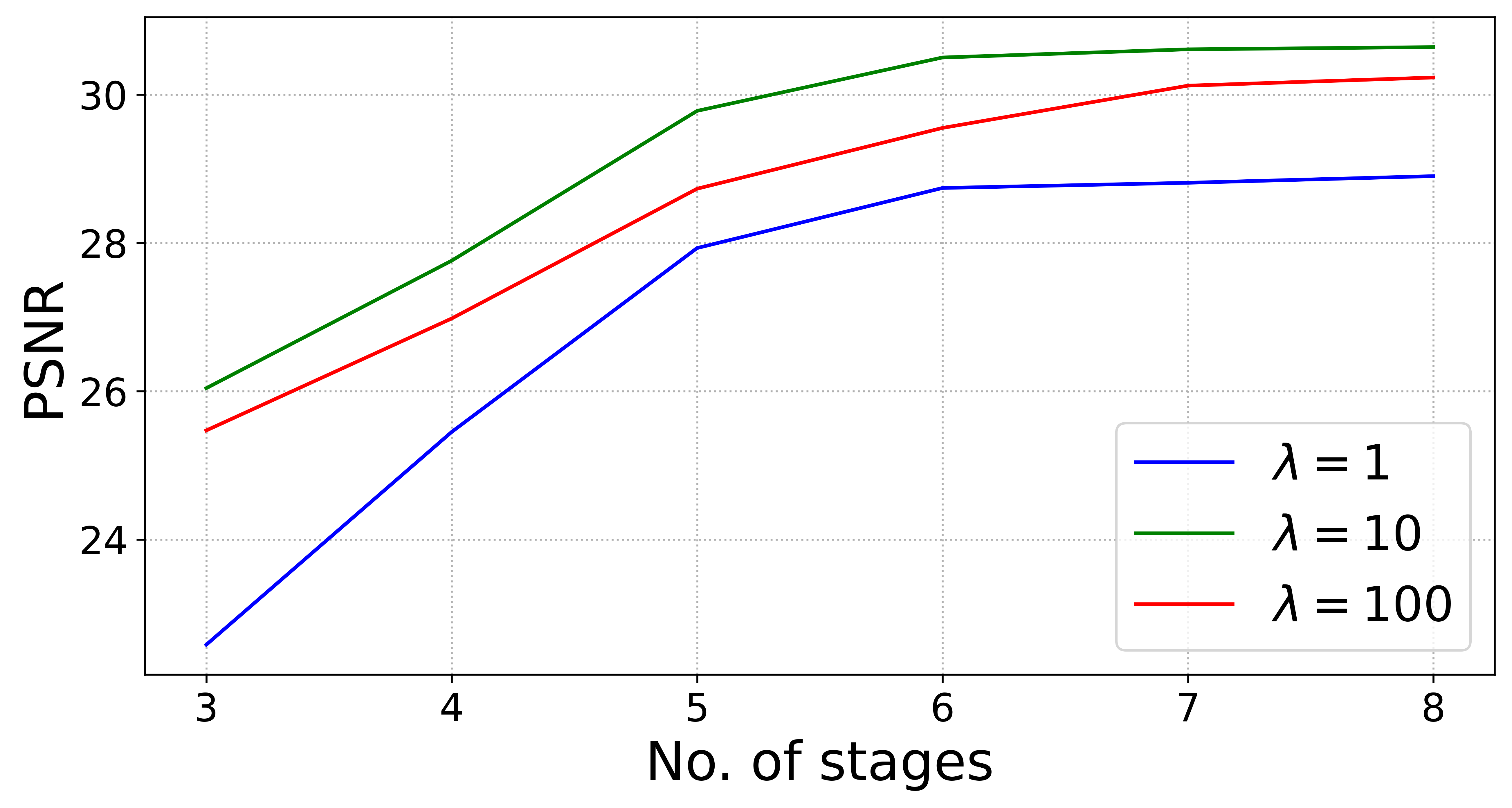}
  \end{center}
  \caption{Ablation study on $\lambda$ and $T$.}
  \vspace{-5mm}
  \label{fig:ablation}
\end{figure*}

\section{Experiments}
\subsection{Implementation details}
\label{sec:implementation}

\textbf{Training data and kernel pool.} Following previous BISR works~\cite{zhang2020deep,luo2020unfolding,wang2021unsupervised}, we use the DIV2K~\cite{Agustsson_2017_CVPR_Workshops} and Flickr2K~\cite{lim2017enhanced} datasets to train our UDKE based framework. The ground truth images $\mathbf{Y}^{gt}$ are obtained by randomly cropping patches of size $128\times 128$ from the original images, and the LR images $\mathbf{Y}$ are obtained by randomly selecting kernels from the DEPD-training kernel pool~\cite{zhou2019kernel} and applying the degradation model in Eq.~\ref{eq:degradation_model}. The DEPD kernel pool adopts dark channel priors~\cite{pan2016blind} to analyze BISR kernels from real-world LR images captured by low-end camera phones. Specifically, the DEPD-training subset consists of 1,000 BISR kernels analyzed from photos captured by Blackberry Passport phone and Sony Xperia Z. The DEPD-evaluation subset consists of 300 BISR kernels analyzed from photos captured by iPhone 3GS.

\textbf{Training details.}
The $L_1$ loss is used as the loss function for each stage of UDKE. The weight on the loss of the last stage is set as $1$, and the weights on all the other stages are set as $\tfrac{1}{T-1}$. The Adam optimizer~\cite{kingma2014adam} is used for updating the parameters of $\text{Net}_{\mathbf{K}}$ and $\text{Net}_{\mathbf{X}}$. The batch size is 32. We train the network for $10^5$ iterations. The learning rate starts from $10^{\text{-}4}$ and decays by a factor of 0.5 for every $2^4$ iterations. In order to speed up and stabilize the training process, we first train a 1-stage model and reload its weights into the $T$-stage model for fine-tuning. All the $T$ stages share the same parameters. The size $k$ for BISR kernel is set to 11 and the scale factor $s$ is set to $\{2,3,4\}$, following previous BISR methods~\cite{zhang2020deep,luo2020unfolding,wang2021unsupervised}. For each scale factor, we train a model for all noise levels, which are set to $\{0,2.55,7.65\}$ as in~\cite{zhang2020deep}.

\begin{table*}[!t]\tiny
  \captionsetup{font=small}
  \caption{2$\times$ BISR results (PSNR(dB)/SSIM). Best results are in red.}
  \setlength\tabcolsep{1pt}
  \centering
  \begin{tabu}{c|X[l]X[l]X[l]|X[l]X[l]X[l]|X[l]X[l]X[l]}
    \toprule

    Datasets                       & \multicolumn{3}{c|}{Set5}      & \multicolumn{3}{c|}{BSD100}    & \multicolumn{3}{c}{Urban100}                                                                                                                                                                                          \\
    $\sigma$                       & 0                              & 2.55                           & 7.65                           & 0                              & 2.55                           & 7.65                           & 0                              & 2.55                           & 7.65            \\
    \hline\hline
    Bicubic                        &
    26.56/\!.8196                  & 26.53/.\!8111                  & 25.87/.\!7569                  & 24.95/.\!7071                  & 24.79/.\!6402                  & 24.49/.\!6402                  & 22.18/.\!7032                  & 22.07/.\!6895                  & 21.86/.\!6327                                    \\
    \toprule
    RCAN                           &
    27.35/.\!8578                  & 27.04/.\!8153                  & 24.73/.\!6113                  & 25.70/.\!7620                  & 25.29/.\!7170                  & 23.69/.\!5345                  & 23.14/.\!7650                  & 22.84/.\!7192                  & 21.75/.\!5442                                    \\
    \toprule
    ZSSR                           & 27.03/.\!8803                  & 26.53/.\!8143                  & 25.41/.\!5840                  & 25.31/.\!7498                  & 25.10/.\!7122                  & 23.29/.\!5449                  & 22.66/.\!7443                  & 22.34/.\!7121                  & 21.89/.\!5579   \\
    \midrule
    DASR                           &
    27.48/.\!8788                  & 27.00/.\!8188                  & 24.95/.\!6310                  & 25.82/.\!7624                  & 25.27/.\!7143                  & 23.81/.\!5442                  & 23.37/.\!7803                  & 22.80/.\!7181                  & 21.86/.\!5611                                    \\
    \midrule
    KGAN                           &
    25.70/.\!8448                  & 24.56/.\!6748                  & 19.50/.\!3765                  & 24.02/.\!7540                  & 22.67/.\!5776                  & 18.56/.\!3149                  & 21.85/.\!7575                  & 21.14/.\!6007                  & 18.02/.\!3706                                    \\
    \midrule
    KFKP                           &
    17.28/.\!5279                  & 14.55/.\!2766                  & 12.25/.\!1829                  & 21.16/.\!6608                  & 19.79/.\!4934                  & 16.27/.\!2890                  & 19.56/.\!6725                  & 18.61/.\!5361                  & 16.02/.\!3260                                    \\
    \midrule
    DFKP                           &
    27.42/.\!8795                  & 26.16/.\!7633                  & 21.44/.\!4464                  & 25.60/.\!7968                  & 24.79/.\!6925                  & 19.99/.\!3687                  & 22.91/.\!7860                  & 22.63/.\!7155                  & 19.76/.\!4414                                    \\
    \toprule
    IKC                            & 27.02/.\!8807                  & 26.36/.\!8095                  & 24.81/.\!6199                  & 25.42/.\!8029                  & 25.26/.\!7141                  & 23.75/.\!5412                  & 23.28/.\!8030                  & 22.86/.\!7168                  & 21.79/.\!55\.39 \\
    \midrule
    DAN                            &
    27.40/.\!8705                  & 27.01/.\!8163                  & 24.93/.\!6277                  & 25.70/.\!7620                  & 25.29/.\!7179                  & 23.79/.\!5438                  & 23.14/.\!7647                  & 22.84/.\!7201                  & 21.79/.\!5500                                    \\
    \midrule
    Ours                           &
    \textcolor{red}{30.50/.\!8964} & \textcolor{red}{30.11/.\!8786} & \textcolor{red}{28.93/.\!8416} & \textcolor{red}{27.68/.\!8282} & \textcolor{red}{27.29/.\!8017} & \textcolor{red}{26.46/.\!7516} & \textcolor{red}{25.51/.\!8341} & \textcolor{red}{25.19/.\!8138} & \textcolor{red}{24.57/.\!7748}                   \\
    \toprule
    UBound                         &
    34.65/.\!9424                  & 33.01/.\!9145                  & 31.07/.\!9004                  & 30.02/.\!8840                  & 29.11/.\!8754                  & 27.94/.\!8433                  & 28.01/.\!8468                  & 27.45/.\!8371                  & 25.87/.\!8210                                    \\

    \bottomrule
  \end{tabu}
  \label{table:x2}
  \vspace{-5mm}
\end{table*}

\begin{table*}[!t]\tiny
  \captionsetup{font=small}
  \caption{3$\times$ BISR results (PSNR(dB)/SSIM). Best results are in red.}
  \setlength\tabcolsep{1pt}
  \centering
  \begin{tabu}{c|X[l]X[l]X[l]|X[l]X[l]X[l]|X[l]X[l]X[l]}
    \toprule
    Datasets                       & \multicolumn{3}{c|}{Set5}      & \multicolumn{3}{c|}{BSD100}    & \multicolumn{3}{c}{Urban100}                                                                                                                                                                        \\
    $\sigma$                       &
    0                              & 2.55                           & 7.65                           & 0                              & 2.55                           & 7.65                           & 0                              & 2.55                           & 7.65                           \\
    \hline\hline
    Bicubic                        &
    24.56/.\!7570                  & 24.51/.\!7463                  & 24.12/.\!6819                  & 23.61/.\!6329                  & 23.56/.\!6243                  & 23.23/.\!5684                  & 20.80/.\!6258                  & 20.76/.\!6163                  & 20.62/.\!5601                  \\
    \toprule
    RCAN                           &
    25.26/.\!7911                  & 25.13/.\!7582                  & 23.94/.\!6680                  & 23.90/.\!7209                  & 23.73/.\!6826                  & 21.43/.\!5088                  & 21.89/.\!6791                  & 21.65/.\!6788                  & 20.01/.\!5020                  \\
    \toprule
    ZSSR                           &
    25.22/.\!7810                  & 24.71/.\!7419                  & 23.47/.\!6213                  & 24.10/.\!7464                  & 23.91/.\!7098                  & 21.29/.\!4971                  & 22.66/.\!7043                  & 22.34/.\!6921                  & 21.09/.\!5179                  \\
    \midrule
    DASR                           &
    25.97/.\!8110                  & 25.37/.\!7833                  & 23.91/.\!6452                  & 24.62/.\!7552                  & 24.55/.\!7254                  & 21.65/.\!5149                  & 22.71/.\!7350                  & 22.37/.\!7152                  & 20.94/.\!5016                  \\
    \midrule
    DFKP                           &
    25.51/.\!7825                  & 25.03/.\!7512                  & 22.32/.\!5917                  & 23.79/.\!7343                  & 23.62/.\!7043                  & 20.03/.\!4543                  & 20.31/.\!7040                  & 19.98/.\!6949                  & 19.02/.\!4654                  \\
    \toprule
    IKC                            &
    24.98/.\!7656                  & 24.81/.\!7413                  & 24.01/.\!6690                  & 24.21/.\!7460                  & 24.07/.\!7154                  & 21.87/.\!5029                  & 22.05/.\!7032                  & 21.68/.\!7010                  & 20.77/.\!5025                  \\
    \midrule
    DAN                            &
    25.80/.\!8012                  & 25.22/.\!7637                  & 23.88/.\!6170                  & 24.42/.\!7554                  & 24.31/.\!7213                  & 21.65/.\!5289                  & 22.71/.\!7335                  & 22.48/.\!7227                  & 20.65/.\!5230                  \\

    \midrule
    Ours                           &
    \textcolor{red}{29.12/.\!8617} & \textcolor{red}{28.92/.\!8478} & \textcolor{red}{27.86/.\!8161} & \textcolor{red}{26.19/.\!7768} & \textcolor{red}{26.04/.\!7426} & \textcolor{red}{25.43/.\!6999} & \textcolor{red}{24.02/.\!7715} & \textcolor{red}{23.92/.\!7630} & \textcolor{red}{23.56/.\!7416} \\
    \toprule
    UBound                         &
    32.61/.\!9004                  & 32.23/.\!8711                  & 31.10/.\!8537                  & 29.22/.\!8313                  & 29.02/.\!7870                  & 28.45/.\!7636                  & 27.99/.\!8010                  & 26.89/.\!7862                  & 26.42/.\!7260                  \\
    \bottomrule
  \end{tabu}
  \label{table:x3}
  \vspace{-5mm}
\end{table*}

\begin{table*}[!t]\tiny
  \captionsetup{font=small}
  \caption{4$\times$ BISR results (PSNR(dB)/SSIM). Best results are in red.}
  \setlength\tabcolsep{1pt}
  \centering
  \begin{tabu}{c|X[l]X[l]X[l]|X[l]X[l]X[l]|X[l]X[l]X[l]}
    \toprule
    Datasets                       & \multicolumn{3}{c|}{Set5}      & \multicolumn{3}{c|}{BSD100}    & \multicolumn{3}{c}{Urban100}                                                                                                                                                                        \\
    $\sigma$                       &
    0                              & 2.55                           & 7.65                           & 0                              & 2.55                           & 7.65                           & 0                              & 2.55                           & 7.65                           \\
    \hline\hline
    Bicubic                        &
    23.05/.\!6844                  & 22.93/.\!6782                  & 22.70/.\!6335                  & 22.60/.\!5680                  & 22.58/.\!5622                  & 22.35/.\!5231                  & 19.70/.\!5524                  & 19.69/.\!5452                  & 19.56/.\!5043                  \\
    \toprule
    RCAN                           &
    24.01/.\!7196                  & 23.87/.\!7132                  & 21.78/.\!4960                  & 23.11/.\!5969                  & 22.50/.\!5740                  & 20.33/.\!4853                  & 19.88/.\!5849                  & 19.48/.\!5742                  & 19.00/.\!4513                  \\
    \toprule
    ZSSR                           &
    24.33/.\!7369                  & 22.88/.\!6563                  & 22.50/.\!5063                  & 23.03/.\!6070                  & 22.79/.\!5933                  & 20.21/.\!5024                  & 20.95/.\!6048                  & 20.19/.\!5824                  & 19.37/.\!4769                  \\
    \midrule
    DASR                           &
    24.33/.\!7243                  & 24.16/.\!7008                  & 22.31/.\!5321                  & 23.23/.\!6490                  & 23.12/.\!6171                  & 21.14/.\!5192                  & 21.43/.\!6360                  & 21.32/.\!6031                  & 19.43/.\!4902                  \\
    \midrule
    KGAN                           &
    22.65/.\!6613                  & 22.03/.\!6325                  & 17.4/.\!3820                   & 22.12/.\!6032                  & 21.76/.\!5832                  & 17.65/.\!3326                  & 18.23/.\!6032                  & 17.65/.\!4723                  & 14.25/.\!3027                  \\
    \midrule
    KFKP                           &
    15.32/.\!4859                  & 14.21/.\!3332                  & 12.75/.\!2199                  & 18.23/.\!3765                  & 17.65/.\!3321                  & 15.64/.\!2818                  & 17.64/.\!4083                  & 17.53/.\!3125                  & 15.13/.\!3022                  \\
    \midrule
    DFKP                           &
    24.15/.\!7318                  & 23.83/.\!6927                  & 19.34/.\!4467                  & 23.12/.\!6530                  & 22.88/.\!6311                  & 18.21/.\!3831                  & 20.23/.\!6320                  & 20.13/.\!5923                  & 18.02/.\!3445                  \\
    \toprule
    IKC                            &
    24.03/.\!7152                  & 22.96/.\!6938                  & 21.39/.\!5062                  & 23.93/.\!6123                  & 23.11/.\!6035                  & 21.12/.\!4767                  & 21.33/.\!6265                  & 21.14/.\!6125                  & 19.34/.\!4532                  \\
    \midrule
    DAN                            &
    24.25/.\!7165                  & 24.13/.\!6817                  & 22.14/.\!5157                  & 23.14/.\!6254                  & 23.01/.\!5963                  & 20.98/.\!4943                  & 21.10/.\!6030                  & 21.01/.\!5977                  & 19.29/.\!4433                  \\
    \midrule
    Ours                           &
    \textcolor{red}{27.33/.\!8118} & \textcolor{red}{27.22/.\!8048} & \textcolor{red}{26.51/.\!7638} & \textcolor{red}{24.99/.\!6813} & \textcolor{red}{24.92/.\!6738} & \textcolor{red}{24.46/.\!6412} & \textcolor{red}{22.47/.\!6937} & \textcolor{red}{22.42/.\!6872} & \textcolor{red}{22.13/.\!6589} \\
    \toprule
    UBound                         &
    31.01/.\!8810                  & 30.43/.\!8793                  & 29.43/.\!8603                  & 26.85/.\!7543                  & 26.21/.\!7432                  & 25.92/.\!7136                  & 24.98/.\!7632                  & 24.39/.\!7412                  & 23.88/.\!6960                  \\
    \bottomrule
  \end{tabu}
  \label{table:x4}
  \vspace{-5mm}
\end{table*}

\textbf{The selection of $\lambda$ and $T$.}
There are mainly two parameters to set in our method, the trade-off parameter $\lambda$ and the no. of stages $T$. We perform ablation studies to select them. The ablation study is done on the Set5~\cite{bevilacqua2012low} dataset with $s=2$ and $\sigma=0$. The PSNR results \wrt. $\lambda$ and $T$ are illustrated in Fig.~\ref{fig:ablation}. It can be seen that the PSNR index increases with the increase of $T$, and the highest PSNR is achieved when $\lambda=10$. However, the improvement becomes minor when $T=6$. Hence, we set $\lambda=10$ and $T=6$ in all the following experiments.

\begin{table}[!t]\tiny
  \captionsetup{font=small}
  \caption{Kernel estimation results (PSNR(dB)). Best results are in red.}
  \setlength\tabcolsep{1pt}
  \centering
  \vspace{2mm}
  \resizebox{1\textwidth}{!}{
    \begin{tabu}{c|c|X[c]X[c]X[c]|X[c]X[c]X[c]|X[c]X[c]X[c]}
      \toprule
      Scale                      & Datasets              & \multicolumn{3}{c|}{Set5} & \multicolumn{3}{c|}{BSD100} & \multicolumn{3}{c}{Urban100}                                                                                                 \\
                                 & $\sigma$              &
      0                          & 2.55                  & 7.65                      & 0                           & 2.55                         & 7.65                  & 0                     & 2.55                  & 7.65                  \\
      \hline
      \hline
      \multirow{4}{*}{$\times$2} & KGAN                  &
      41.0                       & 40.9                  & 39.2                      & 40.4                        & 40.2                         & 39.7                  & 40.3                  & 40.2                  & 39.9                  \\
                                 & KFKP                  &
      37.8                       & 38.0                  & 37.5                      & 38.7                        & 38.9                         & 38.5                  & 38.6                  & 38.7                  & 38.5                  \\
                                 & DFKP                  &
      39.4                       & 39.6                  & 39.4                      & 39.0                        & 39.3                         & 38.9                  & 38.6                  & 38.8                  & 38.5                  \\
                                 & Ours                  &
      \textcolor{red}{51.0}      & \textcolor{red}{49.7} & \textcolor{red}{48.9}     & \textcolor{red}{49.3}       & \textcolor{red}{47.2}        & \textcolor{red}{47.3} & \textcolor{red}{48.8} & \textcolor{red}{47.7} & \textcolor{red}{47.0} \\

      \midrule
      \multirow{2}{*}{$\times$3} & DFKP                  &
      39.2                       & 39.0                  & 37.8                      & 38.7                        & 38.2                         & 37.5                  & 37.2                  & 37.2                  & 37.0                  \\
                                 & Ours                  &
      \textcolor{red}{46.8}      & \textcolor{red}{47.0} & \textcolor{red}{45.5}     & \textcolor{red}{47.1}       & \textcolor{red}{46.7}        & \textcolor{red}{45.1} & \textcolor{red}{47.3} & \textcolor{red}{47.0} & \textcolor{red}{46.1} \\
      \midrule
      \multirow{4}{*}{$\times$4} & KGAN                  &
      40.1                       & 39.1                  & 37.1                      & 39.0                        & 38.2                         & 37.2                  & 38.7                  & 37.1                  & 36.9                  \\
                                 & KFKP                  &
      37.7                       & 37.2                  & 36.8                      & 37.3                        & 36.4                         & 36.1                  & 37.0                  & 36.0                  & 35.8                  \\
                                 & DFKP                  &
      39.1                       & 38.4                  & 37.7                      & 37.8                        & 37.8                         & 37.1                  & 37.1                  & 36.8                  & 36.2                  \\
                                 & Ours                  &
      \textcolor{red}{45.1}      & \textcolor{red}{45.5} & \textcolor{red}{42.2}     & \textcolor{red}{43.8}       & \textcolor{red}{43.6}        & \textcolor{red}{43.5} & \textcolor{red}{44.4} & \textcolor{red}{44.7} & \textcolor{red}{44.0} \\
      \bottomrule
    \end{tabu}
    \label{table:kernel}}
\end{table}

\begin{table*}[!t]\tiny
  \captionsetup{font=small}

  \caption{BISR results (PSNR(dB)/SSIM) on different kernels. Best results are in red.}
  \setlength\tabcolsep{1pt}
  \centering
  \begin{tabu}{c|X[c]X[c]X[c]X[c]X[c]X[c]X[c]X[c]X[c]}
    \toprule
                                                                       &
    \includegraphics[width=0.04\textwidth]{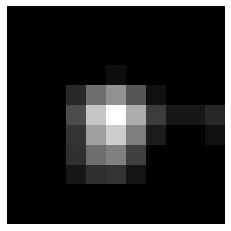} & \includegraphics[width=0.04\textwidth]{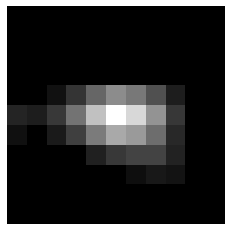} & \includegraphics[width=0.04\textwidth]{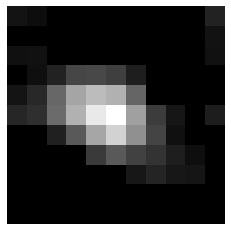} & \includegraphics[width=0.04\textwidth]{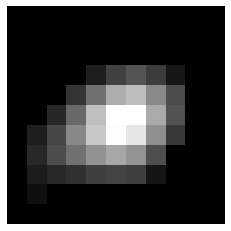} & \includegraphics[width=0.04\textwidth]{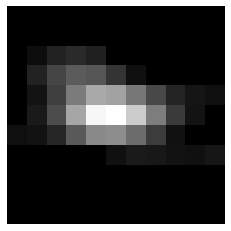} & \includegraphics[width=0.04\textwidth]{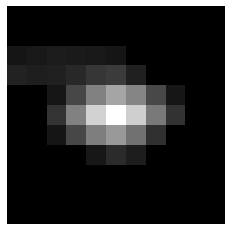} & \includegraphics[width=0.04\textwidth]{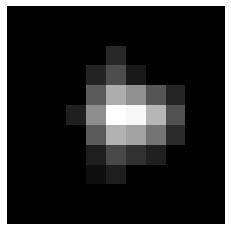} & \includegraphics[width=0.04\textwidth]{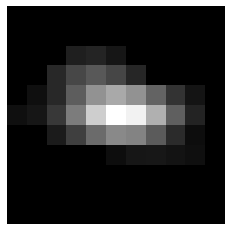} & \includegraphics[width=0.04\textwidth]{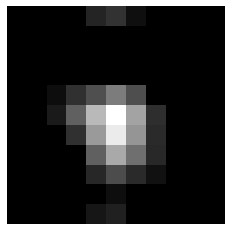} \\
    \hline\hline
    Bicubic                                                            &
    24.52/.\!6411                                                      & 24.58/.\!6025                                                      & 24.02/.\!6123                                                       & 24.04/.\!6433                                                       & 24.85/.\!6768                                                      & 24.44/.\!6328                                                      & 24.52/.\!6470                                                      & 24.03/.\!5979                                                       & 24.68/.\!6708                                                      \\
    \toprule
    RCAN                                                               &
    23.72/.\!5365                                                      & 23.23/.\!5165                                                      & 23.47/.\!5359                                                       & 23.01/.\!5011                                                       & 23.67/.\!5235                                                      & 23.43/.\!5109                                                      & 23.73/.\!5498                                                      & 23.13/.\!5298                                                       & 23.86/.\!5525                                                      \\
    \toprule
    ZSSR                                                               &
    24.71/.\!6834                                                      & 25.02/.\!7043                                                      & 25.24/.\!7323                                                       & 24.83/.\!7233                                                       & 24.89/.\!7310                                                      & 25.20/.\!7321                                                      & 24.99/.\!7432                                                      & 25.45/.\!7236                                                       & 25.20/.\!7235                                                      \\
    \midrule
    DASR                                                               &
    24.01/.\!5632                                                      & 23.92/.\!5594                                                      & 23.45/.\!5293                                                       & 23.77/.\!5695                                                       & 23.99/.\!5860                                                      & 23.04/.\!5202                                                      & 23.57/.\!5905                                                      & 23.77/.\!5342                                                       & 23.12/.\!5231                                                      \\
    \midrule
    KGAN                                                               &
    18.45/.\!3928                                                      & 18.14/.\!3948                                                      & 18.83/.\!3485                                                       & 18.10/.\!3101                                                       & 18.62/.\!3290                                                      & 18.54/.\!3123                                                      & 18.65/.\!3483                                                      & 18.04/.\!2982                                                       & 18.55/.\!3342                                                      \\
    \midrule
    KFKP                                                               &
    16.32/.\!2985                                                      & 16.10/.\!2583                                                      & 16.75/.\!2685                                                       & 16.12/.\!2950                                                       & 16.01/.\!2409                                                      & 16.47/.\!2938                                                      & 16.73/.\!2192                                                      & 16.95/.\!2940                                                       & 15.94/.\!2394                                                      \\
    \midrule
    DFKP                                                               &
    20.18/.\!3711                                                      & 19.57/.\!3634                                                      & 19.85/.\!3621                                                       & 20.01/.\!3582                                                       & 20.13/.\!3364                                                      & 19.03/.\!3531                                                      & 19.85/.\!3651                                                      & 20.10/.\!3427                                                       & 19.57/.\!3922                                                      \\
    \toprule
    IKC                                                                &
    25.10/.\!7343                                                      & 25.12/.\!7256                                                      & 25.01/.\!7074                                                       & 25.32/.\!7368                                                       & 25.03/.\!7266                                                      & 24.98/.\!6950                                                      & 25.23/.\!7654                                                      & 25.09/.\!7079                                                       & 25.20/.\!7354                                                      \\
    \midrule
    DAN                                                                &
    23.62/.\!5543                                                      & 23.15/.\!5125                                                      & 23.54/.\!5540                                                       & 23.92/.\!5783                                                       & 23.11/.\!5089                                                      & 23.62/.\!5798                                                      & 23.35/.\!5610                                                      & 23.80/.\!5910                                                       & 23.88/.\!5910                                                      \\
    \midrule
    Ours                                                               &
    \textcolor{red}{26.56/.\!7636}                                     & \textcolor{red}{26.54/.\!7591}                                     & \textcolor{red}{26.19/.\!7397}                                      & \textcolor{red}{25.41/.\!7052}                                      & \textcolor{red}{26.56/.\!7607}                                     & \textcolor{red}{26.61/.\!7635}                                     & \textcolor{red}{26.67/.\!7702}                                     & \textcolor{red}{25.14/.\!7046}                                      & \textcolor{red}{26.66/.\!7673}                                     \\
    \toprule
    UBound                                                             &
    28.01/.\!8409                                                      & 27.96/.\!8520                                                      & 27.85/.\!8254                                                       & 27.67/.\!8175                                                       & 28.00/.\!8508                                                      & 27.77/.\!8243                                                      & 27.91/.\!8378                                                      & 27.53/.\!8165                                                       & 28.12/.\!8680                                                      \\
    \bottomrule
  \end{tabu}
  \label{table:specifig-kernel}
\end{table*}

\label{sec:stages}
\begin{figure*}[!t]
  \captionsetup{font=small}
  \captionsetup[subfigure]{justification=centering,font=scriptsize}
  \setlength{\abovecaptionskip}{+4pt}
  \setlength{\belowcaptionskip}{-10pt}
  \centering
  \begin{subfigure}{.14\textwidth}
    \setlength{\abovecaptionskip}{0pt}
    \setlength{\belowcaptionskip}{0pt}
    \centering
    \includegraphics[width=\textwidth]{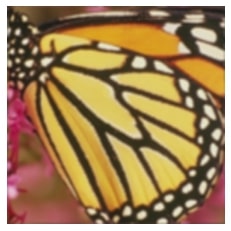}
    \caption[short]{$\mathbf{X}_{\left(1\right)}$: 22.47dB}
  \end{subfigure}\hfill%
  \centering
  \begin{subfigure}{.14\textwidth}
    \setlength{\abovecaptionskip}{0pt}
    \setlength{\belowcaptionskip}{0pt}
    \centering
    \includegraphics[width=\textwidth]{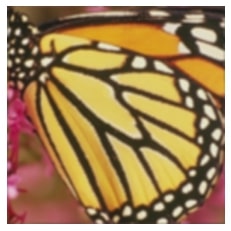}
    \caption[short]{$\mathbf{X}_{\left(1\right)}$: 24.22dB}
  \end{subfigure}\hfill%
  \centering
  \begin{subfigure}{.14\textwidth}
    \setlength{\abovecaptionskip}{0pt}
    \setlength{\belowcaptionskip}{0pt}
    \centering
    \includegraphics[width=\textwidth]{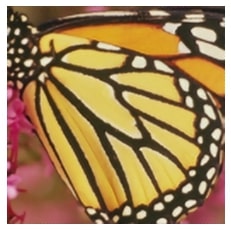}
    \caption[short]{$\mathbf{X}_{\left(3\right)}$: 26.60dB}
  \end{subfigure}\hfill%
  \centering
  \begin{subfigure}{.14\textwidth}
    \setlength{\abovecaptionskip}{0pt}
    \setlength{\belowcaptionskip}{0pt}
    \centering
    \includegraphics[width=\textwidth]{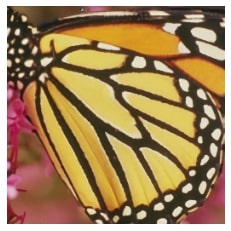}
    \caption[short]{$\mathbf{X}_{\left(4\right)}$: 28.52dB}
  \end{subfigure}\hfill%
  \centering
  \begin{subfigure}{.14\textwidth}
    \setlength{\abovecaptionskip}{0pt}
    \setlength{\belowcaptionskip}{0pt}
    \centering
    \includegraphics[width=\textwidth]{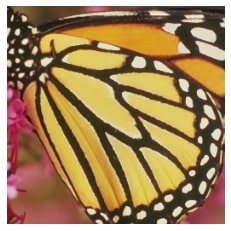}
    \caption[short]{$\mathbf{X}_{\left(5\right)}$: 30.04dB}
  \end{subfigure}\hfill%
  \centering
  \begin{subfigure}{.14\textwidth}
    \setlength{\abovecaptionskip}{0pt}
    \setlength{\belowcaptionskip}{0pt}
    \centering
    \includegraphics[width=\textwidth]{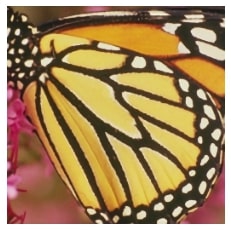}
    \caption[short]{$\mathbf{X}_{\left(6\right)}$: 30.23dB}
  \end{subfigure}\hfill%
  \centering
  \begin{subfigure}{.14\textwidth}
    \setlength{\abovecaptionskip}{0pt}
    \setlength{\belowcaptionskip}{0pt}
    \centering
    \includegraphics[width=\textwidth]{fig/stage/s6.jpg}
    \caption[short]{Ground truth}
  \end{subfigure}\hfill%

  \centering
  \begin{subfigure}{.14\textwidth}
    \setlength{\abovecaptionskip}{0pt}
    \setlength{\belowcaptionskip}{0pt}
    \centering
    \includegraphics[width=\textwidth]{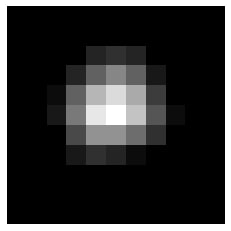}
    \caption[short]{$\mathbf{K}_{\left(1\right)}$: 45.6dB}
  \end{subfigure}\hfill%
  \centering
  \begin{subfigure}{.14\textwidth}
    \setlength{\abovecaptionskip}{0pt}
    \setlength{\belowcaptionskip}{0pt}
    \centering
    \includegraphics[width=\textwidth]{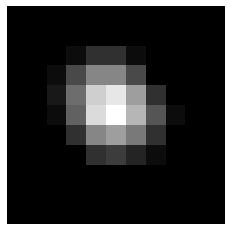}
    \caption[short]{$\mathbf{K}_{\left(2\right)}$: 46.1dB}
  \end{subfigure}\hfill%
  \centering
  \begin{subfigure}{.14\textwidth}
    \setlength{\abovecaptionskip}{0pt}
    \setlength{\belowcaptionskip}{0pt}
    \centering
    \includegraphics[width=\textwidth]{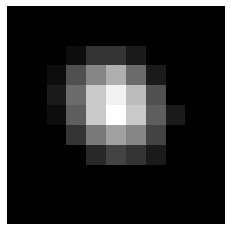}
    \caption[short]{$\mathbf{K}_{\left(3\right)}$: 47.7dB}
  \end{subfigure}\hfill%
  \centering
  \begin{subfigure}{.14\textwidth}
    \setlength{\abovecaptionskip}{0pt}
    \setlength{\belowcaptionskip}{0pt}
    \centering
    \includegraphics[width=\textwidth]{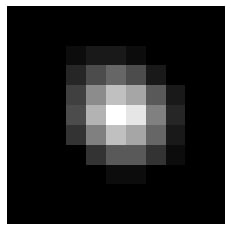}
    \caption[short]{$\mathbf{K}_{\left(4\right)}$: 52.1B}
  \end{subfigure}\hfill%
  \centering
  \begin{subfigure}{.14\textwidth}
    \setlength{\abovecaptionskip}{0pt}
    \setlength{\belowcaptionskip}{0pt}
    \centering
    \includegraphics[width=\textwidth]{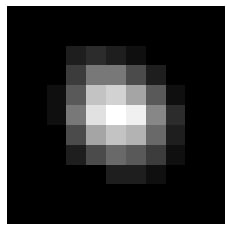}
    \caption[short]{$\mathbf{K}_{\left(5\right)}$: 55.6dB}
  \end{subfigure}\hfill%
  \centering
  \begin{subfigure}{.14\textwidth}
    \setlength{\abovecaptionskip}{0pt}
    \setlength{\belowcaptionskip}{0pt}
    \centering
    \includegraphics[width=\textwidth]{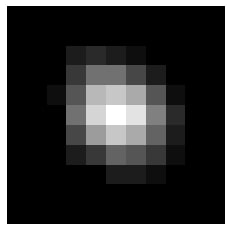}
    \caption[short]{$\mathbf{K}_{\left(6\right)}$: 55.9dB}
  \end{subfigure}\hfill%
  \centering
  \begin{subfigure}{.14\textwidth}
    \setlength{\abovecaptionskip}{0pt}
    \setlength{\belowcaptionskip}{0pt}
    \centering
    \includegraphics[width=\textwidth]{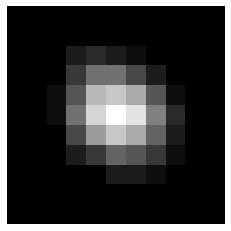}
    \caption[short]{Ground truth}
  \end{subfigure}\hfill%
  \caption[short]{Outputs in each stage of UDKE on ``butterfly'' in Set5 with $s=2$ and $\sigma=0$.}
  \label{fig:stages}
\end{figure*}

\subsection{Comparison with state-of-the-arts}
In this section, we compare the proposed UDKE based framework with state-of-the-art BISR methods. Three sets of experiments are conducted to comprehensively evaluate UDKE. First, we evaluate UDKE by using the DEPD-evaluation kernel pool, which is more complex than the widely used Gaussian kernels and more similar to the degradation in real-world images~\cite{zhou2019kernel}. Second, we compare UDKE, which does not impose presumptions on blur kernels, with the methods that presume Gaussian blur kernels, by using just Gaussian kernels. Third, we evaluate UDKE on real-world images whose degradation kernels can be more complex than Gaussian kernels and those in the DEPD-evaluation kernel pool.

\begin{figure*}[!t]
  \captionsetup[subfigure]{justification=centering,font=scriptsize}
  \centering
  \begin{subfigure}{.198\textwidth}
    \setlength{\abovecaptionskip}{0pt}
    \setlength{\belowcaptionskip}{0pt}
    \centering
    \includegraphics[width=\textwidth]{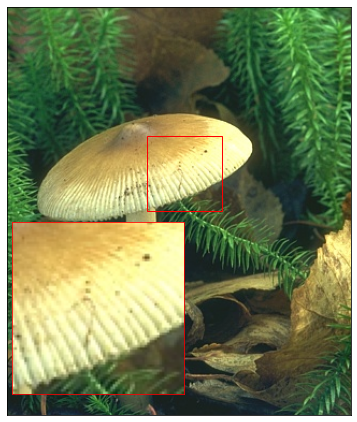}
    \caption[short]{Ground truth\\ \ }
  \end{subfigure}\hfill%
  \begin{subfigure}{.198\textwidth}
    \setlength{\abovecaptionskip}{0pt}
    \setlength{\belowcaptionskip}{0pt}
    \centering
    \includegraphics[width=\textwidth]{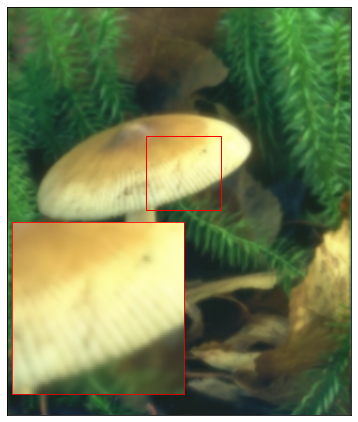}
    \caption[centering]{RCAN \\(26.83dB/.9001)}
  \end{subfigure}\hfill%
  \begin{subfigure}{.198\textwidth}
    \setlength{\abovecaptionskip}{0pt}
    \setlength{\belowcaptionskip}{0pt}
    \centering
    \includegraphics[width=\textwidth]{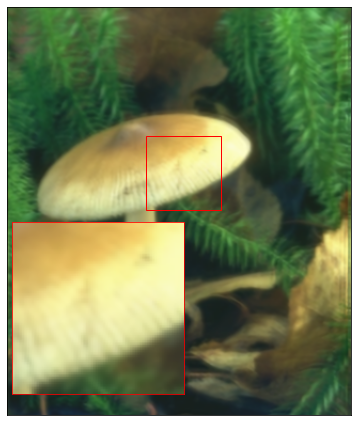}
    \caption[centering]{ZSSR \\(26.86dB/.8996)}
  \end{subfigure}\hfill%
  \begin{subfigure}{.198\textwidth}
    \setlength{\abovecaptionskip}{0pt}
    \setlength{\belowcaptionskip}{0pt}
    \centering
    \includegraphics[width=\textwidth]{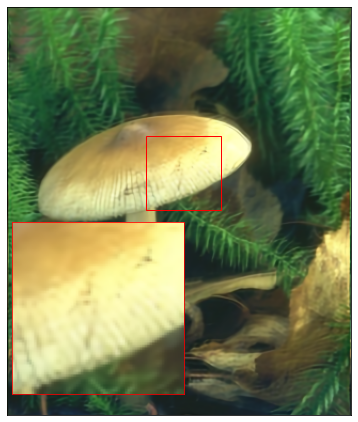}
    \caption[centering]{IKC \\(27.04dB/.9295)}
  \end{subfigure}\hfill%
  \begin{subfigure}{.198\textwidth}
    \setlength{\abovecaptionskip}{0pt}
    \setlength{\belowcaptionskip}{0pt}
    \centering
    \includegraphics[width=\textwidth]{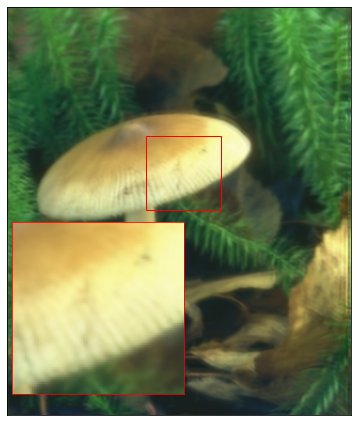}
    \caption[short]{DAN \\(27.34dB/.9167)}
  \end{subfigure}\hfill%
  \begin{subfigure}{.198\textwidth}
    \setlength{\abovecaptionskip}{0pt}
    \setlength{\belowcaptionskip}{0pt}
    \centering
    \includegraphics[width=\textwidth]{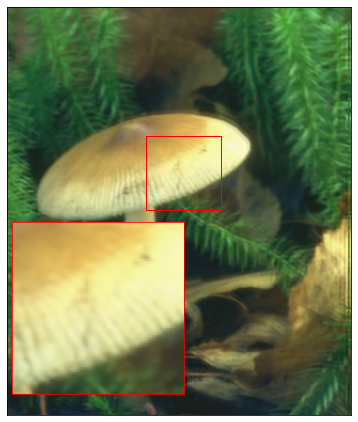}
    \caption[short]{DASR \\(27.42dB/.9101)}
  \end{subfigure}\hfill%
  \begin{subfigure}{.198\textwidth}
    \setlength{\abovecaptionskip}{0pt}
    \setlength{\belowcaptionskip}{0pt}
    \centering
    \includegraphics[width=\textwidth]{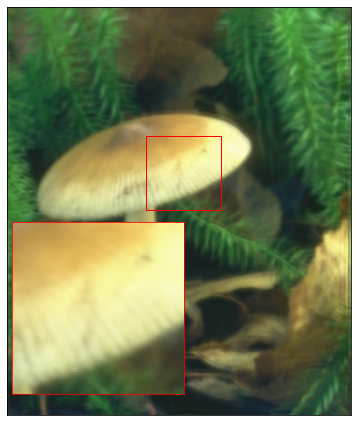}
    \caption[short]{KGAN (24.94dB/.8865)}
  \end{subfigure}\hfill%
  \begin{subfigure}{.198\textwidth}
    \setlength{\abovecaptionskip}{0pt}
    \setlength{\belowcaptionskip}{0pt}
    \centering
    \includegraphics[width=\textwidth]{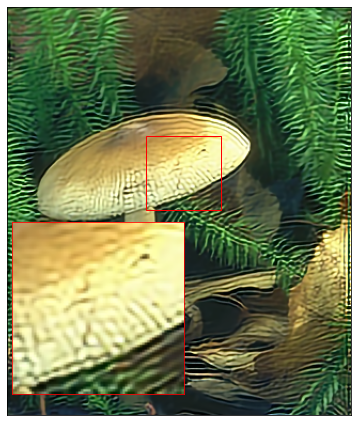}
    \caption[short]{KFKP (14.20dB/.5833)}
  \end{subfigure}\hfill%
  \begin{subfigure}{.198\textwidth}
    \setlength{\abovecaptionskip}{0pt}
    \setlength{\belowcaptionskip}{0pt}
    \centering
    \includegraphics[width=\textwidth]{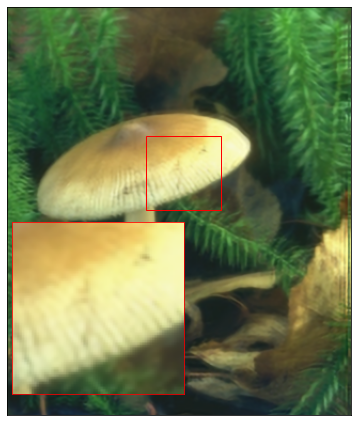}
    \caption[centering]{DFKP \\(26.00dB/.9040)}
  \end{subfigure}\hfill%
  \begin{subfigure}{.198\textwidth}
    \setlength{\abovecaptionskip}{0pt}
    \setlength{\belowcaptionskip}{0pt}
    \centering
    \includegraphics[width=\textwidth]{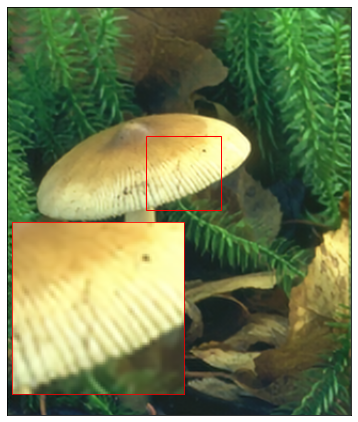}
    \caption[short]{UDKE \\(30.42dB/.9393)}
  \end{subfigure}\hfill%
  \caption[short]{Results on ``208001'' in BSD100 with $s=2$ and $\sigma=0$ (PSNR/SSIM).}
  \label{fig:vis}
\end{figure*}

\begin{figure}[!t]
  \captionsetup[subfigure]{justification=centering,font=scriptsize}
  \begin{subfigure}{0.18\textwidth}
    \includegraphics[width=\linewidth]{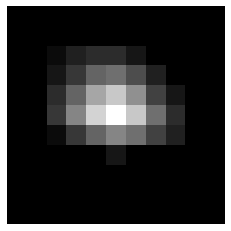}
    \caption{Ground truth\\ \ } \label{subfig:left}
  \end{subfigure}\hfill
  \begin{subfigure}{0.18\linewidth}
    \includegraphics[width=\linewidth]{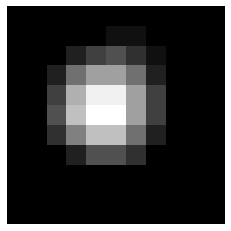}
    \caption{KGAN\\(43.7dB)}
  \end{subfigure}\hfill
  \begin{subfigure}{0.18\linewidth}
    \includegraphics[width=\linewidth]{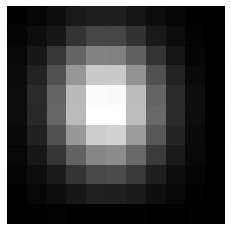}
    \caption{KFKP\\(39.6dB)}
  \end{subfigure}\hfill
  \begin{subfigure}{0.18\linewidth}
    \includegraphics[width=\linewidth]{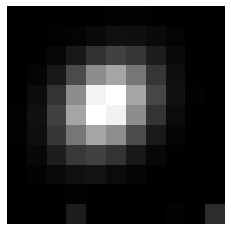}
    \caption{DFKP\\(43.3dB)}
  \end{subfigure}\hfill
  \begin{subfigure}{0.18\linewidth}
    \includegraphics[width=\linewidth]{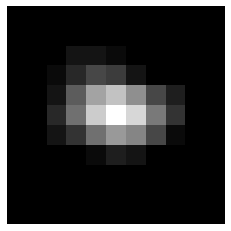}
    \caption{UDKE\\(47.6dB)}
  \end{subfigure}
  \caption[short]{Kernel estimation results on ``208001'' in BSD100 with $s=2$ and $\sigma=0$.}
  \label{fig:kernel}
\end{figure}

\textbf{Experiments with DEPD-evaluation kernel pool.}
Following prior works \cite{liang2021flow,luo2020unfolding}, the widely used Set5, BSD100~\cite{MartinFTM01} and Urban100~\cite{huang2015single} datasets are used in the experiments. Testing LR images are obtained by applying randomly selected BISR kernels from the DEPD-evaluation pool to HR images according to the degradation process described in Eq.~\ref{eq:degradation_model}. We compare our UDKE based framework with bicubic interpolation, SOTA non-blind SR method RCAN~\cite{zhang2018image}, SOTA direct DNN based BISR methods, including ZSSR~\cite{shocher2018zero}, DASR~\cite{wang2021unsupervised}, KGAN~\cite{bell2019blind}, DFKP (DIP-FKP)~\cite{liang2021flow} and KFKP (KernelGAN-FKP)~\cite{liang2021flow}, and SOTA deep unfolding BISR methods, including IKC~\cite{gu2019blind} and DAN~\cite{luo2020unfolding}. Note that some methods (\eg, KernelNet~\cite{yamac2021kernelnet}) are inapplicable for comparison. This is because that their source code is not disclosed and we cannot reproduce them; and they reported performance on non-standard testing sets so that we cannot compare with the reported results either.

The experimental results of $\times 2$, $\times 3$ and $\times 4$ BISR are shown in Tables~\ref{table:x2}$\sim$\ref{table:x4}, respectively. Note that some methods cannot be applied to $\times 3$ SR task and are omitted in the comparison. We also provide an ``upper bound'' (UBound) of the BISR methods, which is obtained by feeding the ground truth kernel to the non-blind SISR method USRNet~\cite{zhang2020deep}, yielding a non-blind ``upper bound'' as a reference for evaluating those blind BISR methods. Besides the PSNR/SSIM of the reconstructed image, for those BISR methods that estimate the blur kernels (\ie, KGAN~\cite{bell2019blind}, DFKP~\cite{liang2021flow}, KFKP~\cite{liang2021flow} and our UDKE), we also compute the PSNR of the estimated kernel (Kernel-PSNR), and list the results in Table~\ref{table:kernel}.

It can be seen that UDKE based framework achieves the best PSNR, SSIM and Kernel-PSNR results under all experiment settings, significantly outperforming the competing methods. Both the direct DNN based methods (ZSSR, DASR, KGAN, KFKP, DFKP) and deep unfolding methods (IKC and DAN) do not surpass the non-blind method RCAN. This is mainly because these methods cannot encode effectively the information of the degradation process in their models, and hence they can not estimate the HR images with unseen degradation parameters during inference. In contrast, our proposed UDKE can estimate the degradation kernel very well (please see the Kernel-PSNR in Table~\ref{table:kernel}) and consequently reproduce the original image with better quality.

Fig.~\ref{fig:stages} shows the outputs in each stage of UDKE. It can be seen that UDKE progressively improves the estimations of BISR kernel and super-resolution results, which validates the motivation and effectiveness of our architecture design. Fig.~\ref{fig:vis} visualizes the BISR results on an image by the competing methods. It can be seen that the image reconstructed by UDKE has the best visual quality with sharp edges and rich textures. Other methods generate blurry results as they fail to well estimate and exploit the BISR kernel information. KFKP, which enforces strong Gaussian priors on kernel estimation and employs unstable adversarial training, results in distorted image textures due to kernel mismatching. Fig.~\ref{fig:kernel} visualizes the predicted kernels by those BISR methods with explicit kernel estimation. It can be seen that only UDKE predicts the kernel with high fidelity, while others yield unreliable results because they generally impose the Gaussian-shape assumption on the kernels. More visualization results can be found in the \textbf{Appendix}.

To further demonstrate the performance of UDKE, we also build several testing sets, each of them is built with a specific kernel from the DEPD-evaluation pool, as shown in Table~\ref{table:specifig-kernel}. The experiment is done on BSD100 images with $s=2$ and $\sigma=7.65$. It can be seen that UDKE based BISR framework consistently shows the SOTA performance on different testing blur kernels.

\begin{table}[!t]\tiny
  \centering
  \captionsetup{font=small}
  \setlength\tabcolsep{3pt}
  \caption{BISR results (PSNR/SSIM/Kernel-PSNR) on isotropic Gaussian kernels.}
  \setlength\tabcolsep{1pt}
  \begin{tabu}{l|l|X[c]X[c]X[c]}
    \toprule
    $s$                &      & $\sigma=0$         & $\sigma=2.55$      & $\sigma=7.65$      \\
    \hline\hline
    \multirow{4}{*}{2} & DAN  & 29.78/.\!8531/-    & 29.25/.\!8358/-    & 28.31/.\!7510/-    \\
                       & KGAN & 29.76/.\!8420/47.9 & 29.13/.\!7378/46.5 & 28.10/.\!7026/45.8 \\
                       & KFKP & 29.79/.\!8530/49.4 & 29.27/.\!8429/47.1 & 28.24/.\!7450/46.2 \\
                       & DFKP & 30.12/.\!8940/49.8 & 29.40/.\!8611/47.4 & 28.45/.\!7570/46.2 \\
                       & Ours & 29.80/.\!8522/48.8 & 29.24/.\!8435/47.0 & 28.57/.\!7700/46.4 \\
    \midrule
    \multirow{4}{*}{4} & DAN  & 26.34/.\!7849/-    & 25.92/.\!7320/-    & 25.42/.\!5820/-    \\
                       & KGAN & 26.21/.\!7810/46.2 & 25.73/.\!7243/44.2 & 25.30/.\!5575/41.9 \\
                       & KFKP & 26.48/.\!7911/46.3 & 25.94/.\!7310/44.9 & 25.43/.\!5801/42.4 \\
                       & DFKP & 26.53/.\!8012/47.5 & 26.10/.\!7592/45.3 & 25.68/.\!6013/44.4 \\
                       & Ours & 26.43/.\!7834/46.2 & 25.98/.\!7324/44.8 & 25.72/.\!6187/44.5 \\
    \bottomrule
  \end{tabu}
  \label{table:i-gaussian}
\end{table}

\begin{table}[t]\tiny
  \centering
  \captionsetup{font=small}
  \caption{BISR results (PSNR/SSIM/Kernel-PSNR) on anisotropic Gaussian kernels.}
  \setlength\tabcolsep{1pt}
  \centering
  \resizebox{1\textwidth}{!}{
    \begin{tabu}{l|l|X[c]X[c]X[c]}
      \toprule
      $s$                &      & $\sigma=0$         & $\sigma=2.55$      & $\sigma=7.65$      \\
      \hline\hline
      \multirow{4}{*}{2} & DAN  & 28.12/.\!8557/-    & 27.13/.\!8028/-    & 26.35/.\!7443/-    \\
                         & KGAN & 26.33/.\!7751/44.3 & 26.01/.\!7432/43.8 & 25.79/.\!7026/42.1 \\
                         & KFKP & 27.62/.\!8512/49.1 & 27.11/.\!8010/46.9 & 26.34/.\!7450/45.2 \\
                         & DFKP & 28.42/.\!8863/49.8 & 27.32/.\!8218/47.2 & 26.71/.\!7570/45.8 \\
                         & Ours & 27.56/.\!8112/48.5 & 27.10/.\!8003/46.9 & 26.73/.\!7700/46.1 \\
      \midrule
      \multirow{4}{*}{4} & DAN  & 24.98/.\!6813/-    & 24.42/.\!6348/-    & 23.01/.\!5575/-    \\
                         & KGAN & 23.97/.\!6211/43.2 & 23.70/.\!5987/42.1 & 23.01/.\!5575/41.8 \\
                         & KFKP & 24.87/.\!6612/44.2 & 24.49/.\!6353/43.0 & 24.01/.\!5801/42.7 \\
                         & DFKP & 25.47/.\!7233/46.4 & 24.62/.\!6891/44.0 & 24.10/.\!6013/44.0 \\
                         & Ours & 24.89/.\!6612/44.3 & 24.52/.\!6348/43.1 & 24.32/.\!6187/44.0 \\
      \bottomrule
    \end{tabu}
    \label{table:ani-gaussian}}
\end{table}

\textbf{Experiments with Gaussian kernels.} Although UDKE is not designed and trained for Gaussian kernel degradation, we also test its effectiveness on Gaussian kernels in comparison with those BISR methods designed and trained for Gaussian blur kernels, \ie, DAN~\cite{luo2020unfolding}, KGAN~\cite{bell2019blind}, KFKP~\cite{liang2021flow}, and DFKP~\cite{liang2021flow}. The Gaussian kernels are obtained by the process adopted in Gaussian BISR methods~\cite{bell2019blind,liang2021flow}, which randomly samples the length and rotation angle to generate Gaussian kernels. The experiment is done on BSD100 set. Tables~\ref{table:i-gaussian} and~\ref{table:ani-gaussian} show the BISR results on isotropic and anisotropic Gaussian degradation, respectively. One can see that UDKE achieves competitive results with those Gaussian BISR methods but without hardcoding any Gaussian priors. With the increase of noise level, it can even surpass DFKP, showing better robustness to noise.

\begin{figure*}[!t]
  \captionsetup[subfigure]{justification=centering,font=scriptsize}
  \centering
  \begin{subfigure}{.198\textwidth}
    \setlength{\abovecaptionskip}{0pt}
    \setlength{\belowcaptionskip}{0pt}
    \centering
    \includegraphics[width=\textwidth]{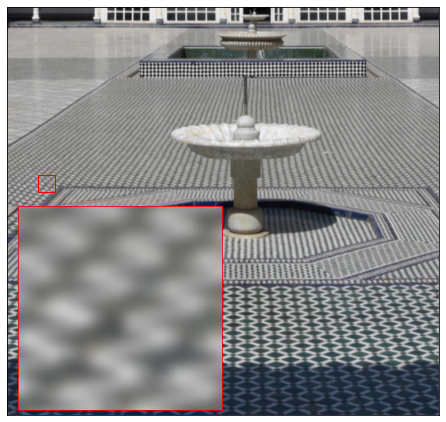}
    \caption[short]{LR image}
  \end{subfigure}\hfill%
  \begin{subfigure}{.198\textwidth}
    \setlength{\abovecaptionskip}{0pt}
    \setlength{\belowcaptionskip}{0pt}
    \centering
    \includegraphics[width=\textwidth]{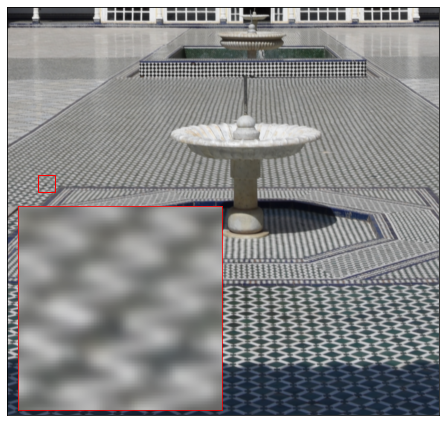}
    \caption[short]{RCAN}
  \end{subfigure}\hfill%
  \begin{subfigure}{.198\textwidth}
    \setlength{\abovecaptionskip}{0pt}
    \setlength{\belowcaptionskip}{0pt}
    \centering
    \includegraphics[width=\textwidth]{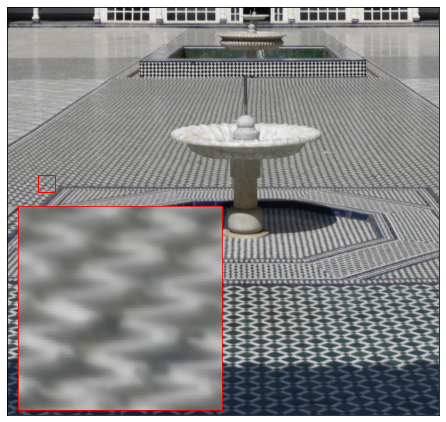}
    \caption[centering]{IKC}
  \end{subfigure}\hfill%
  \begin{subfigure}{.198\textwidth}
    \setlength{\abovecaptionskip}{0pt}
    \setlength{\belowcaptionskip}{0pt}
    \centering
    \includegraphics[width=\textwidth]{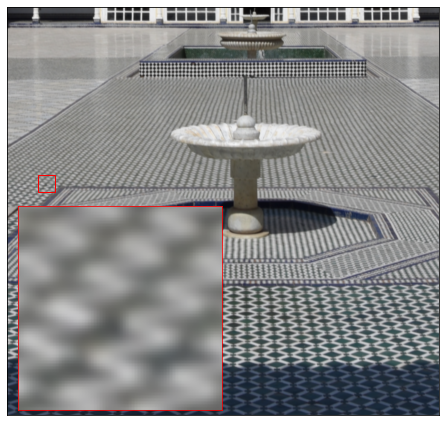}
    \caption[centering]{DAN}
  \end{subfigure}\hfill%
  \begin{subfigure}{.198\textwidth}
    \setlength{\abovecaptionskip}{0pt}
    \setlength{\belowcaptionskip}{0pt}
    \centering
    \includegraphics[width=\textwidth]{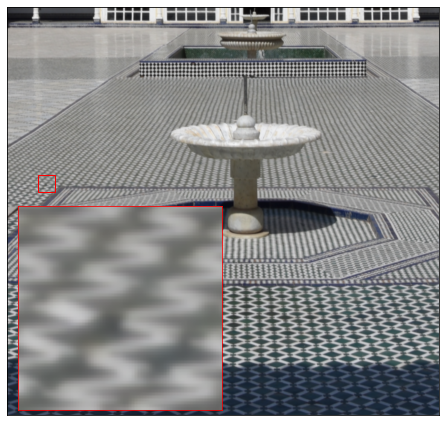}
    \caption[centering]{DASR}
  \end{subfigure}\hfill%
  \begin{subfigure}{.198\textwidth}
    \setlength{\abovecaptionskip}{0pt}
    \setlength{\belowcaptionskip}{0pt}
    \centering
    \includegraphics[width=\textwidth]{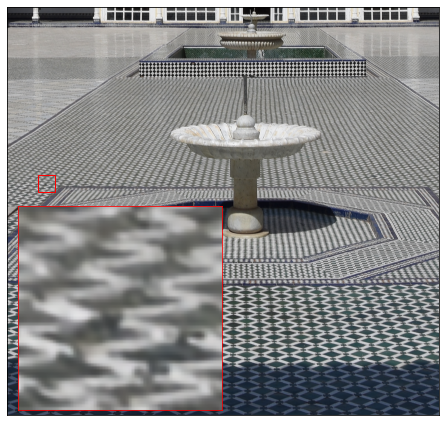}
    \caption[short]{KFKP}
  \end{subfigure}\hfill%
  \begin{subfigure}{.198\textwidth}
    \setlength{\abovecaptionskip}{0pt}
    \setlength{\belowcaptionskip}{0pt}
    \centering
    \includegraphics[width=\textwidth]{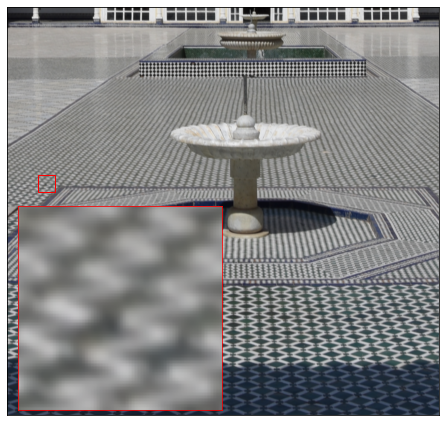}
    \caption[short]{DFKP}
  \end{subfigure}\hfill%
  \begin{subfigure}{.198\textwidth}
    \setlength{\abovecaptionskip}{0pt}
    \setlength{\belowcaptionskip}{0pt}
    \centering
    \includegraphics[width=\textwidth]{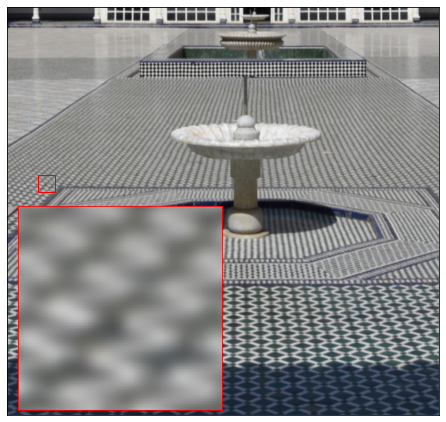}
    \caption[centering]{BSRGAN}
  \end{subfigure}\hfill%
  \begin{subfigure}{.198\textwidth}
    \setlength{\abovecaptionskip}{0pt}
    \setlength{\belowcaptionskip}{0pt}
    \centering
    \includegraphics[width=\textwidth]{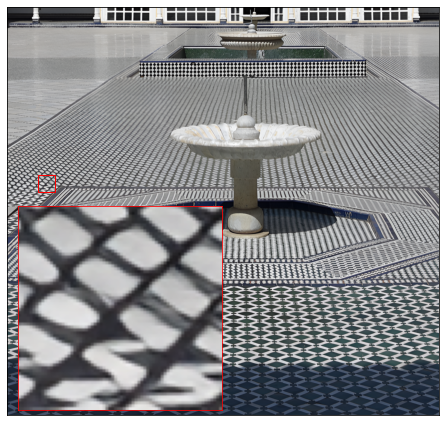}
    \caption[short]{RESRGAN}
  \end{subfigure}\hfill%
  \begin{subfigure}{.198\textwidth}
    \setlength{\abovecaptionskip}{0pt}
    \setlength{\belowcaptionskip}{0pt}
    \centering
    \includegraphics[width=\textwidth]{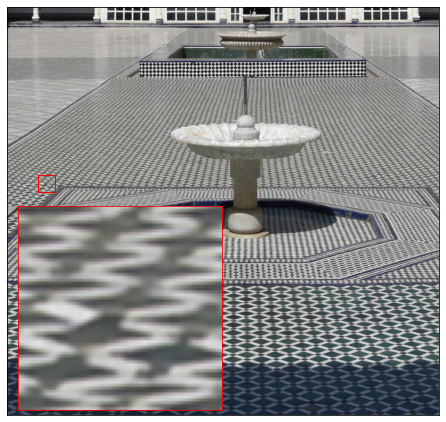}
    \caption[short]{UDKE}
  \end{subfigure}\hfill%
  \caption[short]{Result on a real-world image (better viewed on screen).}
  \label{fig:real}
\end{figure*}

\textbf{Experiments on real-world images.}
We further test UDKE on real-world LR images whose degradation kernels can be more complex than Gaussian as well as the DEPD-evaluation pool. We add two state-of-the-art real-world SISR methods, \ie, RESRGAN (Real-ESRGAN)~\cite{wang2021realesrgan} and BSRGAN~\cite{zhang2021designing}, for more comprehensive comparison. The results are shown in Fig.~\ref{fig:real}. It can be seen that UDKE produces the best result, with not only superior perceptual quality but also more accurate fidelity (see the ground tiles). KFKP generates many noisy artifacts and distorted edges, while RESRGAN generates false patterns on the ground tiles. All the other competing methods output blurry results. This experiment demonstrates the robust kernel estimation capability of UDKE in real-world scenarios. More examples can be found in the \textbf{Appendix}.

\begin{figure*}[t]
  \begin{center}
    \includegraphics[width=0.6\linewidth]{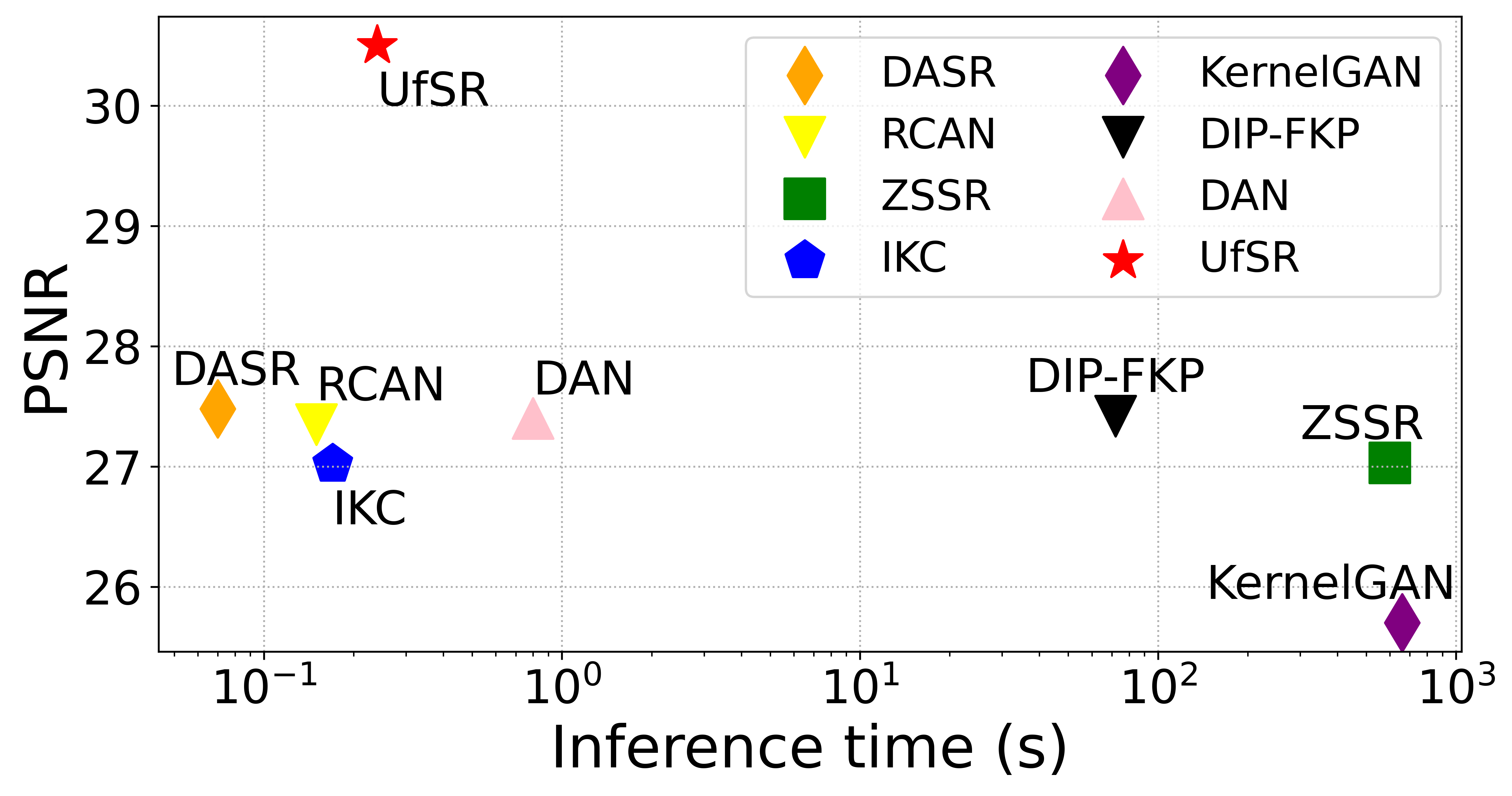}
  \end{center}
  \caption{Ablation study on $\lambda$ and $T$.}
  \vspace{-5mm}
  \label{fig:time}
\end{figure*}
\textbf{Inference time.}
We further compare the inference time of UDKE and the competing methods. The experiments are conducted on Set5 ($s=2$ and $\sigma=0$) with a GTX 2080Ti GPU. The results are shown in Fig.~\ref{fig:time}. It can be seen that UDKE is slightly slower than DASR and has a similar speed to RCAN and IKC. It is over $\times 100$ times faster than the leading online training based BISR method DFKP, which trains a deep model during the inference, and $\times 1000$ times faster than ZSSR and KGAN. Meanwhile, UDKE achieves significantly higher PSNR (about 3dB) than all competing methods.

\section{Conclusion}
We proposed a novel unfolded deep kernel estimation method, namely UDKE, to explicitly solve the data term of the unfolding objective function for effective BISR. Equipped with the designed memory-efficient algorithm, UDKE is free of the im2col operation required in traditional Least Squares Method and hence saves over $17000\times$ memory overhead, providing an efficient solution to explicitly solve the data term under the deep learning framework. UDKE addresses the challenging BISR problem by efficiently utilizing the information of degradation model during inference. Extensive experiments on both synthetic and real-world images validated that UDKE could faithfully predict non-Gaussian blur kernels, and reproduce high quality images with sharp structures and rich textures, surpassing existing BISR methods by a large margin. Meanwhile, UDKE has good efficiency, making it an attractive choice for BISR in practice.

\clearpage
%
%
\bibliographystyle{splncs04}
\bibliography{UDKE}

\begin{thebibliography}{10}
\providecommand{\url}[1]{\texttt{#1}}
\providecommand{\urlprefix}{URL }
\providecommand{\doi}[1]{https://doi.org/#1}

\bibitem{abdi2007method}
Abdi, H., et~al.: The method of least squares. Encyclopedia of Measurement and
  Statistics. CA, USA: Thousand Oaks  (2007)

\bibitem{Agustsson_2017_CVPR_Workshops}
Agustsson, E., Timofte, R.: Ntire 2017 challenge on single image
  super-resolution: Dataset and study. In: The IEEE Conference on Computer
  Vision and Pattern Recognition (CVPR) Workshops (July 2017)

\bibitem{barbu2009training}
Barbu, A.: Training an active random field for real-time image denoising. IEEE
  Transactions on Image Processing  \textbf{18}(11),  2451--2462 (2009)

\bibitem{begin2004blind}
Begin, I., Ferrie, F.: Blind super-resolution using a learning-based approach.
  In: Proceedings of the 17th International Conference on Pattern Recognition,
  2004. ICPR 2004. vol.~2, pp. 85--89. IEEE (2004)

\bibitem{bell2019blind}
Bell-Kligler, S., Shocher, A., Irani, M.: Blind super-resolution kernel
  estimation using an internal-gan. arXiv preprint arXiv:1909.06581  (2019)

\bibitem{bevilacqua2012low}
Bevilacqua, M., Roumy, A., Guillemot, C., Alberi-Morel, M.L.: Low-complexity
  single-image super-resolution based on nonnegative neighbor embedding  (2012)

\bibitem{boyd2011distributed}
Boyd, S., Parikh, N., Chu, E., Peleato, B., Eckstein, J., et~al.: Distributed
  optimization and statistical learning via the alternating direction method of
  multipliers. Foundations and Trends{\textregistered} in Machine learning
  \textbf{3}(1),  1--122 (2011)

\bibitem{bulat2018learn}
Bulat, A., Yang, J., Tzimiropoulos, G.: To learn image super-resolution, use a
  gan to learn how to do image degradation first. In: Proceedings of the
  European conference on computer vision (ECCV). pp. 185--200 (2018)

\bibitem{capel2000super}
Capel, D., Zisserman, A.: Super-resolution enhancement of text image sequences.
  In: Proceedings 15th International Conference on Pattern Recognition.
  ICPR-2000. vol.~1, pp. 600--605. IEEE (2000)

\bibitem{corduneanu2005learning}
Corduneanu, A., Platt, J.C.: Learning spatially-variable filters for
  super-resolution of text. In: IEEE International Conference on Image
  Processing 2005. vol.~1, pp. I--849. IEEE (2005)

\bibitem{dong2014learning}
Dong, C., Loy, C.C., He, K., Tang, X.: Learning a deep convolutional network
  for image super-resolution. In: European conference on computer vision. pp.
  184--199. Springer (2014)

\bibitem{glasner2009super}
Glasner, D., Bagon, S., Irani, M.: Super-resolution from a single image. In:
  2009 IEEE 12th international conference on computer vision. pp. 349--356.
  IEEE (2009)

\bibitem{gu2019blind}
Gu, J., Lu, H., Zuo, W., Dong, C.: Blind super-resolution with iterative kernel
  correction. In: Proceedings of the IEEE/CVF Conference on Computer Vision and
  Pattern Recognition. pp. 1604--1613 (2019)

\bibitem{he2009soft}
He, Y., Yap, K.H., Chen, L., Chau, L.P.: A soft map framework for blind
  super-resolution image reconstruction. Image and Vision Computing
  \textbf{27}(4),  364--373 (2009)

\bibitem{huang2015single}
Huang, J.B., Singh, A., Ahuja, N.: Single image super-resolution from
  transformed self-exemplars. In: Proceedings of the IEEE conference on
  computer vision and pattern recognition. pp. 5197--5206 (2015)

\bibitem{kingma2014adam}
Kingma, D.P., Ba, J.: Adam: A method for stochastic optimization. arXiv
  preprint arXiv:1412.6980  (2014)

\bibitem{liang2021flow}
Liang, J., Zhang, K., Gu, S., Van~Gool, L., Timofte, R.: Flow-based kernel
  prior with application to blind super-resolution. In: Proceedings of the
  IEEE/CVF Conference on Computer Vision and Pattern Recognition. pp.
  10601--10610 (2021)

\bibitem{lim2017enhanced}
Lim, B., Son, S., Kim, H., Nah, S., Mu~Lee, K.: Enhanced deep residual networks
  for single image super-resolution. In: Proceedings of the IEEE conference on
  computer vision and pattern recognition workshops. pp. 136--144 (2017)

\bibitem{liu2020blind}
Liu, Y., Sun, W.: Blind super-resolution for single remote sensing image via
  sparse representation and transformed self-similarity. In: Journal of
  Physics: Conference Series. vol.~1575, p. 012115. IOP Publishing (2020)

\bibitem{luo2020unfolding}
Luo, Z., Huang, Y., Li, S., Wang, L., Tan, T.: Unfolding the alternating
  optimization for blind super resolution. arXiv preprint arXiv:2010.02631
  (2020)

\bibitem{MartinFTM01}
Martin, D., Fowlkes, C., Tal, D., Malik, J.: A database of human segmented
  natural images and its application to evaluating segmentation algorithms and
  measuring ecological statistics. In: Proc. 8th Int'l Conf. Computer Vision.
  vol.~2, pp. 416--423 (July 2001)

\bibitem{michaeli2013nonparametric}
Michaeli, T., Irani, M.: Nonparametric blind super-resolution. In: Proceedings
  of the IEEE International Conference on Computer Vision. pp. 945--952 (2013)

\bibitem{pan2016blind}
Pan, J., Sun, D., Pfister, H., Yang, M.H.: Blind image deblurring using dark
  channel prior. In: Proceedings of the IEEE Conference on Computer Vision and
  Pattern Recognition. pp. 1628--1636 (2016)

\bibitem{ronneberger2015u}
Ronneberger, O., Fischer, P., Brox, T.: U-net: Convolutional networks for
  biomedical image segmentation. In: International Conference on Medical image
  computing and computer-assisted intervention. pp. 234--241. Springer (2015)

\bibitem{samuel2009learning}
Samuel, K.G., Tappen, M.F.: Learning optimized map estimates in
  continuously-valued mrf models. In: 2009 IEEE Conference on Computer Vision
  and Pattern Recognition. pp. 477--484. IEEE (2009)

\bibitem{shao2015simple}
Shao, W.Z., Elad, M.: Simple, accurate, and robust nonparametric blind
  super-resolution. In: International Conference on Image and Graphics. pp.
  333--348. Springer (2015)

\bibitem{shocher2018zero}
Shocher, A., Cohen, N., Irani, M.: “zero-shot” super-resolution using deep
  internal learning. In: Proceedings of the IEEE conference on computer vision
  and pattern recognition. pp. 3118--3126 (2018)

\bibitem{wang2021unsupervised}
Wang, L., Wang, Y., Dong, X., Xu, Q., Yang, J., An, W., Guo, Y.: Unsupervised
  degradation representation learning for blind super-resolution. In:
  Proceedings of the IEEE/CVF Conference on Computer Vision and Pattern
  Recognition. pp. 10581--10590 (2021)

\bibitem{wang2005patch}
Wang, Q., Tang, X., Shum, H.: Patch based blind image super resolution. In:
  Tenth IEEE International Conference on Computer Vision (ICCV'05) Volume 1.
  vol.~1, pp. 709--716. IEEE (2005)

\bibitem{wang2012semi}
Wang, S., Zhang, L., Liang, Y., Pan, Q.: Semi-coupled dictionary learning with
  applications to image super-resolution and photo-sketch synthesis. In: 2012
  IEEE Conference on Computer Vision and Pattern Recognition. pp. 2216--2223.
  IEEE (2012)

\bibitem{wang2021realesrgan}
Wang, X., Xie, L., Dong, C., Shan, Y.: Real-esrgan: Training real-world blind
  super-resolution with pure synthetic data. In: International Conference on
  Computer Vision Workshops (ICCVW)

\bibitem{yamac2021kernelnet}
Yamac, M., Ataman, B., Nawaz, A.: Kernelnet: A blind super-resolution kernel
  estimation network. In: Proceedings of the IEEE/CVF Conference on Computer
  Vision and Pattern Recognition. pp. 453--462 (2021)

\bibitem{yang2019deep}
Yang, W., Zhang, X., Tian, Y., Wang, W., Xue, J.H., Liao, Q.: Deep learning for
  single image super-resolution: A brief review. IEEE Transactions on
  Multimedia  \textbf{21}(12),  3106--3121 (2019)

\bibitem{yuan2018unsupervised}
Yuan, Y., Liu, S., Zhang, J., Zhang, Y., Dong, C., Lin, L.: Unsupervised image
  super-resolution using cycle-in-cycle generative adversarial networks. In:
  Proceedings of the IEEE Conference on Computer Vision and Pattern Recognition
  Workshops. pp. 701--710 (2018)

\bibitem{zhang2020deep}
Zhang, K., Gool, L.V., Timofte, R.: Deep unfolding network for image
  super-resolution. In: Proceedings of the IEEE/CVF Conference on Computer
  Vision and Pattern Recognition. pp. 3217--3226 (2020)

\bibitem{zhang2021designing}
Zhang, K., Liang, J., Van~Gool, L., Timofte, R.: Designing a practical
  degradation model for deep blind image super-resolution. In: arxiv (2021)

\bibitem{zhang2018learning}
Zhang, K., Zuo, W., Zhang, L.: Learning a single convolutional super-resolution
  network for multiple degradations. In: Proceedings of the IEEE Conference on
  Computer Vision and Pattern Recognition. pp. 3262--3271 (2018)

\bibitem{zhang2018image}
Zhang, Y., Li, K., Li, K., Wang, L., Zhong, B., Fu, Y.: Image super-resolution
  using very deep residual channel attention networks. In: Proceedings of the
  European conference on computer vision (ECCV). pp. 286--301 (2018)

\bibitem{zhou2019kernel}
Zhou, R., Susstrunk, S.: Kernel modeling super-resolution on real
  low-resolution images. In: Proceedings of the IEEE/CVF International
  Conference on Computer Vision. pp. 2433--2443 (2019)

\end{thebibliography}

\clearpage

\title{Appendix}

\author{}
\authorrunning{H. Zheng \etal}

\titlerunning{Unfolded Deep Kernel Estimation for Blind Image Super-resolution}
%
\institute{}

\maketitle
\vspace{-2mm}

In this appendix, we provide the following materials:
\begin{itemize}
  \item More visualization of kernels estimated by UDKE and other kernel estimation methods (referring to Section 4.2 in the main paper).
  \item More visualization results of BISR with DEPD kernel pool (referring to Section 4.2 in the main paper).
  \item More visualization results of BISR on real-world low-resolution images (referring to Section 4.2 in the main paper).
\end{itemize}

\section{More Visualization of Estimated Kernels}
\label{sec:gaussian_experiment}
\begin{figure}[!h]
  \captionsetup{font=small}

  \captionsetup[subfigure]{justification=centering,font=scriptsize}
  \begin{subfigure}{0.18\textwidth}
    \includegraphics[width=\linewidth]{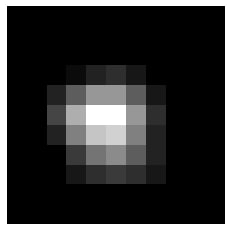}
    \caption{Ground truth\\ \ } \label{subfig:left}
  \end{subfigure}\hfill
  \begin{subfigure}{0.18\linewidth}
    \includegraphics[width=\linewidth]{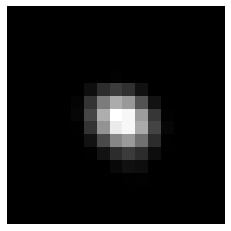}
    \caption{KGAN\\(42.3dB)}
  \end{subfigure}\hfill
  \begin{subfigure}{0.18\linewidth}
    \includegraphics[width=\linewidth]{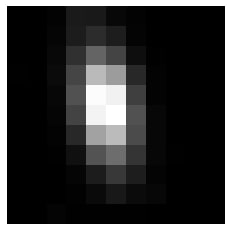}
    \caption{KFKP\\(38.1dB)}
  \end{subfigure}\hfill
  \begin{subfigure}{0.18\linewidth}
    \includegraphics[width=\linewidth]{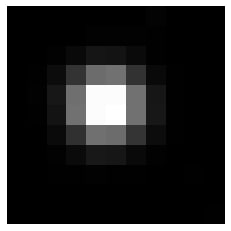}
    \caption{DFKP\\(41.0dB)}
  \end{subfigure}\hfill
  \begin{subfigure}{0.18\linewidth}
    \includegraphics[width=\linewidth]{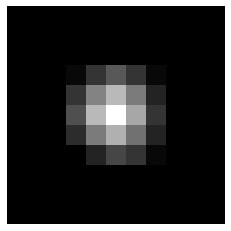}
    \caption{UDKE\\(46.4dB)}
  \end{subfigure}
  \caption[short]{Kernel estimation results by different methods on ``baby'' in Set5 with $s=3$ and $\sigma=2.55$.}
  \label{fig:kernel2}
  \vspace{-1mm}
\end{figure}

\begin{figure}[!h]
  \captionsetup{font=small}

  \captionsetup[subfigure]{justification=centering,font=scriptsize}
  \begin{subfigure}{0.18\textwidth}
    \includegraphics[width=\linewidth]{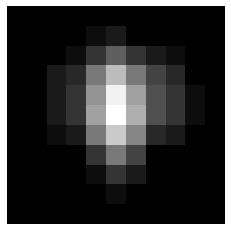}
    \caption{Ground truth\\ \ } \label{subfig:left}
  \end{subfigure}\hfill
  \begin{subfigure}{0.18\linewidth}
    \includegraphics[width=\linewidth]{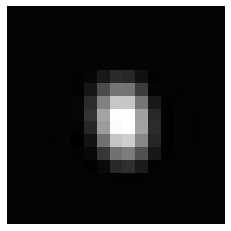}
    \caption{KGAN\\(43.9dB)}
  \end{subfigure}\hfill
  \begin{subfigure}{0.18\linewidth}
    \includegraphics[width=\linewidth]{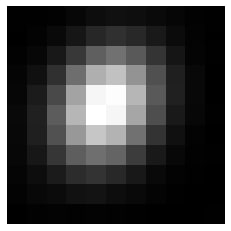}
    \caption{KFKP\\(38.4dB)}
  \end{subfigure}\hfill
  \begin{subfigure}{0.18\linewidth}
    \includegraphics[width=\linewidth]{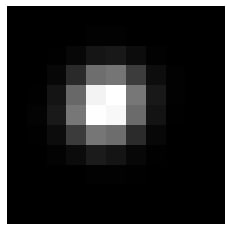}
    \caption{DFKP\\(42.8dB)}
  \end{subfigure}\hfill
  \begin{subfigure}{0.18\linewidth}
    \includegraphics[width=\linewidth]{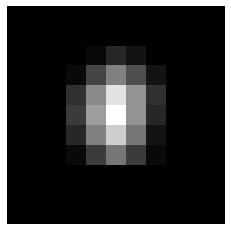}
    \caption{UDKE\\(47.5dB)}
  \end{subfigure}
  \caption[short]{Kernel estimation results by different methods on ``img051'' in Urban100 with $s=4$ and $\sigma=7.65$.}
  \label{fig:kernel3}
  \vspace{-1mm}
\end{figure}

Figs.~\ref{fig:kernel2}$\sim$\ref{fig:kernel3} show more visualization of kernels estimated by UDKE and other kernel estimation methods. It can be seen that UDKE produces more reliable estimation of unseen blur kernels during inference.

\section{More Visualization Results of BISR with DEPD Kernel Pool}
\label{sec:synthesis}
\begin{figure*}[!ht]
  \captionsetup{font=small}

  \captionsetup[subfigure]{justification=centering,font=scriptsize}
  \centering
  \begin{subfigure}{.198\textwidth}
    \setlength{\abovecaptionskip}{0pt}
    \setlength{\belowcaptionskip}{0pt}
    \centering
    \includegraphics[width=\textwidth]{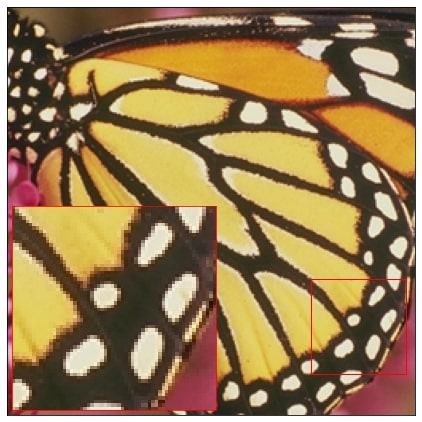}
    \caption[short]{Ground truth\\ \ }
  \end{subfigure}\hfill%
  \begin{subfigure}{.198\textwidth}
    \setlength{\abovecaptionskip}{0pt}
    \setlength{\belowcaptionskip}{0pt}
    \centering
    \includegraphics[width=\textwidth]{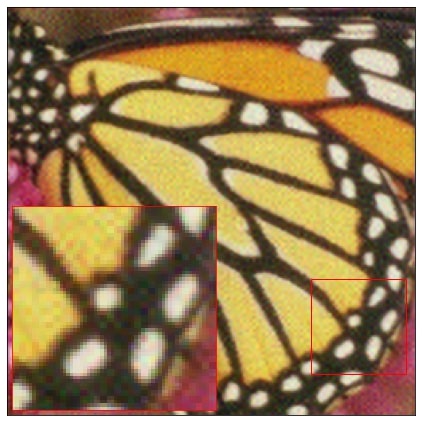}
    \caption[centering]{RCAN \\(19.05dB/.6661)}
  \end{subfigure}\hfill%
  \begin{subfigure}{.198\textwidth}
    \setlength{\abovecaptionskip}{0pt}
    \setlength{\belowcaptionskip}{0pt}
    \centering
    \includegraphics[width=\textwidth]{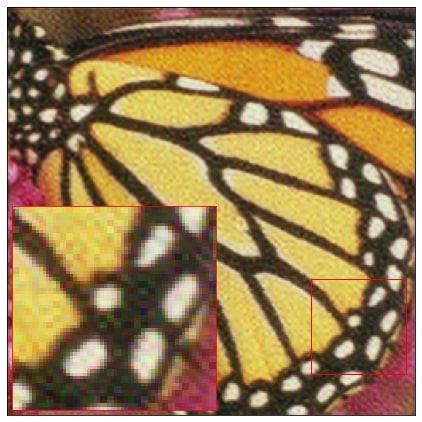}
    \caption[centering]{ZSSR \\(18.92dB/.6489)}
  \end{subfigure}\hfill%
  \begin{subfigure}{.198\textwidth}
    \setlength{\abovecaptionskip}{0pt}
    \setlength{\belowcaptionskip}{0pt}
    \centering
    \includegraphics[width=\textwidth]{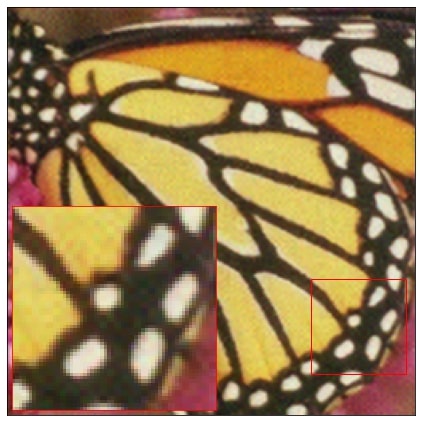}
    \caption[centering]{IKC \\(24.11dB/.7717)}
  \end{subfigure}\hfill%
  \begin{subfigure}{.198\textwidth}
    \setlength{\abovecaptionskip}{0pt}
    \setlength{\belowcaptionskip}{0pt}
    \centering
    \includegraphics[width=\textwidth]{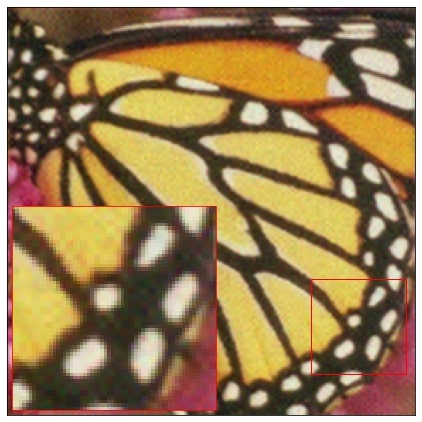}
    \caption[short]{DAN \\(23.62dB/.7476)}
  \end{subfigure}\hfill%
  \begin{subfigure}{.198\textwidth}
    \setlength{\abovecaptionskip}{0pt}
    \setlength{\belowcaptionskip}{0pt}
    \centering
    \includegraphics[width=\textwidth]{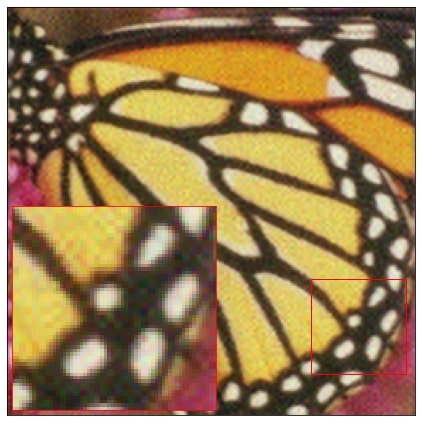}
    \caption[short]{DASR \\(22.97dB/.7267)}
  \end{subfigure}\hfill%
  \begin{subfigure}{.198\textwidth}
    \setlength{\abovecaptionskip}{0pt}
    \setlength{\belowcaptionskip}{0pt}
    \centering
    \includegraphics[width=\textwidth]{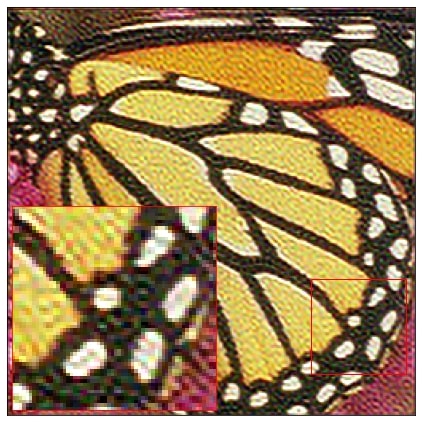}
    \caption[short]{KGAN \\(18.74dB/.4677)}
  \end{subfigure}\hfill%
  \begin{subfigure}{.198\textwidth}
    \setlength{\abovecaptionskip}{0pt}
    \setlength{\belowcaptionskip}{0pt}
    \centering
    \includegraphics[width=\textwidth]{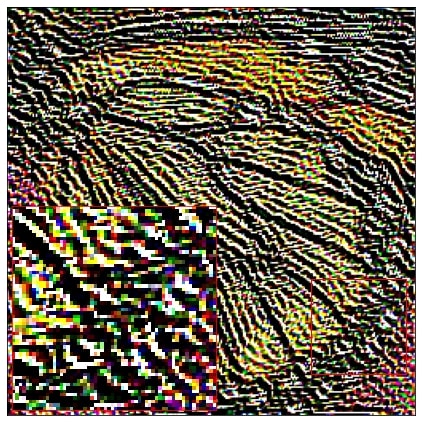}
    \caption[short]{KFKP \\(8.26dB/.1274)}
  \end{subfigure}\hfill%
  \begin{subfigure}{.198\textwidth}
    \setlength{\abovecaptionskip}{0pt}
    \setlength{\belowcaptionskip}{0pt}
    \centering
    \includegraphics[width=\textwidth]{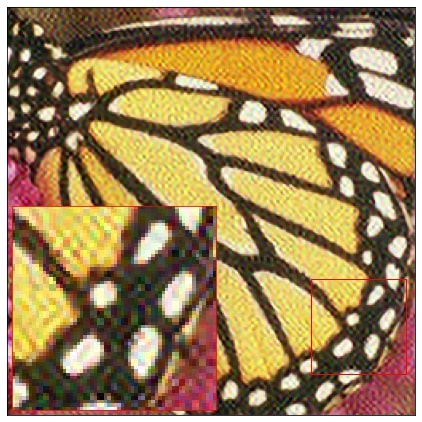}
    \caption[centering]{DFKP \\(19.80dB/.5118)}
  \end{subfigure}\hfill%
  \begin{subfigure}{.198\textwidth}
    \setlength{\abovecaptionskip}{0pt}
    \setlength{\belowcaptionskip}{0pt}
    \centering
    \includegraphics[width=\textwidth]{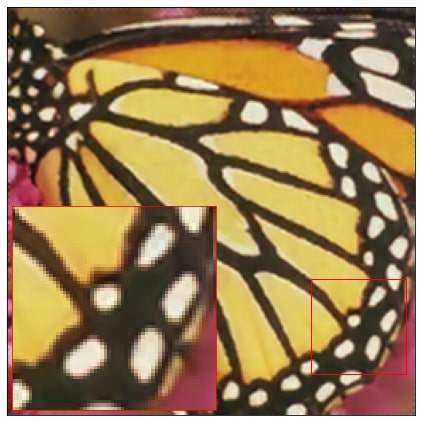}
    \caption[short]{UDKE \\(25.28dB/.8421)}
  \end{subfigure}\hfill%
  \caption[short]{BISR results (PSNR/SSIM) on ``butterfly'' in Set5 with $s=2$ and $\sigma=7.65$.}
  \label{fig:1}
  \vspace{-2mm}
\end{figure*}

\begin{figure*}[!ht]
  \captionsetup{font=small}

  \captionsetup[subfigure]{justification=centering,font=scriptsize}
  \centering
  \begin{subfigure}{.22\textwidth}
    \setlength{\abovecaptionskip}{0pt}
    \setlength{\belowcaptionskip}{0pt}
    \centering
    \includegraphics[width=\textwidth]{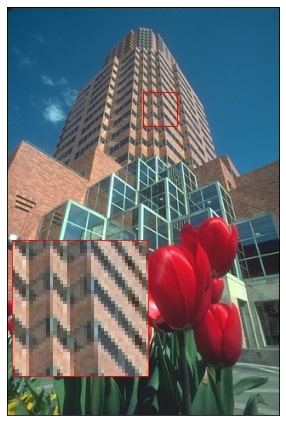}
    \caption[short]{Ground truth\\ \ }
  \end{subfigure}\hfill%
  \begin{subfigure}{.22\textwidth}
    \setlength{\abovecaptionskip}{0pt}
    \setlength{\belowcaptionskip}{0pt}
    \centering
    \includegraphics[width=\textwidth]{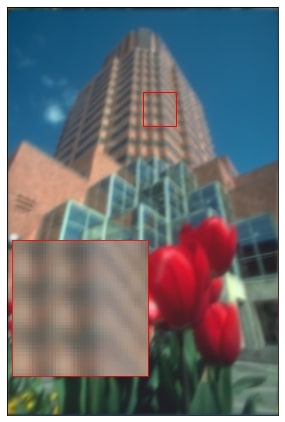}
    \caption[centering]{RCAN \\(19.29dB/.5469)}
  \end{subfigure}\hfill%
  \begin{subfigure}{.22\textwidth}
    \setlength{\abovecaptionskip}{0pt}
    \setlength{\belowcaptionskip}{0pt}
    \centering
    \includegraphics[width=\textwidth]{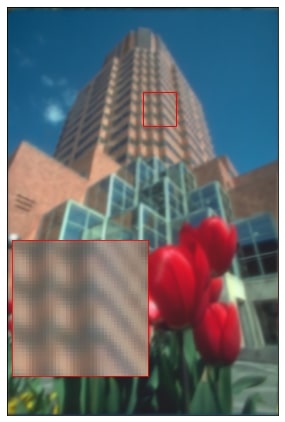}
    \caption[centering]{ZSSR \\(19.86/.6137)}
  \end{subfigure}\hfill%
  \begin{subfigure}{.22\textwidth}
    \setlength{\abovecaptionskip}{0pt}
    \setlength{\belowcaptionskip}{0pt}
    \centering
    \includegraphics[width=\textwidth]{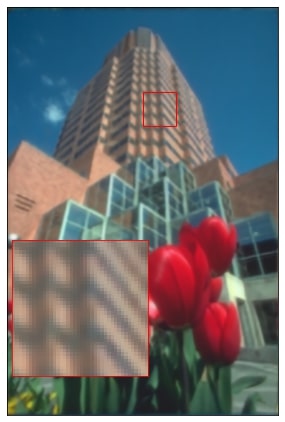}
    \caption[centering]{IKC \\(20.56dB/.6967)}
  \end{subfigure}\hfill%
  \begin{subfigure}{.22\textwidth}
    \setlength{\abovecaptionskip}{0pt}
    \setlength{\belowcaptionskip}{0pt}
    \centering
    \includegraphics[width=\textwidth]{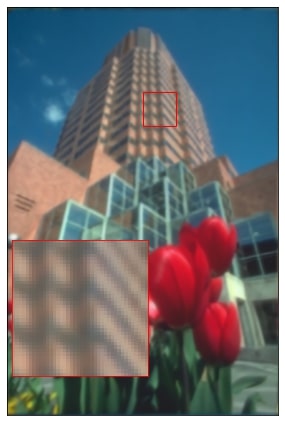}
    \caption[short]{DAN \\(20.49dB/.6903)}
  \end{subfigure}\hfill%
  \begin{subfigure}{.22\textwidth}
    \setlength{\abovecaptionskip}{0pt}
    \setlength{\belowcaptionskip}{0pt}
    \centering
    \includegraphics[width=\textwidth]{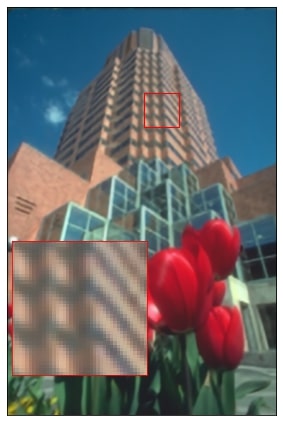}
    \caption[short]{DASR \\(20.30dB/.6178)}
  \end{subfigure}\hfill%
  \begin{subfigure}{.22\textwidth}
    \setlength{\abovecaptionskip}{0pt}
    \setlength{\belowcaptionskip}{0pt}
    \centering
    \includegraphics[width=\textwidth]{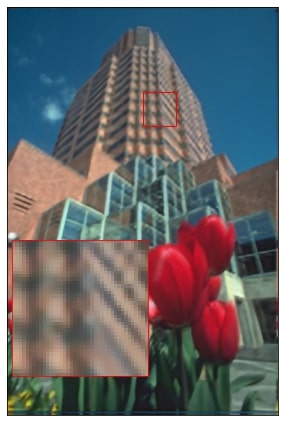}
    \caption[centering]{DFKP \\(18.31dB/.5529)}
  \end{subfigure}\hfill%
  \begin{subfigure}{.22\textwidth}
    \setlength{\abovecaptionskip}{0pt}
    \setlength{\belowcaptionskip}{0pt}
    \centering
    \includegraphics[width=\textwidth]{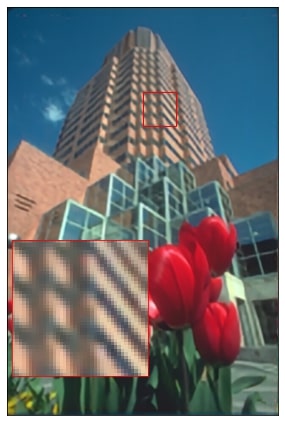}
    \caption[short]{UDKE \\(22.38dB/.7540)}
  \end{subfigure}\hfill%
  \caption[short]{BISR results (PSNR/SSIM) on ``86000'' in BSD100 with $s=3$ and $\sigma=0$.}
  \label{fig:4}
  \vspace{-2mm}
\end{figure*}

\begin{figure*}[!ht]
  \captionsetup{font=small}
  \captionsetup[subfigure]{justification=centering,font=scriptsize}
  \centering
  \begin{subfigure}{.22\textwidth}
    \setlength{\abovecaptionskip}{0pt}
    \setlength{\belowcaptionskip}{0pt}
    \centering
    \includegraphics[width=\textwidth]{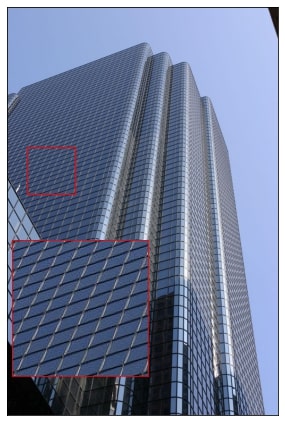}
    \caption[short]{Ground truth\\ \ }
  \end{subfigure}\hfill%
  \begin{subfigure}{.22\textwidth}
    \setlength{\abovecaptionskip}{0pt}
    \setlength{\belowcaptionskip}{0pt}
    \centering
    \includegraphics[width=\textwidth]{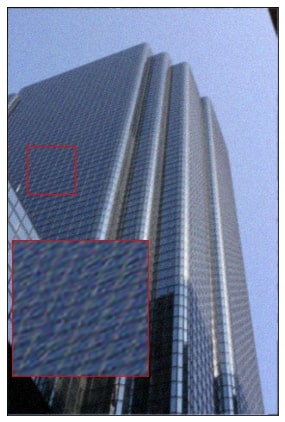}
    \caption[centering]{RCAN \\(24.60dB/.7035)}
  \end{subfigure}\hfill%
  \begin{subfigure}{.22\textwidth}
    \setlength{\abovecaptionskip}{0pt}
    \setlength{\belowcaptionskip}{0pt}
    \centering
    \includegraphics[width=\textwidth]{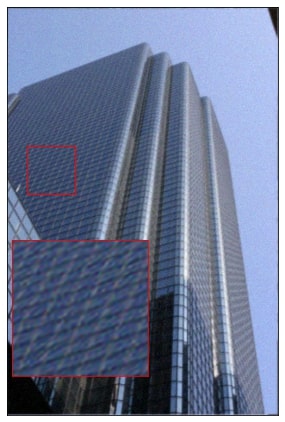}
    \caption[centering]{ZSSR \\(24.39dB/.7164)}
  \end{subfigure}\hfill%
  \begin{subfigure}{.22\textwidth}
    \setlength{\abovecaptionskip}{0pt}
    \setlength{\belowcaptionskip}{0pt}
    \centering
    \includegraphics[width=\textwidth]{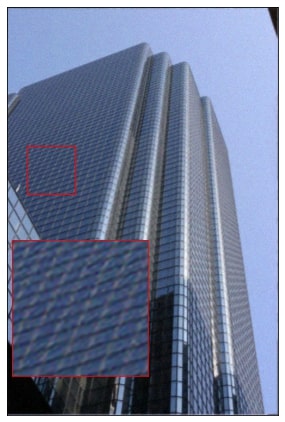}
    \caption[centering]{IKC \\(24.94dB/.7665)}
  \end{subfigure}\hfill%
  \begin{subfigure}{.22\textwidth}
    \setlength{\abovecaptionskip}{0pt}
    \setlength{\belowcaptionskip}{0pt}
    \centering
    \includegraphics[width=\textwidth]{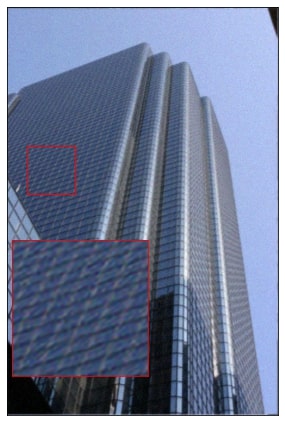}
    \caption[short]{DAN \\(24.71dB/.7442)}
  \end{subfigure}\hfill%
  \begin{subfigure}{.22\textwidth}
    \setlength{\abovecaptionskip}{0pt}
    \setlength{\belowcaptionskip}{0pt}
    \centering
    \includegraphics[width=\textwidth]{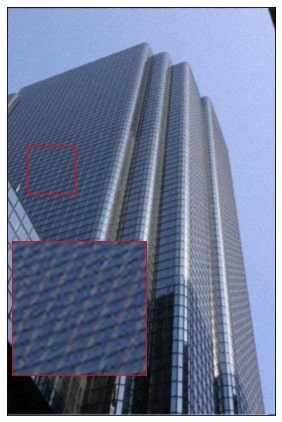}
    \caption[short]{DASR \\(24.29dB/.7631)}
  \end{subfigure}\hfill%
  \begin{subfigure}{.22\textwidth}
    \setlength{\abovecaptionskip}{0pt}
    \setlength{\belowcaptionskip}{0pt}
    \centering
    \includegraphics[width=\textwidth]{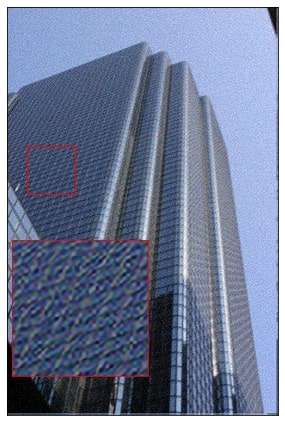}
    \caption[centering]{DFKP \\(20.62dB/.3900)}
  \end{subfigure}\hfill%
  \begin{subfigure}{.22\textwidth}
    \setlength{\abovecaptionskip}{0pt}
    \setlength{\belowcaptionskip}{0pt}
    \centering
    \includegraphics[width=\textwidth]{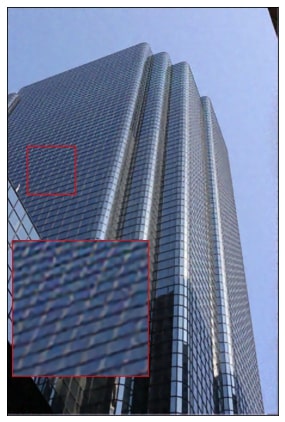}
    \caption[short]{UDKE \\(25.98dB/.8240)}
  \end{subfigure}\hfill%
  \caption[short]{BISR results (PSNR/SSIM) on ``074'' in Urban100 with $s=3$ and $\sigma=7.65$.}
  \label{fig:5}
  \vspace{-2mm}
\end{figure*}

\begin{figure*}[!ht]
  \captionsetup{font=small}

  \captionsetup[subfigure]{justification=centering,font=scriptsize}
  \centering
  \begin{subfigure}{.198\textwidth}
    \setlength{\abovecaptionskip}{0pt}
    \setlength{\belowcaptionskip}{0pt}
    \centering
    \includegraphics[width=\textwidth]{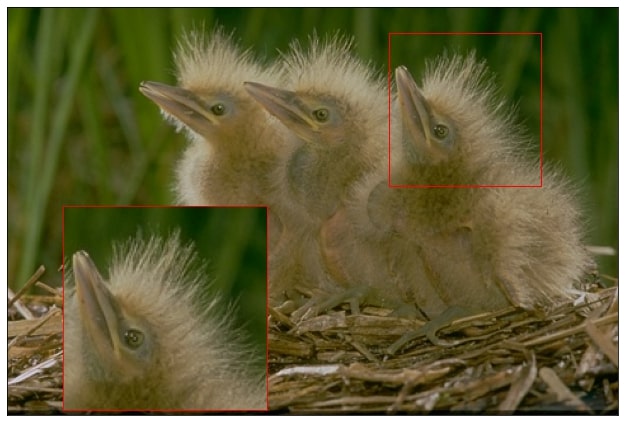}
    \caption[short]{Ground truth\\ \ }
  \end{subfigure}\hfill%
  \begin{subfigure}{.198\textwidth}
    \setlength{\abovecaptionskip}{0pt}
    \setlength{\belowcaptionskip}{0pt}
    \centering
    \includegraphics[width=\textwidth]{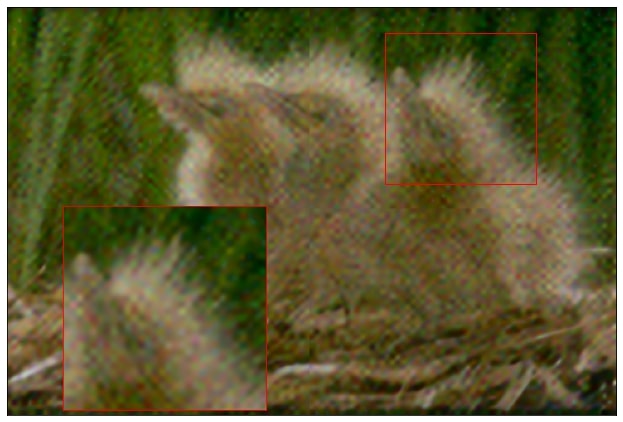}
    \caption[centering]{RCAN \\(21.81dB/.5349)}
  \end{subfigure}\hfill%
  \begin{subfigure}{.198\textwidth}
    \setlength{\abovecaptionskip}{0pt}
    \setlength{\belowcaptionskip}{0pt}
    \centering
    \includegraphics[width=\textwidth]{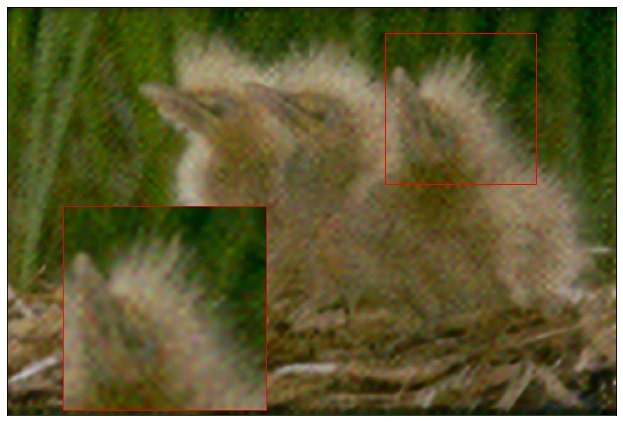}
    \caption[centering]{ZSSR \\(19.96dB/.5129)}
  \end{subfigure}\hfill%
  \begin{subfigure}{.198\textwidth}
    \setlength{\abovecaptionskip}{0pt}
    \setlength{\belowcaptionskip}{0pt}
    \centering
    \includegraphics[width=\textwidth]{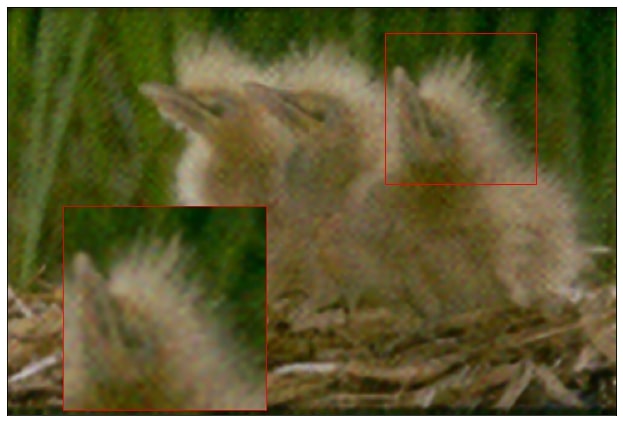}
    \caption[centering]{IKC \\(21.81dB/.5806)}
  \end{subfigure}\hfill%
  \begin{subfigure}{.198\textwidth}
    \setlength{\abovecaptionskip}{0pt}
    \setlength{\belowcaptionskip}{0pt}
    \centering
    \includegraphics[width=\textwidth]{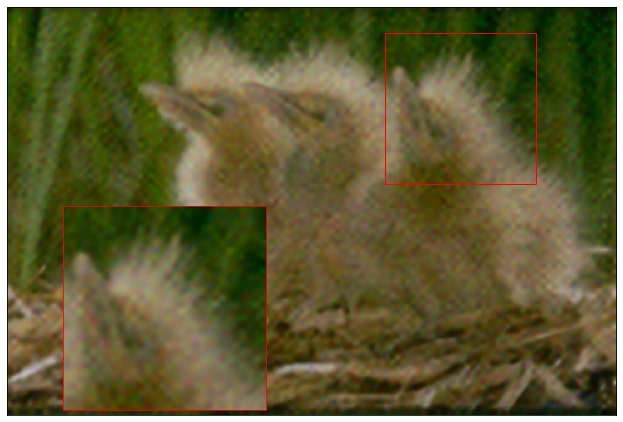}
    \caption[short]{DAN \\(22.00dB/.5943)}
  \end{subfigure}\hfill%
  \begin{subfigure}{.198\textwidth}
    \setlength{\abovecaptionskip}{0pt}
    \setlength{\belowcaptionskip}{0pt}
    \centering
    \includegraphics[width=\textwidth]{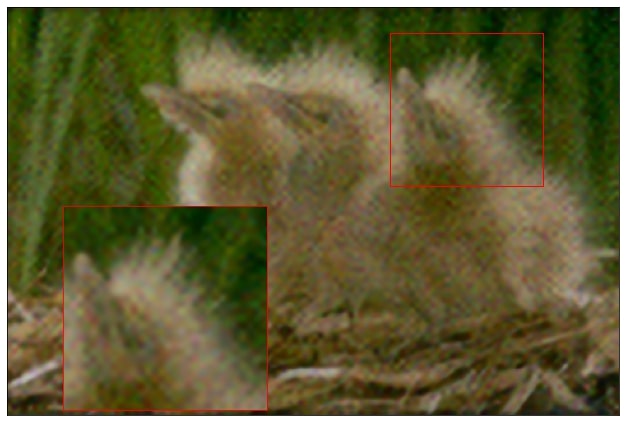}
    \caption[short]{DASR \\(22.70dB/.5850)}
  \end{subfigure}\hfill%
  \begin{subfigure}{.198\textwidth}
    \setlength{\abovecaptionskip}{0pt}
    \setlength{\belowcaptionskip}{0pt}
    \centering
    \includegraphics[width=\textwidth]{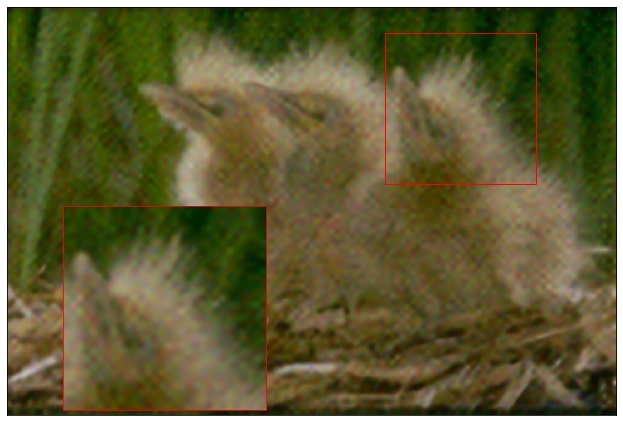}
    \caption[short]{KGAN \\(20.00dB/.5148)}
  \end{subfigure}\hfill%
  \begin{subfigure}{.198\textwidth}
    \setlength{\abovecaptionskip}{0pt}
    \setlength{\belowcaptionskip}{0pt}
    \centering
    \includegraphics[width=\textwidth]{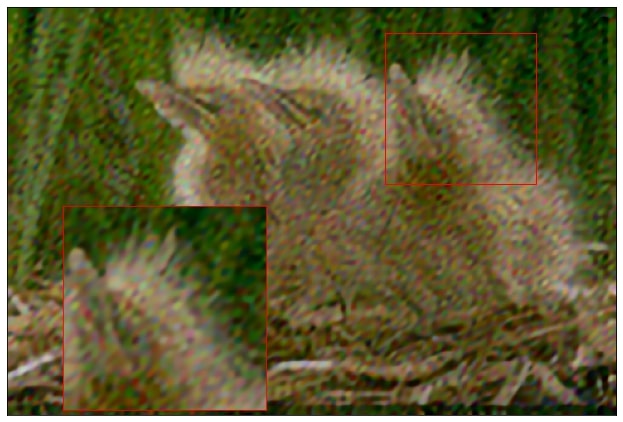}
    \caption[short]{KFKP \\(19.53dB/.3468)}
  \end{subfigure}\hfill%
  \begin{subfigure}{.198\textwidth}
    \setlength{\abovecaptionskip}{0pt}
    \setlength{\belowcaptionskip}{0pt}
    \centering
    \includegraphics[width=\textwidth]{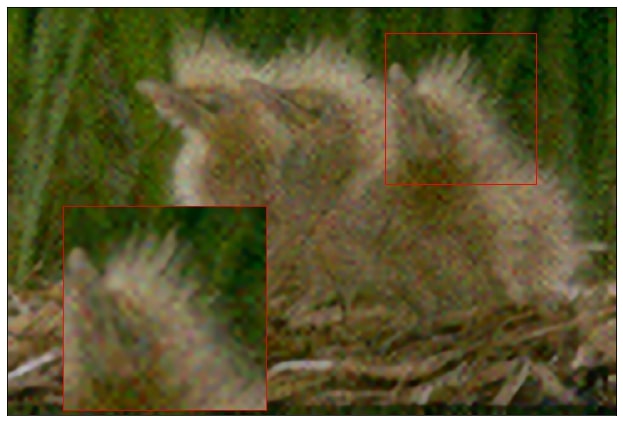}
    \caption[centering]{DFKP \\(21.12dB/.5401)}
  \end{subfigure}\hfill%
  \begin{subfigure}{.198\textwidth}
    \setlength{\abovecaptionskip}{0pt}
    \setlength{\belowcaptionskip}{0pt}
    \centering
    \includegraphics[width=\textwidth]{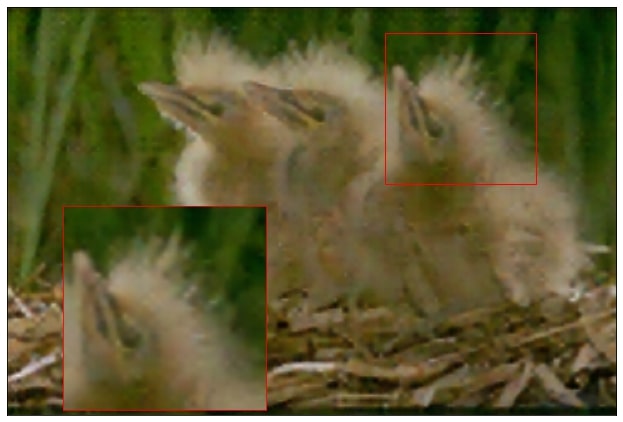}
    \caption[short]{UDKE \\(23.49dB/.6804)}
  \end{subfigure}\hfill%
  \caption[short]{BISR results (PSNR/SSIM) on ``163085'' in BSD100 with $s=4$ and $\sigma=7.65$.}
  \label{fig:7}
  \vspace{-2mm}
\end{figure*}

\begin{figure*}[!ht]
  \captionsetup{font=small}
  \captionsetup[subfigure]{justification=centering,font=scriptsize}
  \centering
  \begin{subfigure}{.198\textwidth}
    \setlength{\abovecaptionskip}{0pt}
    \setlength{\belowcaptionskip}{0pt}
    \centering
    \includegraphics[width=\textwidth]{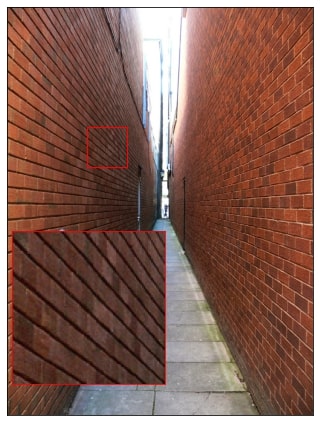}
    \caption[short]{Ground truth\\ \ }
  \end{subfigure}\hfill%
  \begin{subfigure}{.198\textwidth}
    \setlength{\abovecaptionskip}{0pt}
    \setlength{\belowcaptionskip}{0pt}
    \centering
    \includegraphics[width=\textwidth]{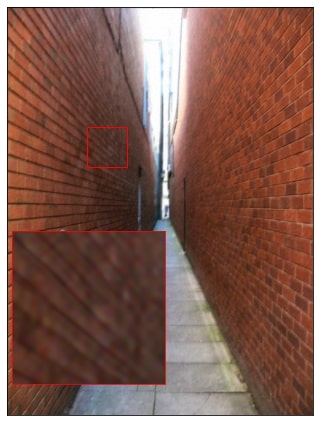}
    \caption[centering]{RCAN \\(26.78dB/.6835)}
  \end{subfigure}\hfill%
  \begin{subfigure}{.198\textwidth}
    \setlength{\abovecaptionskip}{0pt}
    \setlength{\belowcaptionskip}{0pt}
    \centering
    \includegraphics[width=\textwidth]{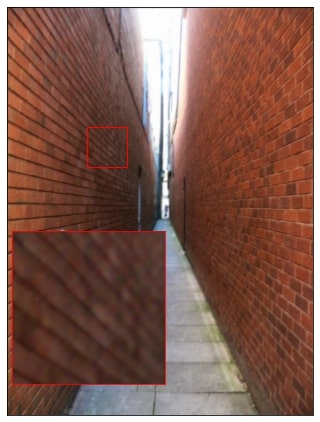}
    \caption[centering]{ZSSR \\(25.16dB/.6843)}
  \end{subfigure}\hfill%
  \begin{subfigure}{.198\textwidth}
    \setlength{\abovecaptionskip}{0pt}
    \setlength{\belowcaptionskip}{0pt}
    \centering
    \includegraphics[width=\textwidth]{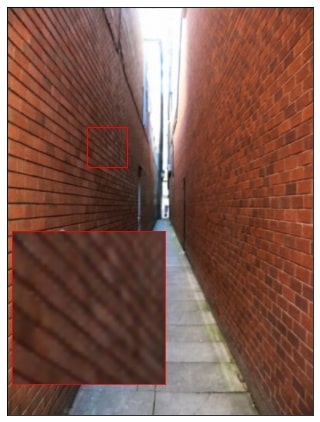}
    \caption[centering]{IKC \\(28.52dB/.7330)}
  \end{subfigure}\hfill%
  \begin{subfigure}{.198\textwidth}
    \setlength{\abovecaptionskip}{0pt}
    \setlength{\belowcaptionskip}{0pt}
    \centering
    \includegraphics[width=\textwidth]{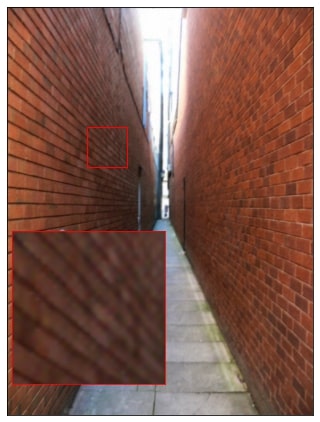}
    \caption[short]{DAN \\(28.04dB/.7319)}
  \end{subfigure}\hfill%
  \begin{subfigure}{.198\textwidth}
    \setlength{\abovecaptionskip}{0pt}
    \setlength{\belowcaptionskip}{0pt}
    \centering
    \includegraphics[width=\textwidth]{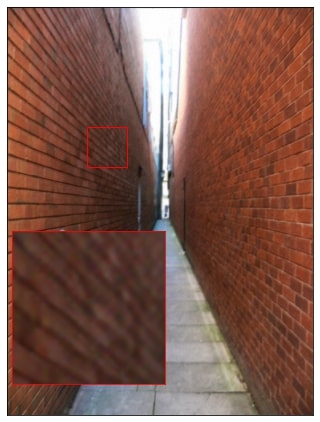}
    \caption[short]{DASR \\(27.68dB/.7330)}
  \end{subfigure}\hfill%
  \begin{subfigure}{.198\textwidth}
    \setlength{\abovecaptionskip}{0pt}
    \setlength{\belowcaptionskip}{0pt}
    \centering
    \includegraphics[width=\textwidth]{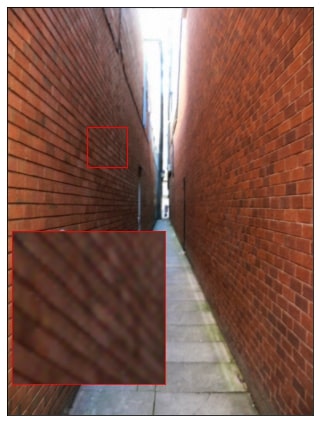}
    \caption[short]{KGAN \\(25.09dB/.6735)}
  \end{subfigure}\hfill%
  \begin{subfigure}{.198\textwidth}
    \setlength{\abovecaptionskip}{0pt}
    \setlength{\belowcaptionskip}{0pt}
    \centering
    \includegraphics[width=\textwidth]{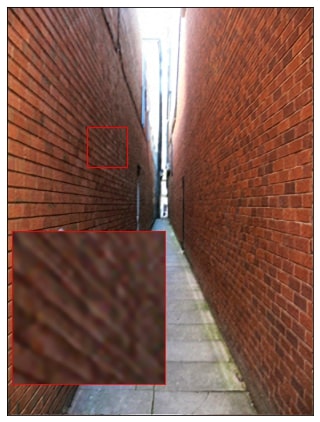}
    \caption[short]{KFKP \\(24.63dB/.6062)}
  \end{subfigure}\hfill%
  \begin{subfigure}{.198\textwidth}
    \setlength{\abovecaptionskip}{0pt}
    \setlength{\belowcaptionskip}{0pt}
    \centering
    \includegraphics[width=\textwidth]{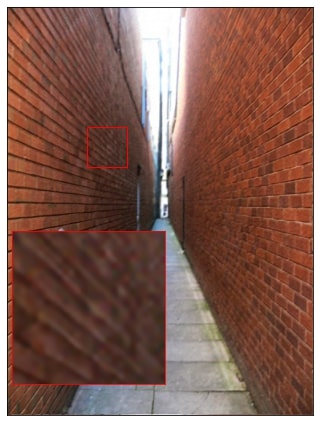}
    \caption[centering]{DFKP \\(25.41dB/.6446)}
  \end{subfigure}\hfill%
  \begin{subfigure}{.198\textwidth}
    \setlength{\abovecaptionskip}{0pt}
    \setlength{\belowcaptionskip}{0pt}
    \centering
    \includegraphics[width=\textwidth]{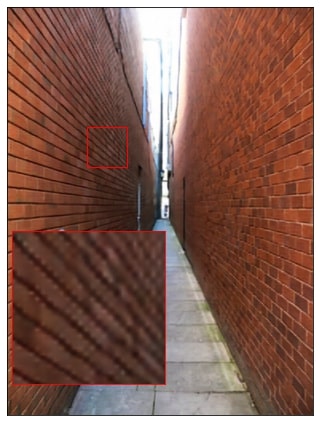}
    \caption[short]{UDKE \\(30.03dB/.8342)}
  \end{subfigure}\hfill%
  \caption[short]{BISR results (PSNR/SSIM) on ``038'' in Urban100 with $s=4$ and $\sigma=2.55$.}
  \label{fig:8}
  \vspace{-2mm}
\end{figure*}

Figs.~\ref{fig:1}$\sim$\ref{fig:8} visualize the BISR results with DEPD kernel pool. It can be seen that UDKE generally recovers richer and clearer textures (see Figs.~\ref{fig:4} and~\ref{fig:8}) and it is more robust to noisy images (see Figs.~\ref{fig:1}, \ref{fig:5}, \ref{fig:7}).

\section{More Visualization Results of BISR on Real-world Images}
\label{sec:real}

\begin{figure*}[!ht]
  \captionsetup{font=small}
  \captionsetup[subfigure]{justification=centering,font=scriptsize}
  \centering
  \begin{subfigure}{.198\textwidth}
    \setlength{\abovecaptionskip}{0pt}
    \setlength{\belowcaptionskip}{0pt}
    \centering
    \includegraphics[width=\textwidth]{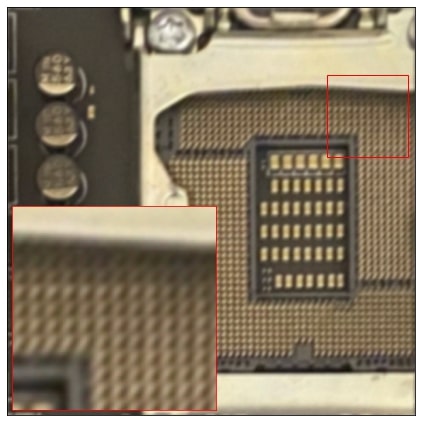}
    \caption[short]{LR image}
  \end{subfigure}\hfill%
  \begin{subfigure}{.198\textwidth}
    \setlength{\abovecaptionskip}{0pt}
    \setlength{\belowcaptionskip}{0pt}
    \centering
    \includegraphics[width=\textwidth]{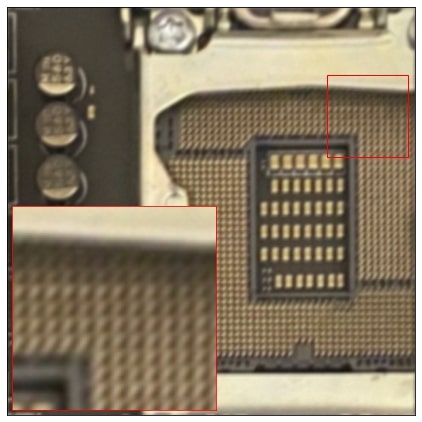}
    \caption[short]{RCAN}
  \end{subfigure}\hfill%
  \begin{subfigure}{.198\textwidth}
    \setlength{\abovecaptionskip}{0pt}
    \setlength{\belowcaptionskip}{0pt}
    \centering
    \includegraphics[width=\textwidth]{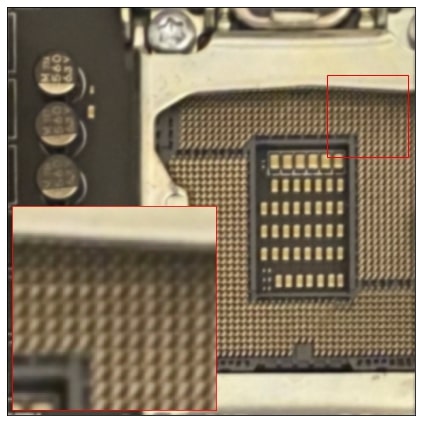}
    \caption[centering]{IKC}
  \end{subfigure}\hfill%
  \begin{subfigure}{.198\textwidth}
    \setlength{\abovecaptionskip}{0pt}
    \setlength{\belowcaptionskip}{0pt}
    \centering
    \includegraphics[width=\textwidth]{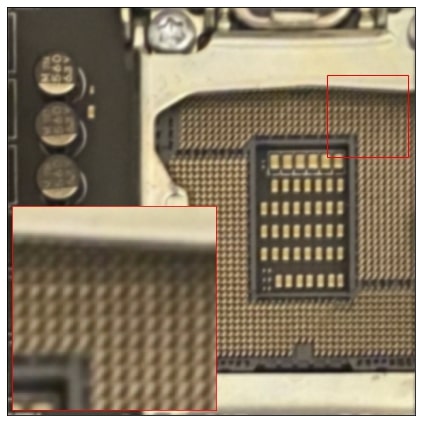}
    \caption[centering]{DAN}
  \end{subfigure}\hfill%
  \begin{subfigure}{.198\textwidth}
    \setlength{\abovecaptionskip}{0pt}
    \setlength{\belowcaptionskip}{0pt}
    \centering
    \includegraphics[width=\textwidth]{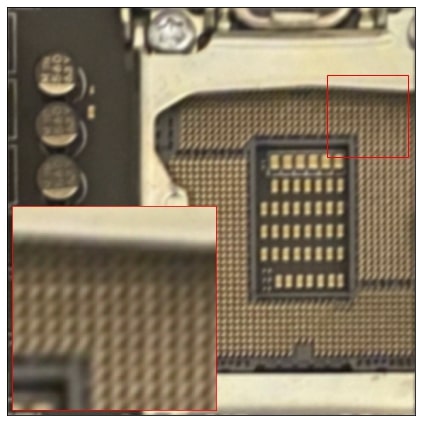}
    \caption[centering]{DASR}
  \end{subfigure}\hfill%
  \begin{subfigure}{.198\textwidth}
    \setlength{\abovecaptionskip}{0pt}
    \setlength{\belowcaptionskip}{0pt}
    \centering
    \includegraphics[width=\textwidth]{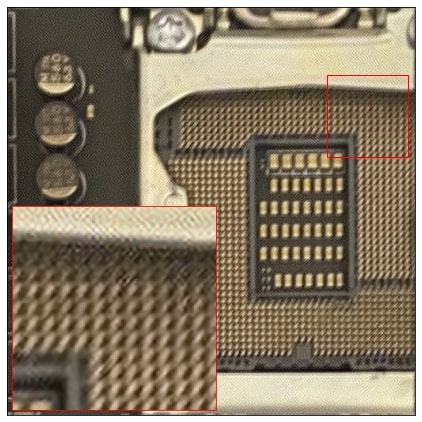}
    \caption[short]{KFKP}
  \end{subfigure}\hfill%
  \begin{subfigure}{.198\textwidth}
    \setlength{\abovecaptionskip}{0pt}
    \setlength{\belowcaptionskip}{0pt}
    \centering
    \includegraphics[width=\textwidth]{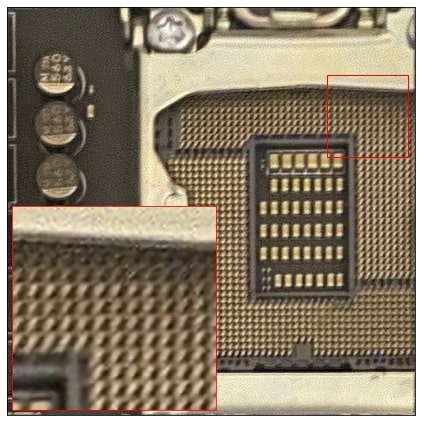}
    \caption[short]{DFKP}
  \end{subfigure}\hfill%
  \begin{subfigure}{.198\textwidth}
    \setlength{\abovecaptionskip}{0pt}
    \setlength{\belowcaptionskip}{0pt}
    \centering
    \includegraphics[width=\textwidth]{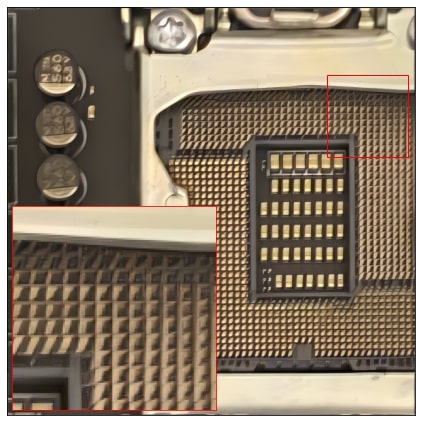}
    \caption[centering]{BSRGAN}
  \end{subfigure}\hfill%
  \begin{subfigure}{.198\textwidth}
    \setlength{\abovecaptionskip}{0pt}
    \setlength{\belowcaptionskip}{0pt}
    \centering
    \includegraphics[width=\textwidth]{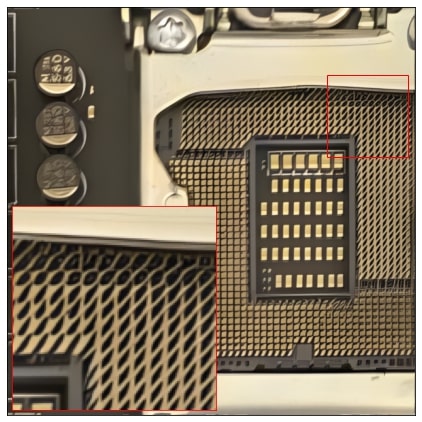}
    \caption[short]{RESRGAN}
  \end{subfigure}\hfill%
  \begin{subfigure}{.198\textwidth}
    \setlength{\abovecaptionskip}{0pt}
    \setlength{\belowcaptionskip}{0pt}
    \centering
    \includegraphics[width=\textwidth]{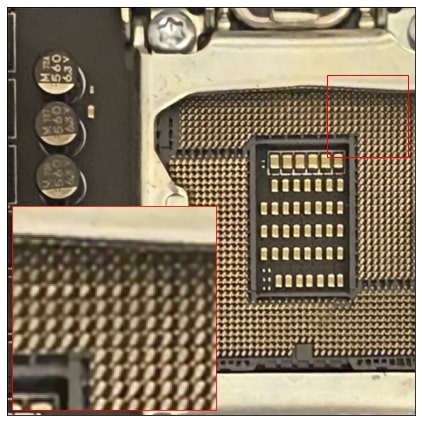}
    \caption[short]{UDKE}
  \end{subfigure}\hfill%
  \caption[short]{$\times 2$ BISR results on a real-world image (better viewed on screen).}
  \label{fig:real2}
  \vspace{-1mm}
\end{figure*}

\begin{figure*}[!ht]
  \captionsetup{font=small}
  \captionsetup[subfigure]{justification=centering,font=scriptsize}
  \centering
  \begin{subfigure}{.198\textwidth}
    \setlength{\abovecaptionskip}{0pt}
    \setlength{\belowcaptionskip}{0pt}
    \centering
    \includegraphics[width=\textwidth]{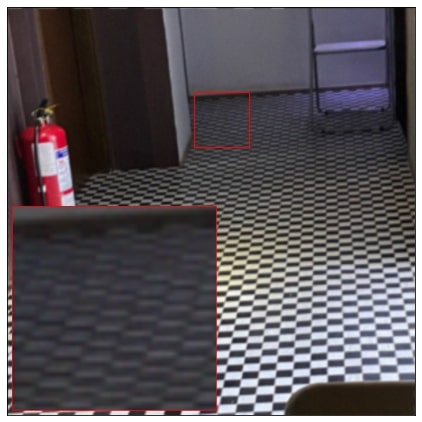}
    \caption[short]{LR image}
  \end{subfigure}\hfill%
  \begin{subfigure}{.198\textwidth}
    \setlength{\abovecaptionskip}{0pt}
    \setlength{\belowcaptionskip}{0pt}
    \centering
    \includegraphics[width=\textwidth]{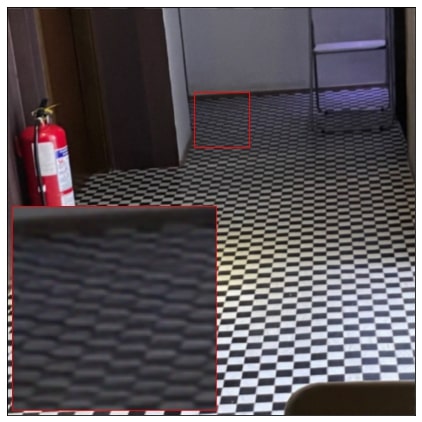}
    \caption[short]{RCAN}
  \end{subfigure}\hfill%
  \begin{subfigure}{.198\textwidth}
    \setlength{\abovecaptionskip}{0pt}
    \setlength{\belowcaptionskip}{0pt}
    \centering
    \includegraphics[width=\textwidth]{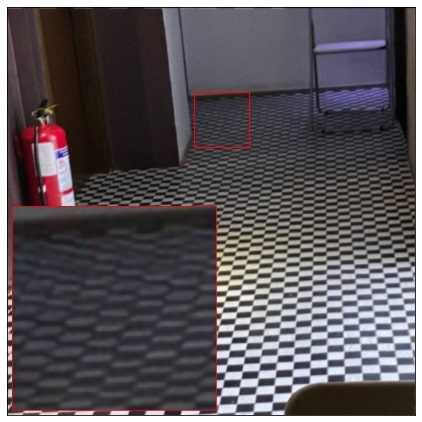}
    \caption[centering]{IKC}
  \end{subfigure}\hfill%
  \begin{subfigure}{.198\textwidth}
    \setlength{\abovecaptionskip}{0pt}
    \setlength{\belowcaptionskip}{0pt}
    \centering
    \includegraphics[width=\textwidth]{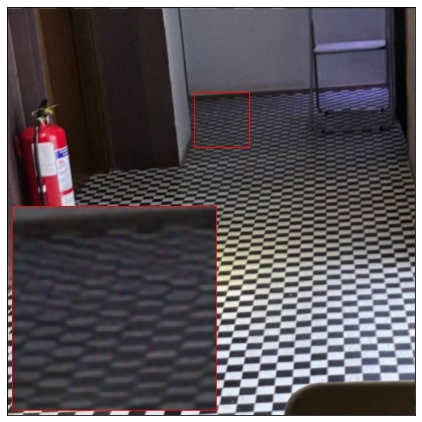}
    \caption[centering]{DAN}
  \end{subfigure}\hfill%
  \begin{subfigure}{.198\textwidth}
    \setlength{\abovecaptionskip}{0pt}
    \setlength{\belowcaptionskip}{0pt}
    \centering
    \includegraphics[width=\textwidth]{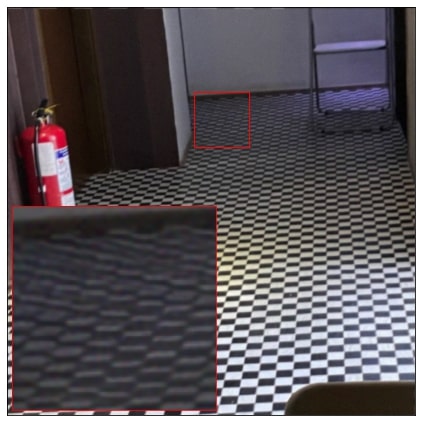}
    \caption[centering]{DASR}
  \end{subfigure}\hfill%
  \begin{subfigure}{.198\textwidth}
    \setlength{\abovecaptionskip}{0pt}
    \setlength{\belowcaptionskip}{0pt}
    \centering
    \includegraphics[width=\textwidth]{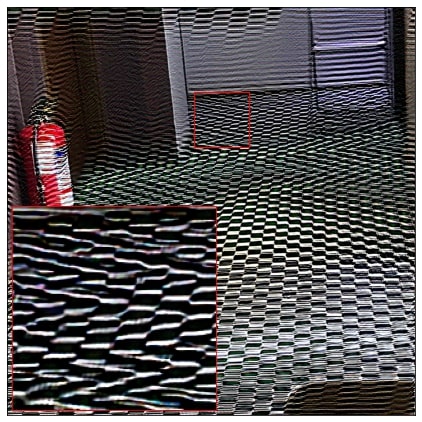}
    \caption[short]{KFKP}
  \end{subfigure}\hfill%
  \begin{subfigure}{.198\textwidth}
    \setlength{\abovecaptionskip}{0pt}
    \setlength{\belowcaptionskip}{0pt}
    \centering
    \includegraphics[width=\textwidth]{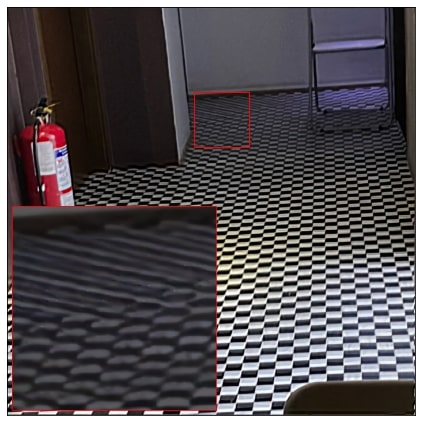}
    \caption[short]{DFKP}
  \end{subfigure}\hfill%
  \begin{subfigure}{.198\textwidth}
    \setlength{\abovecaptionskip}{0pt}
    \setlength{\belowcaptionskip}{0pt}
    \centering
    \includegraphics[width=\textwidth]{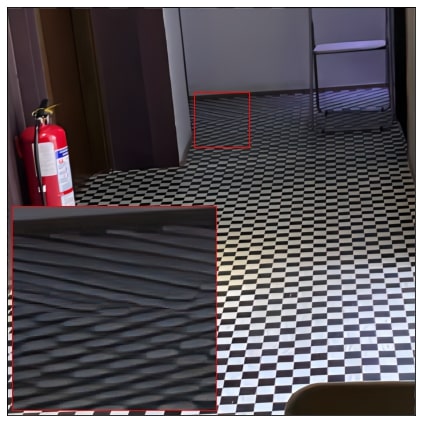}
    \caption[centering]{BSRGAN}
  \end{subfigure}\hfill%
  \begin{subfigure}{.198\textwidth}
    \setlength{\abovecaptionskip}{0pt}
    \setlength{\belowcaptionskip}{0pt}
    \centering
    \includegraphics[width=\textwidth]{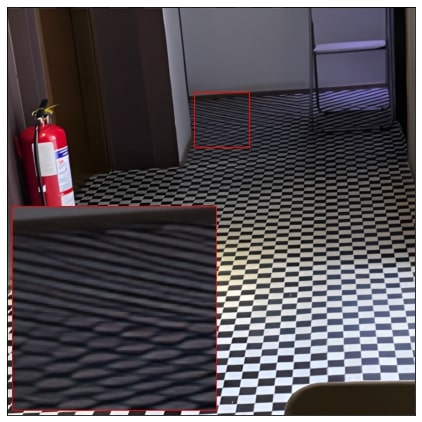}
    \caption[short]{RESRGAN}
  \end{subfigure}\hfill%
  \begin{subfigure}{.198\textwidth}
    \setlength{\abovecaptionskip}{0pt}
    \setlength{\belowcaptionskip}{0pt}
    \centering
    \includegraphics[width=\textwidth]{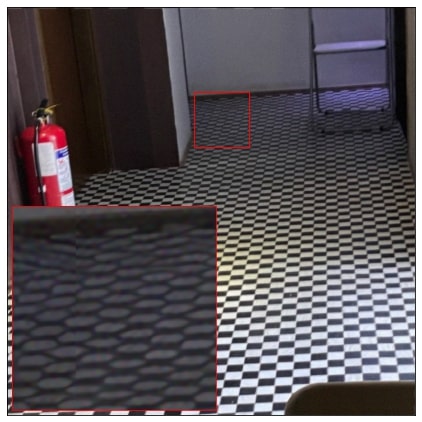}
    \caption[short]{UDKE}
  \end{subfigure}\hfill%
  \caption[short]{$\times 4$ BISR results on a real-world image (better viewed on screen).}
  \label{fig:real4-2}
  \vspace{-1mm}
\end{figure*}

Figs.~\ref{fig:real2}$\sim$\ref{fig:real4-2} visualizes the BISR results on real-world images. Note that KGAN, KGAN-FKP, BSRGAN, and RESRGAN cannot be applied to $3\times$ super-resolution so that their results are not shown for $3\times$ BISR. One can see that UDKE can reproduce much sharper edges and clearer textures without falsely changing their original patterns.

\clearpage

\begin{figure*}[!ht]
  \captionsetup{font=small}

  \captionsetup[subfigure]{justification=centering,font=scriptsize}
  \centering
  \begin{subfigure}{.198\textwidth}
    \setlength{\abovecaptionskip}{0pt}
    \setlength{\belowcaptionskip}{0pt}
    \centering
    \includegraphics[width=\textwidth]{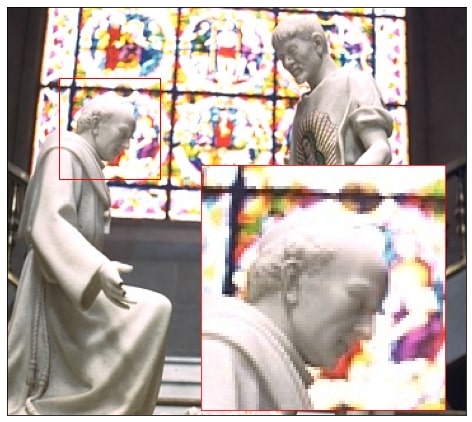}
    \caption[short]{Ground truth\\ \ }
  \end{subfigure}\hfill%
  \begin{subfigure}{.198\textwidth}
    \setlength{\abovecaptionskip}{0pt}
    \setlength{\belowcaptionskip}{0pt}
    \centering
    \includegraphics[width=\textwidth]{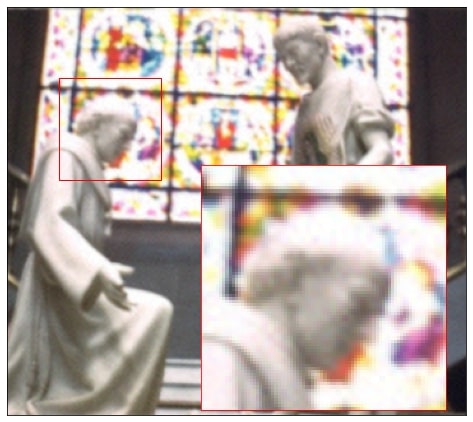}
    \caption[centering]{RCAN \\(19.05dB/.6871)}
  \end{subfigure}\hfill%
  \begin{subfigure}{.198\textwidth}
    \setlength{\abovecaptionskip}{0pt}
    \setlength{\belowcaptionskip}{0pt}
    \centering
    \includegraphics[width=\textwidth]{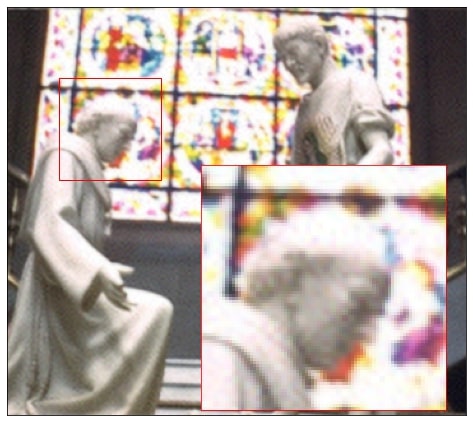}
    \caption[centering]{ZSSR \\(18.92dB/.6489)}
  \end{subfigure}\hfill%
  \begin{subfigure}{.198\textwidth}
    \setlength{\abovecaptionskip}{0pt}
    \setlength{\belowcaptionskip}{0pt}
    \centering
    \includegraphics[width=\textwidth]{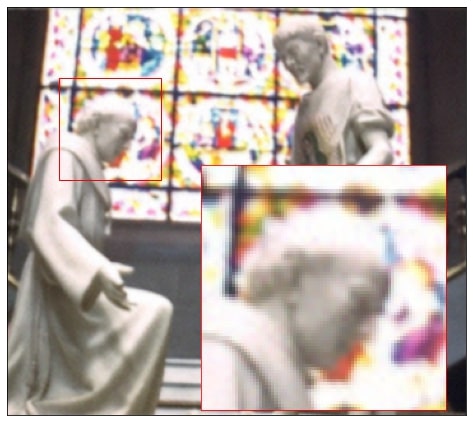}
    \caption[centering]{IKC \\(20.06dB/.7339)}
  \end{subfigure}\hfill%
  \begin{subfigure}{.198\textwidth}
    \setlength{\abovecaptionskip}{0pt}
    \setlength{\belowcaptionskip}{0pt}
    \centering
    \includegraphics[width=\textwidth]{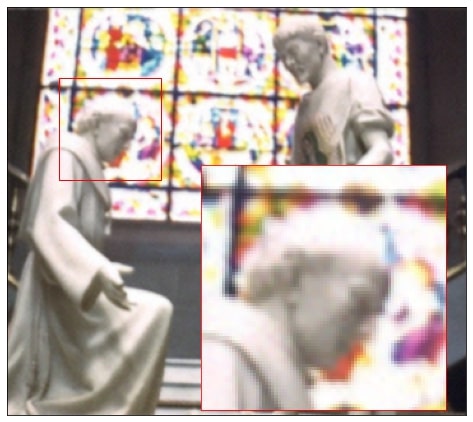}
    \caption[short]{DAN \\(21.90dB/.6748)}
  \end{subfigure}\hfill%
  \begin{subfigure}{.198\textwidth}
    \setlength{\abovecaptionskip}{0pt}
    \setlength{\belowcaptionskip}{0pt}
    \centering
    \includegraphics[width=\textwidth]{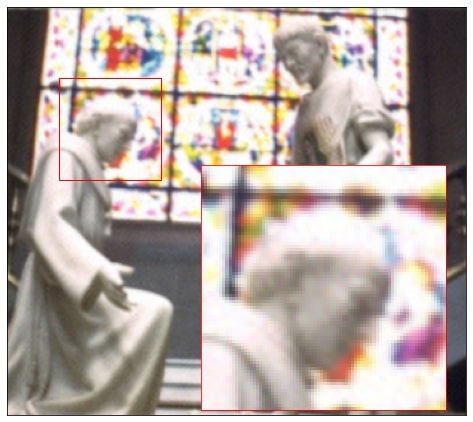}
    \caption[short]{DASR \\(22.05dB/.7166)}
  \end{subfigure}\hfill%
  \begin{subfigure}{.198\textwidth}
    \setlength{\abovecaptionskip}{0pt}
    \setlength{\belowcaptionskip}{0pt}
    \centering
    \includegraphics[width=\textwidth]{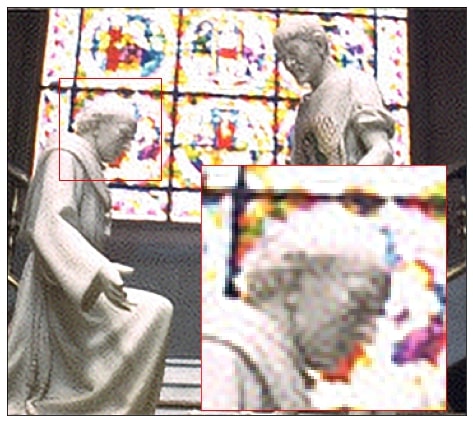}
    \caption[short]{KGAN \\(17.74dB/.5301)}
  \end{subfigure}\hfill%
  \begin{subfigure}{.198\textwidth}
    \setlength{\abovecaptionskip}{0pt}
    \setlength{\belowcaptionskip}{0pt}
    \centering
    \includegraphics[width=\textwidth]{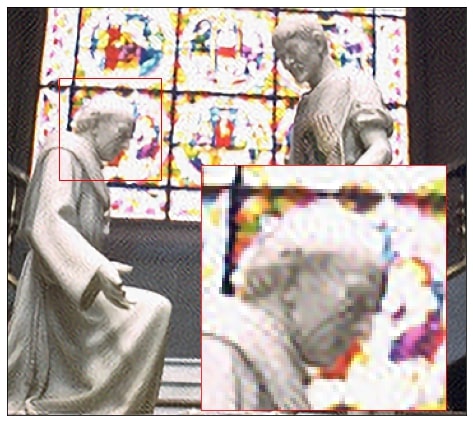}
    \caption[short]{KFKP \\(21.82dB/.7122)}
  \end{subfigure}\hfill%
  \begin{subfigure}{.198\textwidth}
    \setlength{\abovecaptionskip}{0pt}
    \setlength{\belowcaptionskip}{0pt}
    \centering
    \includegraphics[width=\textwidth]{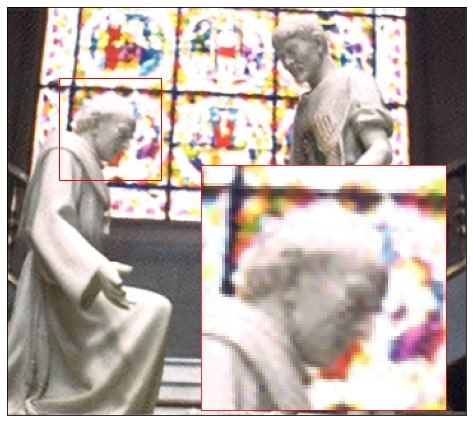}
    \caption[centering]{DFKP \\(22.70dB/.8120)}
  \end{subfigure}\hfill%
  \begin{subfigure}{.198\textwidth}
    \setlength{\abovecaptionskip}{0pt}
    \setlength{\belowcaptionskip}{0pt}
    \centering
    \includegraphics[width=\textwidth]{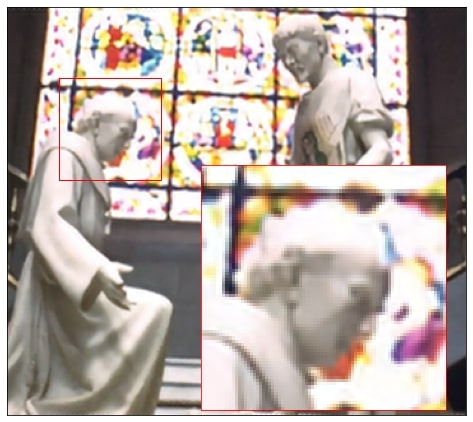}
    \caption[short]{UDKE \\(23.82dB/.8471)}
  \end{subfigure}\hfill%
  \caption[short]{BISR results (PSNR/SSIM) on ``24077'' in BSD100 with $s=2$ and $\sigma=2.55$.}
  \label{fig:2}
  \vspace{-2mm}
\end{figure*}

\begin{figure*}[!ht]
  \captionsetup{font=small}

  \captionsetup[subfigure]{justification=centering,font=scriptsize}
  \centering
  \begin{subfigure}{.22\textwidth}
    \setlength{\abovecaptionskip}{0pt}
    \setlength{\belowcaptionskip}{0pt}
    \centering
    \includegraphics[width=\textwidth]{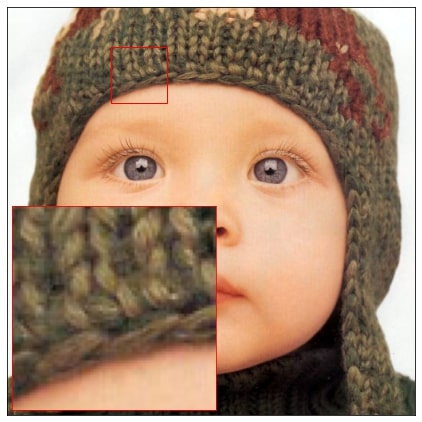}
    \caption[short]{Ground truth\\ \ }
  \end{subfigure}\hfill%
  \begin{subfigure}{.22\textwidth}
    \setlength{\abovecaptionskip}{0pt}
    \setlength{\belowcaptionskip}{0pt}
    \centering
    \includegraphics[width=\textwidth]{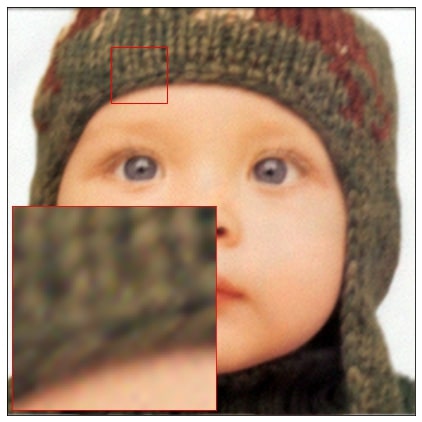}
    \caption[centering]{RCAN \\(27.03dB/.7888)}
  \end{subfigure}\hfill%
  \begin{subfigure}{.22\textwidth}
    \setlength{\abovecaptionskip}{0pt}
    \setlength{\belowcaptionskip}{0pt}
    \centering
    \includegraphics[width=\textwidth]{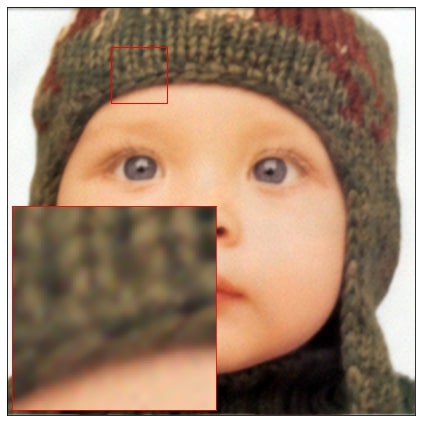}
    \caption[centering]{ZSSR \\(28.21dB/.8267)}
  \end{subfigure}\hfill%
  \begin{subfigure}{.22\textwidth}
    \setlength{\abovecaptionskip}{0pt}
    \setlength{\belowcaptionskip}{0pt}
    \centering
    \includegraphics[width=\textwidth]{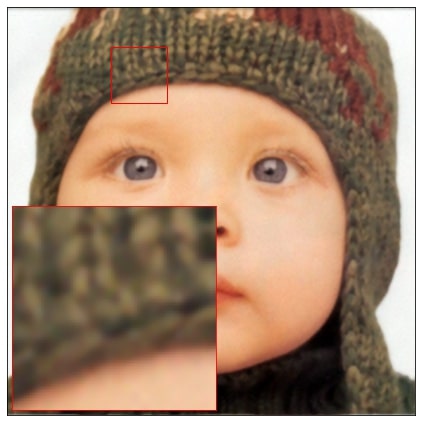}
    \caption[centering]{IKC \\(28.35dB/.8547)}
  \end{subfigure}\hfill%
  \begin{subfigure}{.22\textwidth}
    \setlength{\abovecaptionskip}{0pt}
    \setlength{\belowcaptionskip}{0pt}
    \centering
    \includegraphics[width=\textwidth]{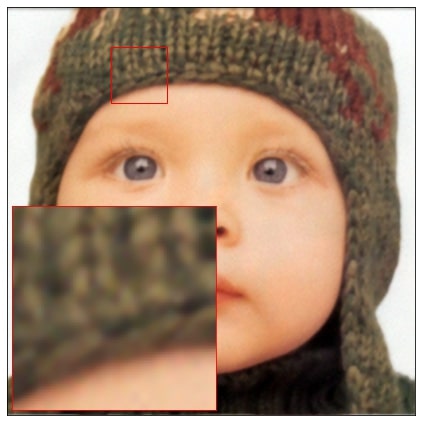}
    \caption[short]{DAN \\(28.79dB/.8418)}
  \end{subfigure}\hfill%
  \begin{subfigure}{.22\textwidth}
    \setlength{\abovecaptionskip}{0pt}
    \setlength{\belowcaptionskip}{0pt}
    \centering
    \includegraphics[width=\textwidth]{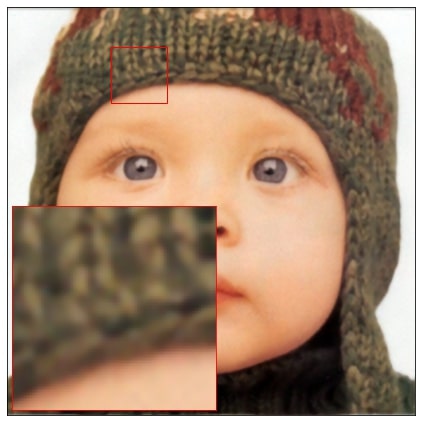}
    \caption[short]{DASR \\(28.86dB/.8065)}
  \end{subfigure}\hfill%
  \begin{subfigure}{.22\textwidth}
    \setlength{\abovecaptionskip}{0pt}
    \setlength{\belowcaptionskip}{0pt}
    \centering
    \includegraphics[width=\textwidth]{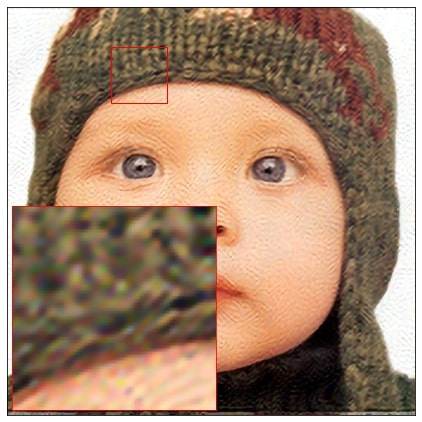}
    \caption[centering]{DFKP \\(24.61dB/.6229)}
  \end{subfigure}\hfill%
  \begin{subfigure}{.22\textwidth}
    \setlength{\abovecaptionskip}{0pt}
    \setlength{\belowcaptionskip}{0pt}
    \centering
    \includegraphics[width=\textwidth]{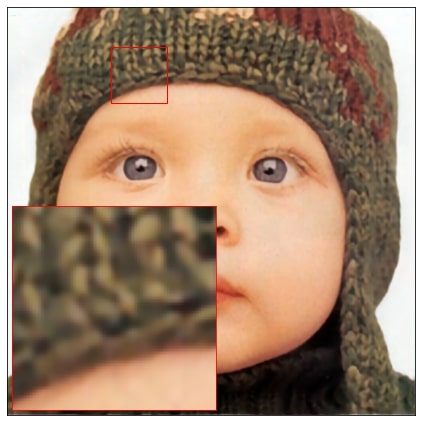}
    \caption[short]{UDKE \\(30.17dB/.8755)}
  \end{subfigure}\hfill%
  \caption[short]{BISR results (PSNR/SSIM) on ``baby'' in Set5 with $s=3$ and $\sigma=2.55$).}
  \label{fig:3}
  \vspace{-2mm}
\end{figure*}

\begin{figure*}[!ht]
  \captionsetup{font=small}
  \captionsetup[subfigure]{justification=centering,font=scriptsize}
  \centering
  \begin{subfigure}{.198\textwidth}
    \setlength{\abovecaptionskip}{0pt}
    \setlength{\belowcaptionskip}{0pt}
    \centering
    \includegraphics[width=\textwidth]{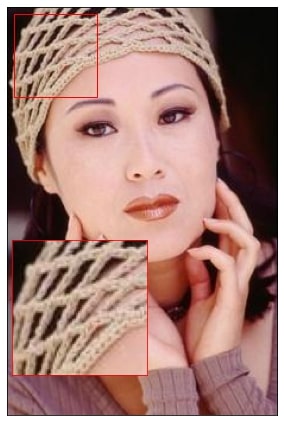}
    \caption[short]{Ground truth\\ \ }
  \end{subfigure}\hfill%
  \begin{subfigure}{.198\textwidth}
    \setlength{\abovecaptionskip}{0pt}
    \setlength{\belowcaptionskip}{0pt}
    \centering
    \includegraphics[width=\textwidth]{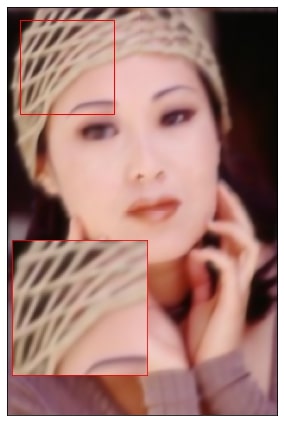}
    \caption[centering]{RCAN \\(18.45dB/.6491)}
  \end{subfigure}\hfill%
  \begin{subfigure}{.198\textwidth}
    \setlength{\abovecaptionskip}{0pt}
    \setlength{\belowcaptionskip}{0pt}
    \centering
    \includegraphics[width=\textwidth]{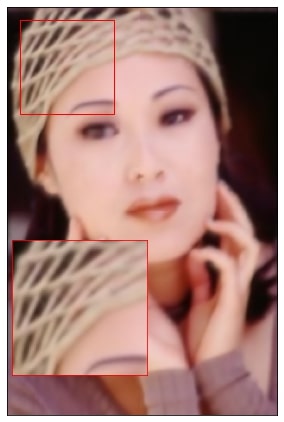}
    \caption[centering]{ZSSR \\(18.80dB/.6764)}
  \end{subfigure}\hfill%
  \begin{subfigure}{.198\textwidth}
    \setlength{\abovecaptionskip}{0pt}
    \setlength{\belowcaptionskip}{0pt}
    \centering
    \includegraphics[width=\textwidth]{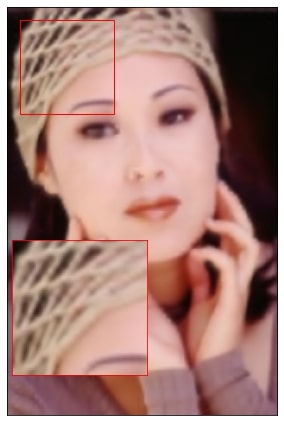}
    \caption[centering]{IKC \\(19.03dB/.6998)}
  \end{subfigure}\hfill%
  \begin{subfigure}{.198\textwidth}
    \setlength{\abovecaptionskip}{0pt}
    \setlength{\belowcaptionskip}{0pt}
    \centering
    \includegraphics[width=\textwidth]{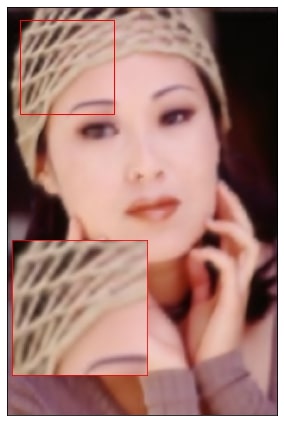}
    \caption[short]{DAN \\(19.67dB/.7021)}
  \end{subfigure}\hfill%
  \begin{subfigure}{.198\textwidth}
    \setlength{\abovecaptionskip}{0pt}
    \setlength{\belowcaptionskip}{0pt}
    \centering
    \includegraphics[width=\textwidth]{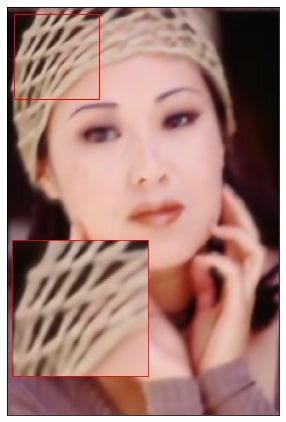}
    \caption[short]{DASR \\(20.38dB/.7221)}
  \end{subfigure}\hfill%
  \begin{subfigure}{.198\textwidth}
    \setlength{\abovecaptionskip}{0pt}
    \setlength{\belowcaptionskip}{0pt}
    \centering
    \includegraphics[width=\textwidth]{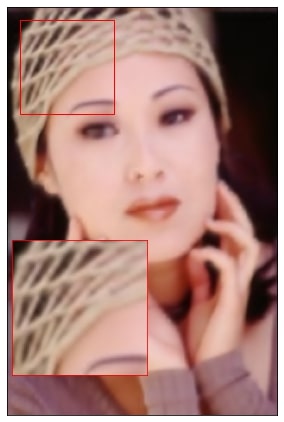}
    \caption[short]{KGAN \\(18.95dB/.6888)}
  \end{subfigure}\hfill%
  \begin{subfigure}{.198\textwidth}
    \setlength{\abovecaptionskip}{0pt}
    \setlength{\belowcaptionskip}{0pt}
    \centering
    \includegraphics[width=\textwidth]{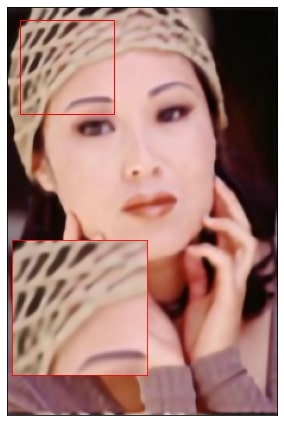}
    \caption[short]{KFKP \\(18.93dB/.6883)}
  \end{subfigure}\hfill%
  \begin{subfigure}{.198\textwidth}
    \setlength{\abovecaptionskip}{0pt}
    \setlength{\belowcaptionskip}{0pt}
    \centering
    \includegraphics[width=\textwidth]{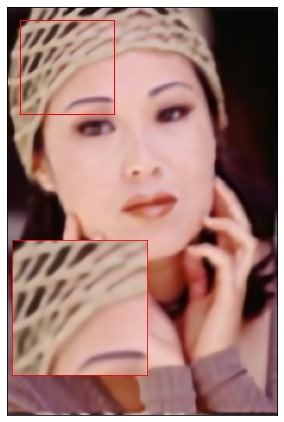}
    \caption[centering]{DFKP \\(19.31dB/.6948)}
  \end{subfigure}\hfill%
  \begin{subfigure}{.198\textwidth}
    \setlength{\abovecaptionskip}{0pt}
    \setlength{\belowcaptionskip}{0pt}
    \centering
    \includegraphics[width=\textwidth]{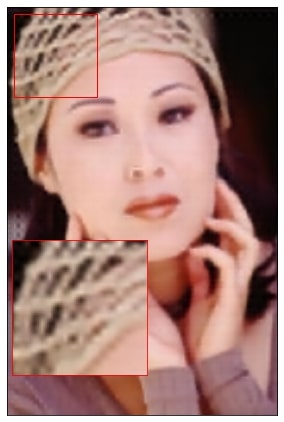}
    \caption[short]{UDKE \\(24.47dB/.8083)}
  \end{subfigure}\hfill%
  \caption[short]{BISR results (PSNR/SSIM) on ``woman'' in Set5 with $s=4$ and $\sigma=0$.}
  \label{fig:6}
  \vspace{-2mm}
\end{figure*}

\begin{figure*}[!ht]
  \captionsetup{font=small}
  \captionsetup[subfigure]{justification=centering,font=scriptsize}
  \centering
  \begin{subfigure}{.198\textwidth}
    \setlength{\abovecaptionskip}{0pt}
    \setlength{\belowcaptionskip}{0pt}
    \centering
    \includegraphics[width=\textwidth]{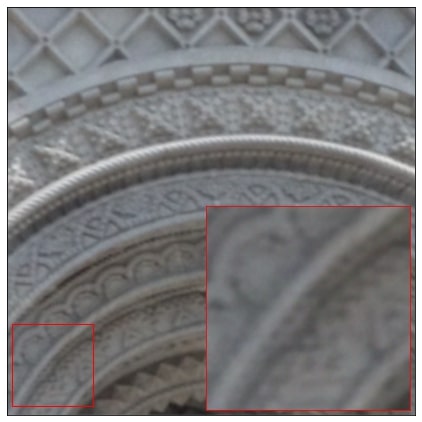}
    \caption[short]{LR image}
  \end{subfigure}\hfill%
  \begin{subfigure}{.198\textwidth}
    \setlength{\abovecaptionskip}{0pt}
    \setlength{\belowcaptionskip}{0pt}
    \centering
    \includegraphics[width=\textwidth]{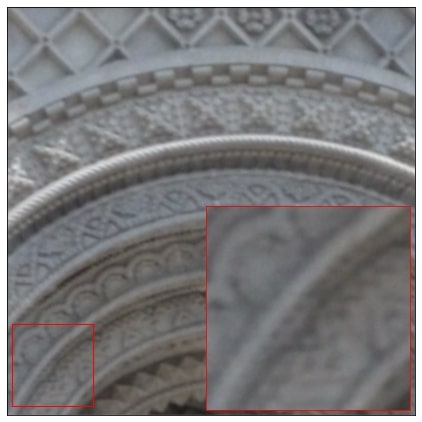}
    \caption[short]{RCAN}
  \end{subfigure}\hfill%
  \begin{subfigure}{.198\textwidth}
    \setlength{\abovecaptionskip}{0pt}
    \setlength{\belowcaptionskip}{0pt}
    \centering
    \includegraphics[width=\textwidth]{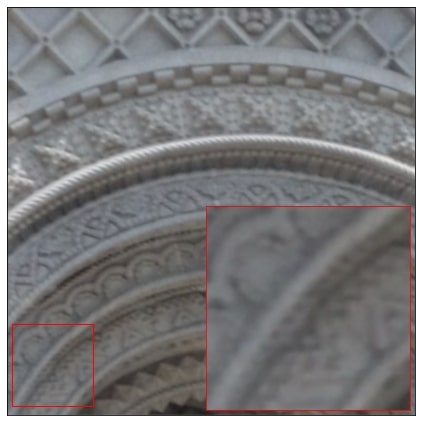}
    \caption[centering]{IKC}
  \end{subfigure}\hfill%
  \begin{subfigure}{.198\textwidth}
    \setlength{\abovecaptionskip}{0pt}
    \setlength{\belowcaptionskip}{0pt}
    \centering
    \includegraphics[width=\textwidth]{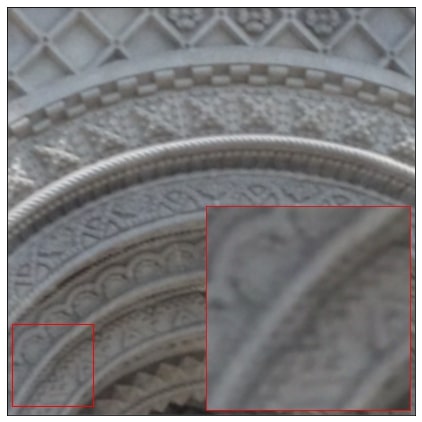}
    \caption[centering]{DAN}
  \end{subfigure}\hfill%
  \begin{subfigure}{.198\textwidth}
    \setlength{\abovecaptionskip}{0pt}
    \setlength{\belowcaptionskip}{0pt}
    \centering
    \includegraphics[width=\textwidth]{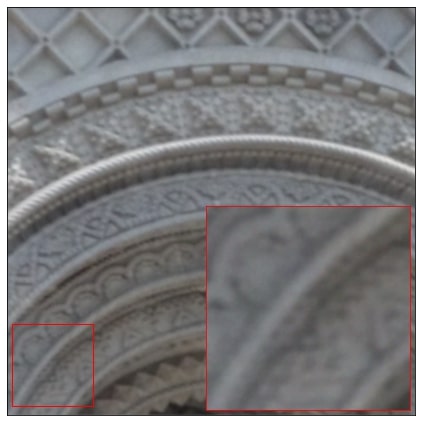}
    \caption[centering]{DASR}
  \end{subfigure}\hfill%
  \begin{subfigure}{.198\textwidth}
    \setlength{\abovecaptionskip}{0pt}
    \setlength{\belowcaptionskip}{0pt}
    \centering
    \includegraphics[width=\textwidth]{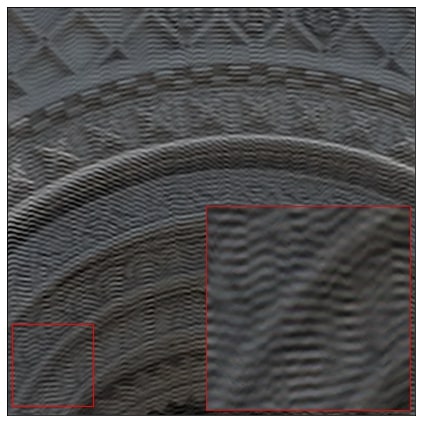}
    \caption[short]{KFKP}
  \end{subfigure}\hfill%
  \begin{subfigure}{.198\textwidth}
    \setlength{\abovecaptionskip}{0pt}
    \setlength{\belowcaptionskip}{0pt}
    \centering
    \includegraphics[width=\textwidth]{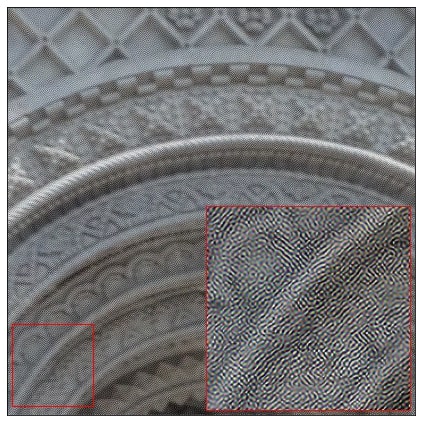}
    \caption[short]{DFKP}
  \end{subfigure}\hfill%
  \begin{subfigure}{.198\textwidth}
    \setlength{\abovecaptionskip}{0pt}
    \setlength{\belowcaptionskip}{0pt}
    \centering
    \includegraphics[width=\textwidth]{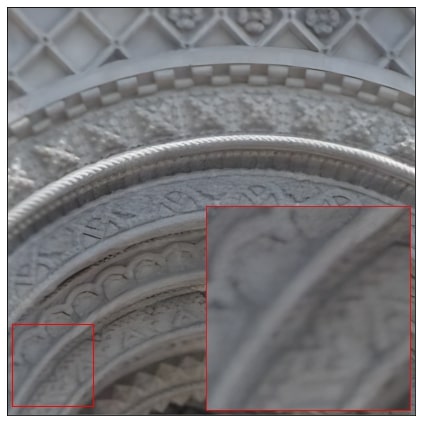}
    \caption[centering]{BSRGAN}
  \end{subfigure}\hfill%
  \begin{subfigure}{.198\textwidth}
    \setlength{\abovecaptionskip}{0pt}
    \setlength{\belowcaptionskip}{0pt}
    \centering
    \includegraphics[width=\textwidth]{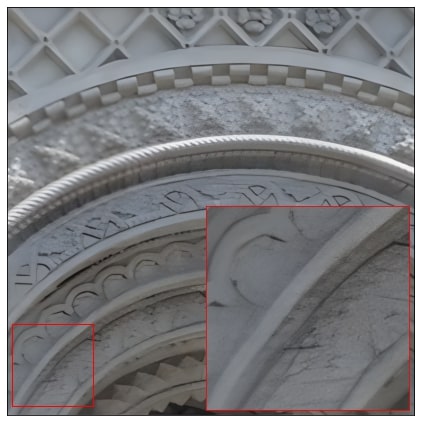}
    \caption[short]{RESRGAN}
  \end{subfigure}\hfill%
  \begin{subfigure}{.198\textwidth}
    \setlength{\abovecaptionskip}{0pt}
    \setlength{\belowcaptionskip}{0pt}
    \centering
    \includegraphics[width=\textwidth]{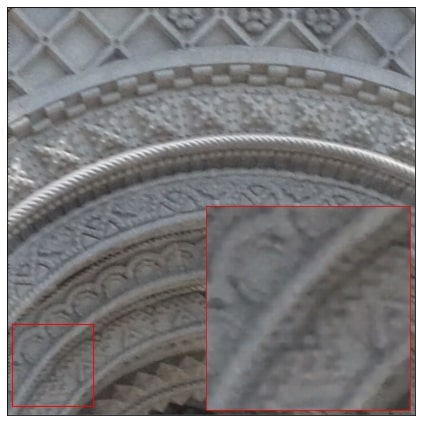}
    \caption[short]{UDKE}
  \end{subfigure}\hfill%
  \caption[short]{$\times 2$ BISR results on a real-world image (better viewed on screen).}
  \label{fig:real2-2}
  \vspace{-1mm}
\end{figure*}

\begin{figure*}[!ht]
  \captionsetup{font=small}
  \captionsetup[subfigure]{justification=centering,font=scriptsize}
  \centering
  \begin{subfigure}{.22\textwidth}
    \setlength{\abovecaptionskip}{0pt}
    \setlength{\belowcaptionskip}{0pt}
    \centering
    \includegraphics[width=\textwidth]{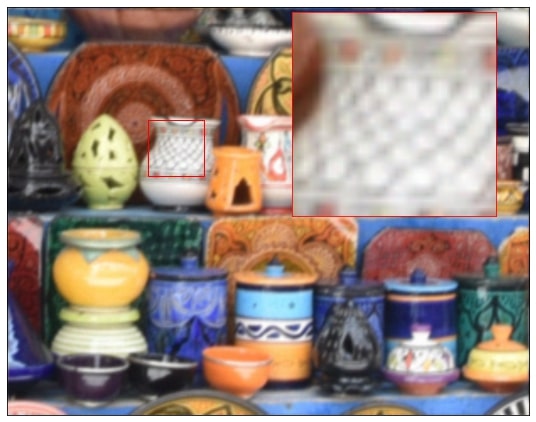}
    \caption[short]{LR image}
  \end{subfigure}\hfill%
  \begin{subfigure}{.22\textwidth}
    \setlength{\abovecaptionskip}{0pt}
    \setlength{\belowcaptionskip}{0pt}
    \centering
    \includegraphics[width=\textwidth]{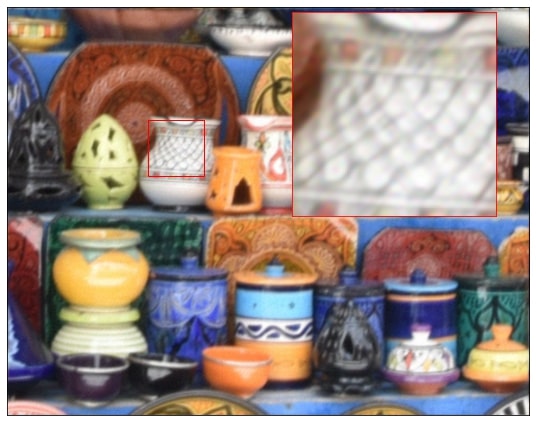}
    \caption[short]{RCAN}
  \end{subfigure}\hfill%
  \begin{subfigure}{.22\textwidth}
    \setlength{\abovecaptionskip}{0pt}
    \setlength{\belowcaptionskip}{0pt}
    \centering
    \includegraphics[width=\textwidth]{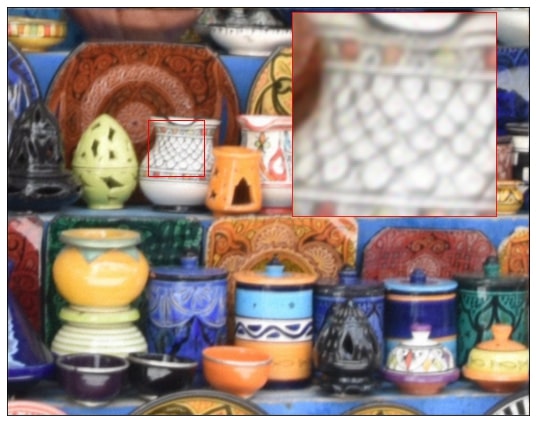}
    \caption[short]{ZSSR}
  \end{subfigure}\hfill%
  \begin{subfigure}{.22\textwidth}
    \setlength{\abovecaptionskip}{0pt}
    \setlength{\belowcaptionskip}{0pt}
    \centering
    \includegraphics[width=\textwidth]{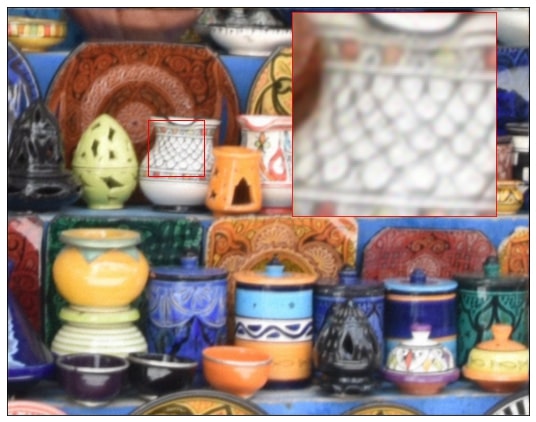}
    \caption[centering]{IKC}
  \end{subfigure}\hfill%
  \begin{subfigure}{.22\textwidth}
    \setlength{\abovecaptionskip}{0pt}
    \setlength{\belowcaptionskip}{0pt}
    \centering
    \includegraphics[width=\textwidth]{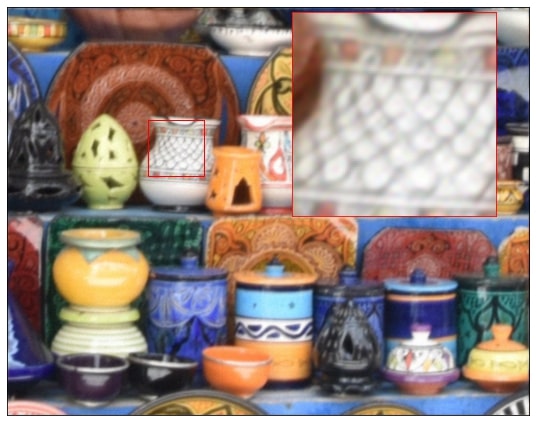}
    \caption[centering]{DAN}
  \end{subfigure}\hfill%
  \begin{subfigure}{.22\textwidth}
    \setlength{\abovecaptionskip}{0pt}
    \setlength{\belowcaptionskip}{0pt}
    \centering
    \includegraphics[width=\textwidth]{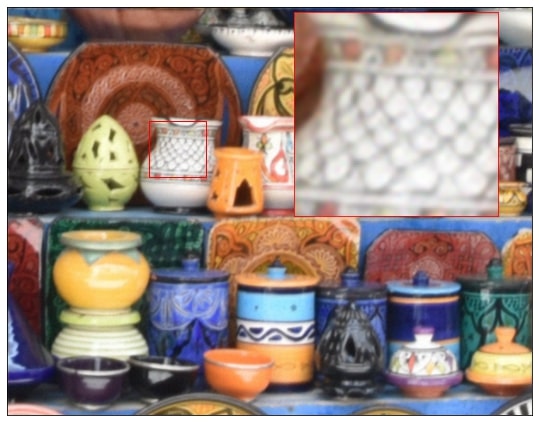}
    \caption[centering]{DASR}
  \end{subfigure}\hfill%
  \begin{subfigure}{.22\textwidth}
    \setlength{\abovecaptionskip}{0pt}
    \setlength{\belowcaptionskip}{0pt}
    \centering
    \includegraphics[width=\textwidth]{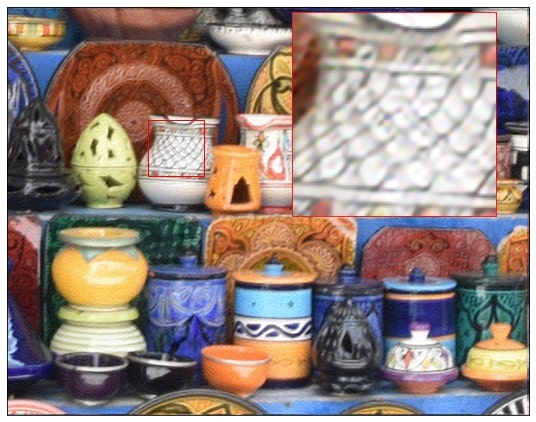}
    \caption[short]{DFKP}
  \end{subfigure}\hfill%
  \begin{subfigure}{.22\textwidth}
    \setlength{\abovecaptionskip}{0pt}
    \setlength{\belowcaptionskip}{0pt}
    \centering
    \includegraphics[width=\textwidth]{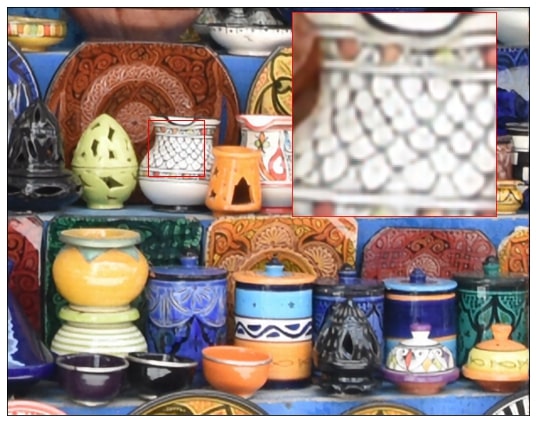}
    \caption[short]{UDKE}
  \end{subfigure}\hfill%
  \caption[short]{$\times 3$ BISR results on a real-world image (better viewed on screen).}
  \label{fig:real3}
  \vspace{-1mm}
\end{figure*}

\begin{figure*}[!ht]
  \captionsetup{font=small}
  \captionsetup[subfigure]{justification=centering,font=scriptsize}
  \centering
  \begin{subfigure}{.22\textwidth}
    \setlength{\abovecaptionskip}{0pt}
    \setlength{\belowcaptionskip}{0pt}
    \centering
    \includegraphics[width=\textwidth]{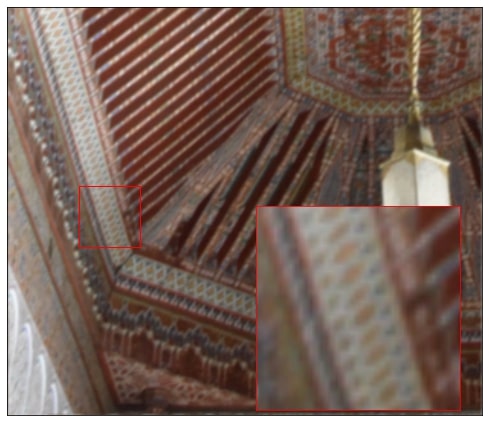}
    \caption[short]{LR image}
  \end{subfigure}\hfill%
  \begin{subfigure}{.22\textwidth}
    \setlength{\abovecaptionskip}{0pt}
    \setlength{\belowcaptionskip}{0pt}
    \centering
    \includegraphics[width=\textwidth]{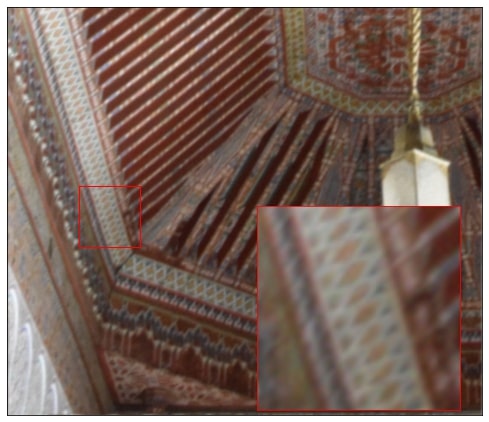}
    \caption[short]{RCAN}
  \end{subfigure}\hfill%
  \begin{subfigure}{.22\textwidth}
    \setlength{\abovecaptionskip}{0pt}
    \setlength{\belowcaptionskip}{0pt}
    \centering
    \includegraphics[width=\textwidth]{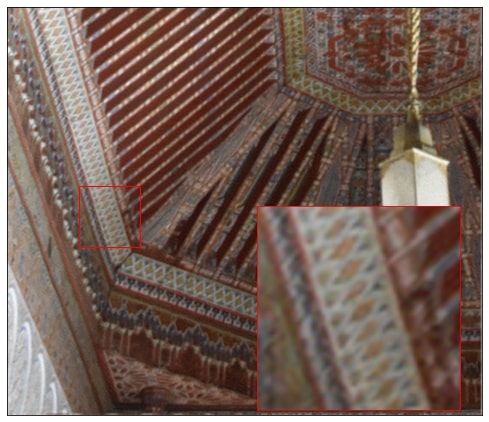}
    \caption[short]{ZSSR}
  \end{subfigure}\hfill%
  \begin{subfigure}{.22\textwidth}
    \setlength{\abovecaptionskip}{0pt}
    \setlength{\belowcaptionskip}{0pt}
    \centering
    \includegraphics[width=\textwidth]{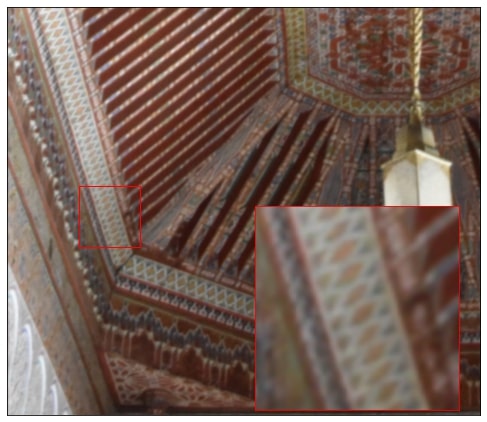}
    \caption[centering]{IKC}
  \end{subfigure}\hfill%
  \begin{subfigure}{.22\textwidth}
    \setlength{\abovecaptionskip}{0pt}
    \setlength{\belowcaptionskip}{0pt}
    \centering
    \includegraphics[width=\textwidth]{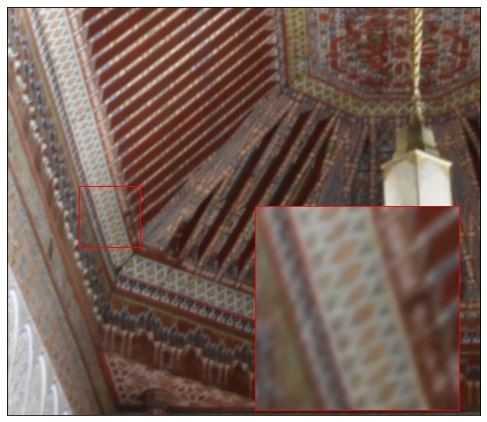}
    \caption[centering]{DAN}
  \end{subfigure}\hfill%
  \begin{subfigure}{.22\textwidth}
    \setlength{\abovecaptionskip}{0pt}
    \setlength{\belowcaptionskip}{0pt}
    \centering
    \includegraphics[width=\textwidth]{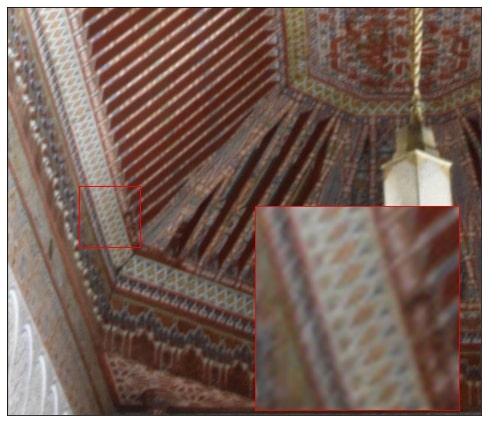}
    \caption[centering]{DASR}
  \end{subfigure}\hfill%
  \begin{subfigure}{.22\textwidth}
    \setlength{\abovecaptionskip}{0pt}
    \setlength{\belowcaptionskip}{0pt}
    \centering
    \includegraphics[width=\textwidth]{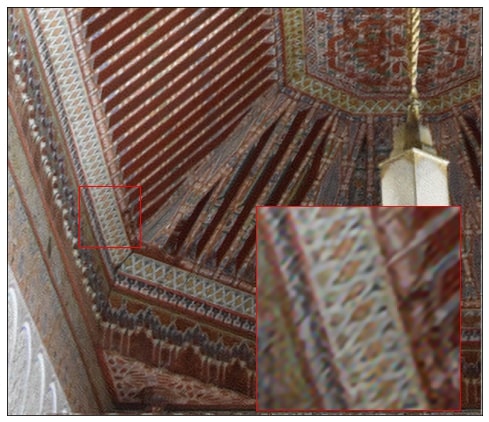}
    \caption[short]{DFKP}
  \end{subfigure}\hfill%
  \begin{subfigure}{.22\textwidth}
    \setlength{\abovecaptionskip}{0pt}
    \setlength{\belowcaptionskip}{0pt}
    \centering
    \includegraphics[width=\textwidth]{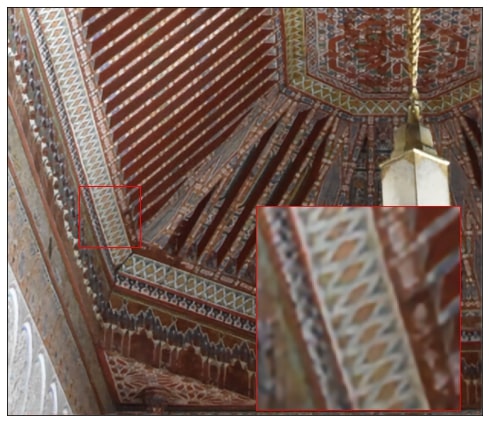}
    \caption[short]{UDKE}
  \end{subfigure}\hfill%
  \caption[short]{$\times 3$ BISR results on a real-world image (better viewed on screen).}
  \label{fig:real3-2}
  \vspace{-1mm}
\end{figure*}

\begin{figure*}[!ht]
  \captionsetup{font=small}
  \captionsetup[subfigure]{justification=centering,font=scriptsize}
  \centering
  \begin{subfigure}{.198\textwidth}
    \setlength{\abovecaptionskip}{0pt}
    \setlength{\belowcaptionskip}{0pt}
    \centering
    \includegraphics[width=\textwidth]{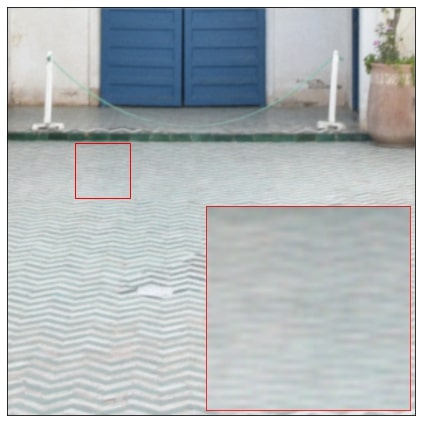}
    \caption[short]{LR image}
  \end{subfigure}\hfill%
  \begin{subfigure}{.198\textwidth}
    \setlength{\abovecaptionskip}{0pt}
    \setlength{\belowcaptionskip}{0pt}
    \centering
    \includegraphics[width=\textwidth]{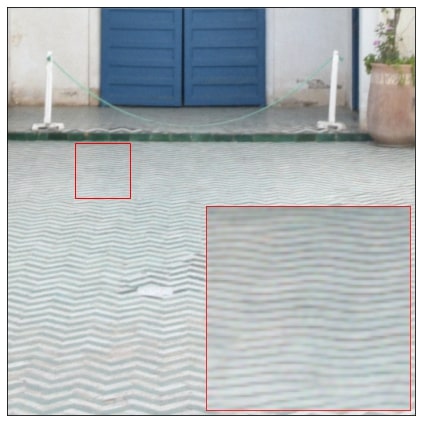}
    \caption[short]{RCAN}
  \end{subfigure}\hfill%
  \begin{subfigure}{.198\textwidth}
    \setlength{\abovecaptionskip}{0pt}
    \setlength{\belowcaptionskip}{0pt}
    \centering
    \includegraphics[width=\textwidth]{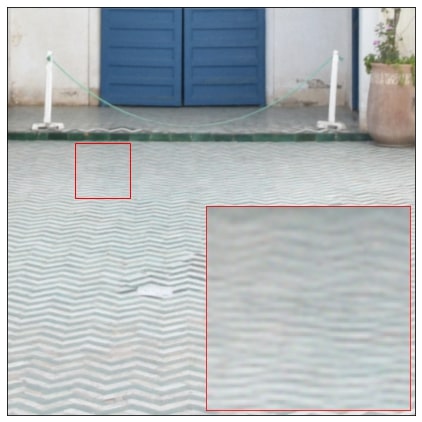}
    \caption[centering]{IKC}
  \end{subfigure}\hfill%
  \begin{subfigure}{.198\textwidth}
    \setlength{\abovecaptionskip}{0pt}
    \setlength{\belowcaptionskip}{0pt}
    \centering
    \includegraphics[width=\textwidth]{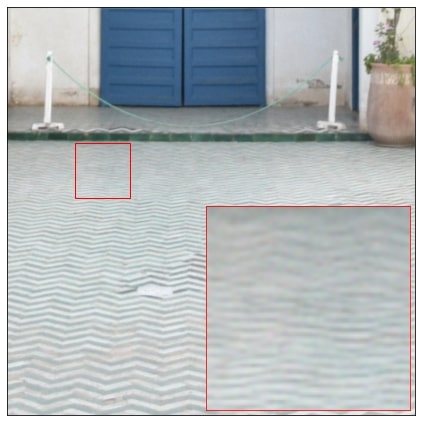}
    \caption[centering]{DAN}
  \end{subfigure}\hfill%
  \begin{subfigure}{.198\textwidth}
    \setlength{\abovecaptionskip}{0pt}
    \setlength{\belowcaptionskip}{0pt}
    \centering
    \includegraphics[width=\textwidth]{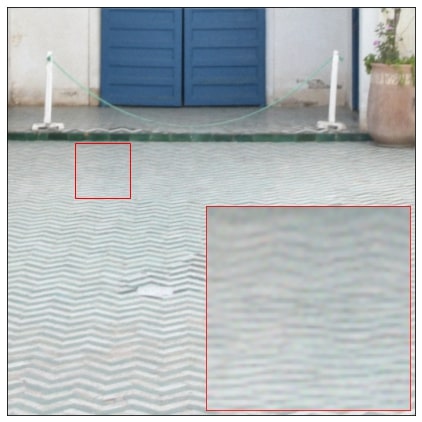}
    \caption[centering]{DASR}
  \end{subfigure}\hfill%
  \begin{subfigure}{.198\textwidth}
    \setlength{\abovecaptionskip}{0pt}
    \setlength{\belowcaptionskip}{0pt}
    \centering
    \includegraphics[width=\textwidth]{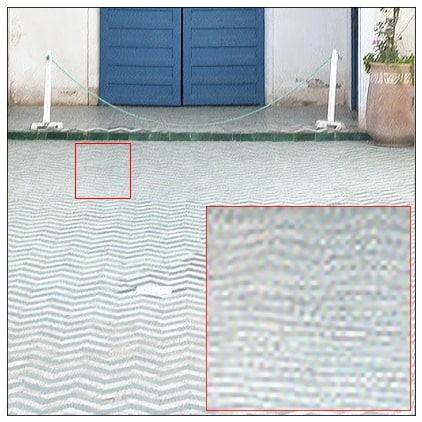}
    \caption[short]{KFKP}
  \end{subfigure}\hfill%
  \begin{subfigure}{.198\textwidth}
    \setlength{\abovecaptionskip}{0pt}
    \setlength{\belowcaptionskip}{0pt}
    \centering
    \includegraphics[width=\textwidth]{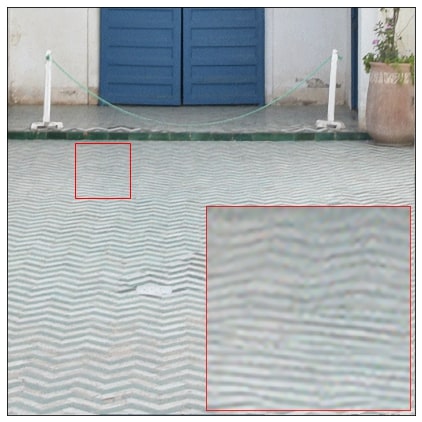}
    \caption[short]{DFKP}
  \end{subfigure}\hfill%
  \begin{subfigure}{.198\textwidth}
    \setlength{\abovecaptionskip}{0pt}
    \setlength{\belowcaptionskip}{0pt}
    \centering
    \includegraphics[width=\textwidth]{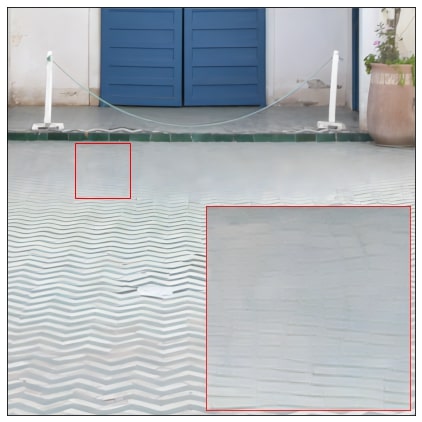}
    \caption[centering]{BSRGAN}
  \end{subfigure}\hfill%
  \begin{subfigure}{.198\textwidth}
    \setlength{\abovecaptionskip}{0pt}
    \setlength{\belowcaptionskip}{0pt}
    \centering
    \includegraphics[width=\textwidth]{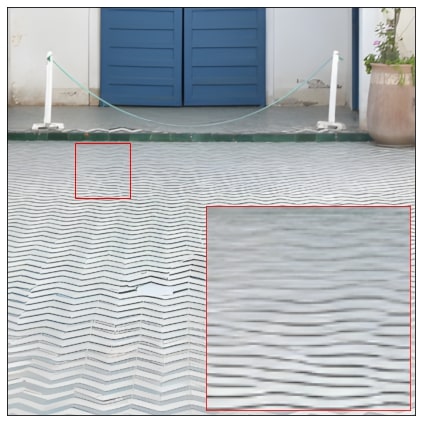}
    \caption[short]{RESRGAN}
  \end{subfigure}\hfill%
  \begin{subfigure}{.198\textwidth}
    \setlength{\abovecaptionskip}{0pt}
    \setlength{\belowcaptionskip}{0pt}
    \centering
    \includegraphics[width=\textwidth]{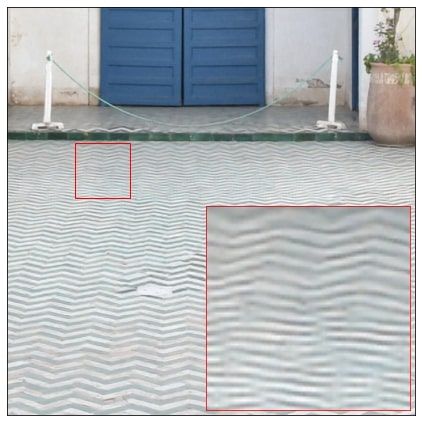}
    \caption[short]{UDKE}
  \end{subfigure}\hfill%
  \caption[short]{$\times 4$ BISR results on a real-world image (better viewed on screen).}
  \label{fig:real4}
  \vspace{-1mm}
\end{figure*}

\end{document}